\documentclass[prx,twocolumn,superscriptaddress,showpacs,notitlepage,floatfix,nofootinbib]{revtex4-2}
\usepackage{amsmath,amsfonts, amssymb, amsthm, dsfont}
\usepackage{yfonts}
\usepackage{bm}
\usepackage{mathrsfs}
\usepackage{graphicx}
\usepackage{verbatim}
\usepackage{wasysym}
\usepackage{xcolor}
\usepackage{tikz}
\usepackage{physics}
\usepackage{floatrow}
\usepackage{diagbox}
\usepackage{footmisc}
\usepackage{ulem}
\usepackage[caption=false,label font={bf,normalsize}]{subfig}
\AtBeginDocument{\tabcolsep=7pt} 
\usepackage{booktabs}

\newcommand{\comments}[1]{}

\newlength{\fighskip} \fighskip=2pt
\newlength{\figvskip} \figvskip=3pt

\makeatletter
\def\l@subsubsection#1#2{}
\makeatother

\definecolor{new}{rgb}{.08,.05,.8}

\usepackage[pdfpagelabels,pdftex,bookmarks,breaklinks]{hyperref}
\definecolor{darkblue}{RGB}{0,0,127} 
\definecolor{darkgreen}{RGB}{0,150,0}
\hypersetup{
	colorlinks, 
	linkcolor=darkblue, 
	citecolor=darkgreen, 
	filecolor=blue, 
	urlcolor=blue,
  pdfauthor = {},
  pdftitle = {}
}

\newcommand{\ZZ}{\mathbb{Z}}

\begin{document}
\title{Engineering 3D Floquet codes by rewinding}

\author{Arpit Dua}
\email{adua@caltech.edu}
\affiliation{Department of Physics, California Institute of Technology, Pasadena, CA 91125, USA}
\affiliation{Institute for Quantum Information and Matter, California Institute of Technology, Pasadena, California 91125, USA}
\author{Nathanan Tantivasadakarn}
\affiliation{Walter Burke Institute for Theoretical Physics, California Institute of Technology, Pasadena, CA 91125, USA}
\affiliation{Department of Physics, California Institute of Technology, Pasadena, CA 91125, USA}
\affiliation{Department of Physics, Harvard University, Cambridge, MA 02138, USA}
\author{Joseph Sullivan}
\affiliation{Stewart Blusson Quantum Matter Institute, University of British Columbia, Vancouver, BC, Canada V6T 1Z1}
\affiliation{Department of Physics, Yale University, New Haven, CT 06511, USA}
\author{Tyler~D. Ellison}
\email{tyler.ellison@yale.edu}
\affiliation{Department of Physics, Yale University, New Haven, CT 06511, USA}

\definecolor{arpit}{RGB}{200,0,0}
\newcommand{\arpit}[1]{{\color{arpit}{#1}}}

\begin{abstract}
{Floquet codes are a novel class of quantum error-correcting codes with dynamically generated logical qubits arising from a periodic schedule of non-commuting measurements. We utilize the interpretation of measurements in terms of condensation of topological excitations and the rewinding of measurement sequences to engineer new examples of Floquet codes. In particular, rewinding is advantageous for obtaining a desired set of instantaneous stabilizer groups on both toric and planar layouts. Our first example is a Floquet code with instantaneous stabilizer codes that have the same topological order as 3D toric code(s). 
This Floquet code also exhibits a splitting of the topological order of the 3D toric code under the associated sequence of measurements, i.e., an instantaneous stabilizer group of a single copy of 3D toric code in one round transforms into an instantaneous stabilizer group of two copies of 3D toric codes up to nonlocal stabilizers in the following round. We further construct boundaries for this 3D code and argue that stacking it with two copies of 3D subsystem toric code allows for a transversal implementation of the logical non-Clifford $CCZ$ gate. We also show that the coupled-layer construction of the X-cube Floquet code can be modified by a rewinding schedule such that each of the instantaneous stabilizer codes is finite-depth-equivalent to the X-cube model up to toric codes; the X-cube Floquet code exhibits a splitting of the X-cube model into a copy of the X-cube model and toric codes under the measurement sequence. Our final 3D example is a generalization of the 2D Floquet toric code on the honeycomb lattice to 3D, which has instantaneous stabilizer codes with the same topological order as the 3D fermionic toric code. }
\end{abstract}

\maketitle
\date{}

\tableofcontents

\section{Introduction}
\label{sec:intro}
Quantum error-correcting codes are a key ingredient for fault-tolerant quantum computation.
There is an active effort to develop new error-correcting codes with better code properties, such as encoding rate, code distance, and circuit-level thresholds. 
For every such error-correcting code, there is an associated quantum dynamics involving quantum gates, errors, and the repeated extraction of the syndrome for decoding. Naturally, the goal 
of developing new error-correcting codes
is to optimize the quantum dynamics to reduce overheads and minimize the noise on the logical information. 

Recently, Hastings and Haah introduced the first example of what has emerged as a new class of codes, dubbed Floquet codes, which exhibit dynamically generated logical qubits~\cite{hhdynamic2021}. In their example, the dynamics of the system is governed by a periodic sequence of non-commuting 2-qubit Pauli measurements and exhibit instantaneous stabilizer codes that are a sequence of topological quantum error-correcting codes. Importantly, the schedule is such that the logical information is preserved from one instantaneous code space to the next.  
Given the low-weight parity checks needed to operate the code and its relatively high error threshold~\cite{Gidney2021faulttolerant, Gidney2022benchmarkingplanar, Paetznick2023Performance}, Floquet codes offer compelling alternatives to the toric code (TC). The current understanding of Floquet codes is still under active development, thus, underscoring the importance of introducing new examples and formalizing the tools to develop new Floquet codes.

In this paper, we introduce new examples of 3D topological Floquet codes, referred to as:
{{(1)} the 3D Floquet toric code (TC), (2) the X-cube Floquet code with a rewinding schedule, and {(3)} the 3D Floquet fermionic toric code (fTC). We also construct an example of a new 2D Floquet code called the Floquet color code, for which we expect the interpretation of measurements as condensation in a parent stabilizer code and rewinding schedules to be useful.  

An essential tool employed in our constructions is the concept of ``rewinding'' a measurement schedule, where the sequence of measurements is reversed at some point within a period. A similar strategy was stated in Ref.~\cite{Haah2022boundarieshoneycomb} to adapt the honeycomb code to a system with boundary. {We adopt a physical description of rewinding in terms of the evolution of measured checks. Firstly, we claim that for every Floquet code discussed in our work, there is an associated nontrivial parent stabilizer code. We can consider the sequence of measurements in a Floquet code as performing a sequence of condensations on this parent stabilizer code. The operators associated with condensation in one step can sometimes survive as finite-weight stabilizers under subsequent measurements; this is the case we are interested in, and we refer to this as the evolution of the condensation checks. Under rewinding, the evolution of condensation checks is reversed before all of them evolve into non-local stabilizers. In our examples of the 3D Floquet TC and X-cube Floquet code, the evolved condensation checks at each step determine the topological order of the instantaneous stabilizer groups (ISGs), and rewinding helps to achieve the desired ISGs by not evolving beyond a certain point and reversing the evolution. 
Sometimes, these non-local stabilizers can be the logical operators of the Floquet code, like in the case of the 2D Floquet TC on a planar layout, and in those cases, rewinding helps to avoid measuring logical information. In summary, our understanding of rewinding in terms of the evolution of condensation checks leads to a physical interpretation for the ISGs and boundaries of Floquet codes constructed using rewinding schedules, including the construction in Ref.~\cite{Haah2022boundarieshoneycomb}.  
Therefore, we argue that rewinding is beneficial for creating a desired set of ISGs on both toric and planar layouts. 
We expect the tools we explicitly utilized for our microscopic constructions of Floquet codes to be useful in constructing a wider class of Floquet codes, including those with quantum low-density parity check (LDPC) codes as ISGs.
}

\textbf{(1)} Our first key example is the 3D Floquet TC, which has ISGs that are equivalent under a finite-depth local quantum circuit with ancilla (FDLQC-equivalent) to the usual 3D TC (or two copies of it). Our construction is inspired by the coupled layer construction of Ref.~\cite{ma2017}, in the sense that we prepare an instantaneous state of stacks of 2D TCs along orthogonal directions, then perform measurements that ``condense''  pairs of $e$ anyons 
along the intersection of two orthogonal layers. In the subsequent rounds of measurements, the stabilizers responsible for this condensation can generically evolve into higher-weight stabilizers. By using a schedule that rewinds, we ensure that not all evolved checks are non-local stabilizers, and that, in turn, ensures that the system does not evolve back into a stack of 2D TCs. This could be useful for specialized decoding tasks such as single-shot decoding of loop-like excitations in the 3D TC ISGs. We note that Ref.~\cite{bauer2023topological} presents a construction of a 3D Floquet TC using the path-integral framework. 

Due to the evolution of checks into higher-weight operators, we observe an exotic splitting of the 3D TC ISG obtained in the preceding round to two copies of 3D TC up to nonlocal stabilizers. We also construct a planar variant of the 3D Floquet TC with boundaries that condense point-like or loop-like excitations for each ISG. Unlike the planar variant of the 2D Floquet TC, the boundaries of the 3D Floquet TC do not undergo an automorphism. 
By stacking this planar variant of the 3D Floquet TC with two copies of the planar variant of ``subsystem toric code''~\cite{kubica2022} (which has 3-qubit checks), we can prepare an instantaneous state of the cubic lattice 3D TC stacked with two copies of 3D checkerboard lattice TC.  
Such a stacked code allows for an implementation of the logical non-Clifford $CCZ$ gate. ~\cite{Vasmer2019transversal}. 

\textbf{(2)} Our second example is the X-cube Floquet code with a rewinding schedule. We note that the construction of the X-cube Floquet code in Ref.~\cite{XcubeFloquet2022} also uses a coupled layer construction. However, in their construction, the ISGs evolve back into decoupled layers of 2D TCs (up to nonlocal stabilizers). We show in Sec.~\ref{sec:Xcube} that this can be avoided by rewinding the schedule. We also work out explicit FDLQCs to map the ISGs to the X-cube model up to 3D TC and stacks of 2D TCs. This, in turn, leads to a result of the splitting of topological order in the X-cube Floquet code under the sequence of measurements: the X-cube model ISG in one round splits into a copy of the X-cube model and TCs in the subsequent rounds.

\textbf{(3)} Lastly, we construct a Floquet code with ISGs that are FDLQC-equivalent to the 3D fermionic TC, i.e., with the same topological order as a 3D $\ZZ_2$ gauge theory with an emergent fermion~\cite{Levin2003Fermions}. Our construction is based on the 3D generalization of Kitaev's honeycomb model of Ref.~\cite{mandal2009}. In this construction, we use {rewinding} to avoid inadvertently measuring \textit{all} logical operators throughout the schedule. 

Besides the above-mentioned 3D examples, we also construct a new 2D Floquet code for which we expect the key principles used in this paper for the 3D Floquet codes i.e., rewinding and the interpretation of condensation in a parent stabilizer code, to be useful. In Appendix \ref{sec:Floquetcolorcode}, we construct the 2D Floquet color code, which exhibits ISGs that are FDLQC-equivalent to the 2D color code. This should not be confused with the CSS honeycomb code of Refs.~\cite{Davydova2022,Kesselring2022condensation,bombin2023unifying}, which also has been referred to as the Floquet color code in Ref.~\cite{Kesselring2022condensation}\footnote{In contrast to our Floquet color code, the ISGs of the CSS honeycomb code are FDLQC-equivalent to the 2D TC.}. The 2D Floquet color code is constructed on the ruby lattice and is the first example of a 2D Floquet code on a non-trivalent lattice. We consider three different measurement schedules for the Floquet color code. The first one is a six-step schedule that exhibits an order three automorphism of the logical operators, the second one is a rewound version of the first one with a trivial automorphism, and the last one is a six-step schedule that is not a rewinding schedule but has a trivial automorphism. For the first two schedules, notably, one of the ISGs is equivalent to the conventional color code up to concatenation with a 3-qubit repetition code -- thus allowing for the transversal implementation of certain logical Clifford gates. The rewound version is expected to yield gapped boundaries by truncating the checks at the boundaries; however, we were not able to construct boundary conditions that maintain the transversality of the above-mentioned logical Clifford gates, and hence, we leave the boundaries for future work. We also explicitly write the parent stabilizer code of the Floquet color code, which is FDQC-equivalent to two copies of the 2D color code. 

In summary, our work establishes new examples of Floquet codes and formalizes rewinding as a tool for designing Floquet codes with beneficial code properties, such as desired ISGs, which could be relevant for decoding and transversal logical gates. We expect that constructing examples such as these will be an important step towards developing a comprehensive classification of Floquet codes and for the construction of novel Floquet codes.  

The paper is organized as follows. {In Sec.~\ref{sec:review}, we state preliminary definitions, formalize the notion of rewinding measurement schedules, and state general properties of rewinding schedules -- such as trivial logical automorphisms after a measurement cycle. We also discuss the basic ideas behind interpreting Floquet codes as sequences of condensations in a parent stabilizer code and the evolution of checks associated with those condensations. In Sec.~\ref{sec:2DFloquetreviewandrewinding}, we review the Floquet code of Ref.~\cite{Paetznick2023Performance}, referred to as the Floquet TC and describe the rewinding schedule for the boundary construction using the evolution of checks.  
Sec.~\ref{sec:3DbosonicFloquet}, Sec.~\ref{sec:Xcube} and Sec.~\ref{sec:3DfermionicFloquet} describe the constructions of the 3D Floquet TC, the rewinding X-cube Floquet code, and the 3D Floquet fTC, respectively. In Appendix~\ref{sec:Floquetcolorcode}, we discuss the Floquet color codes with $\mathbb{Z}_3$ automorphism and trivial automorphisms, respectively. In Appendix~\ref{sec:counting}, we discuss the counting of logical qubits in the 3D Floquet TC. In Appendix~\ref{sec:eff_description_and_counting_BISG_XC}, we describe a particular ISG of the X-cube Floquet code in terms of the effective qubits created by the checks. 
}

\section{Rewinding Floquet code schedules}
\label{sec:review}

In this section, we state preliminary definitions and formalize the notion of a rewinding measurement schedule for a Floquet code. We then argue that a Floquet code with a rewinding schedule exhibits a trivial automorphism of the logical operators after a single period and that the measurement quantum cellular automata (MQCAs) defined at the boundary have a trivial index~\cite{aasen2023measurement}. {Lastly, we describe the interpretation of Floquet codes as a sequence of condensations in a parent stabilizer code and how the checks can sometimes evolve and grow into bigger instantaneous stabilizers.} 


\subsection{Definitions}

In general, a Floquet code is defined by two pieces of data: (i) a set of operators that are measured throughout the dynamics, known as the check operators (or more simply, the checks), and (ii) a periodic measurement schedule, which dictates when the check operators should be measured. We find it convenient to further define the check group as the group generated by the checks. We call the center of the check group, i.e., the subgroup of operators in the check group that commute with every element of the check group, the stabilizer group of the check group. We note that the stabilizers can be interpreted as conserved quantities of the dynamics since their measurement outcomes are not affected by the measurement of the checks. 

We say that a measurement schedule rewinds or that the Floquet code is a rewinding Floquet code if the sequence of measurements is performed in reverse order at some point in the schedule. For example, we consider a set of measurements labeled 0, 1, and 2. If the measurements are performed periodically in the sequence 012021, as in Ref.~\cite{Haah2022boundarieshoneycomb}, then the schedule is in fact, rewinding. This can be seen by writing a few periods of the schedule as: 
\begin{align} 
    \ldots0120\text{-}0210\text{-}0120\text{-}0210\text{-}0120\text{-}0210\ldots,
\end{align}
where we have written a single 0-round with repeated 0s, i.e., as 0-0 to demonstrate the rewinding explicitly. The sequence 0120 is then explicitly followed by the reverse sequence 0210. In our examples below, we find that rewinding is a valuable tool for constructing boundaries of Floquet codes, as acknowledged in Refs.~\cite{Haah2022boundarieshoneycomb,Gidney2022benchmarkingplanar,Paetznick2023Performance} and for obtaining a desired set of ISGs. This may, in turn, be useful for developing Floquet codes with beneficial decoding properties and transversal implementations of logical gates. 

\subsection{Trivial automorphism and MQCA index}
\label{subsec:trivial_automorph_rewinding}

We now state the general properties of such rewinding schedules -- specifically, we emphasize that there is a trivial automorphism of the logical operators, implying that a planar variant of the rewinding Floquet code defines a measurement quantum cellular automata at its boundary with a trivial index \cite{aasen2023measurement, sullivan2023floquet}.

Consider a Floquet code with a rewinding schedule of form 012021 on a torus. We assume that this Floquet code is reversible in the sense of Ref.~\cite{aasen2023measurement}. This means that for each consecutive pair of ISGs, there exists a complete set of shared logical operator representations for every logical operator. That is, there is a complete set of operators that commute with both stabilizer groups. In between rounds, we must update the logical operators by multiplying instantaneous stabilizers of the current round to obtain a logical operator that also commutes with the ISG of the next round. Intuitively speaking, the rewinding ensures that any instantaneous stabilizers that were multiplied to the logical operators are removed when the schedule is run backward. Thus, we arrive back at the same logical representative after the full rewound cycle. 

To see this explicitly, we denote the representations of a logical operator that are shared by ISGs of consecutive rounds $r$ and $r'$ as $\text{LOR}(r,r')$ where round $r'$ follows round $r$. For the 6-round schedule 012021, the logical operator representations of rounds 0 and 1 can be chosen to be LOR(0,1). To evolve to round 2, we can multiply by instantaneous stabilizers of the 1-ISG to obtain LOR(1,2). To go to the second instance of round 0,  we can multiply by instantaneous stabilizers of the 2-ISG to obtain LOR(2,0). We now begin the rewinding process. $\text{LOR}(2,0) = \text{LOR}(0,2)$ is a valid representation for the second instance of round 2. The next rounds are 1 and 0 so we multiply by the same elements of the 2-ISG to obtain LOR(2,1) and the same elements of the 1-ISG to obtain $\text{LOR}(1,0)=\text{LOR}(0,1)$. The net result is that we are back to the original representation LOR(0,1) that we started with. Hence, the automorphism is trivial. Later in Sec.~\ref{sec:square-octagon_RGBRBGschedule}, we explicitly show the shared logical operators for the rewinding sequence of the 2D Floquet TC in Fig.~\ref{fig:square-octagonsubsystemcode}.

The argument that the logical operators “rewind” also implies that the MQCA for the full period, as defined in Ref.~\cite{aasen2023measurement}, is trivial. If the Floquet code is put on a system with a boundary, one can consider the evolution of the logical operators at the boundary as defining an MQCA. Then, since the automorphism of logical operator representations is trivial for the full period, the MQCA is also trivial for the full period. We note however that the MQCA is nontrivial for half a period of the rewinding schedule for the 2D Floquet TC~\cite{aasen2023measurement}.

{\subsection{Condensation in a parent stabilizer code 
and the evolution of checks}
\label{subsec:condensation}
The measurements in the Floquet codes discussed in our work can be interpreted as a sequence of condensations in a parent stabilizer code. Let us first clarify what is meant by condensation. If we start with a topologically-ordered state, condensing a nontrivial topological excitation means that those excitations are proliferated and can now be created without any energetic cost. In other words, the resulting ground state involves a superposition of the presence and absence of that excitation in any location. 
One way to implement this condensation in the Hamiltonian picture is to add the creation operator of the excitation as a term in the Hamiltonian and tune its strength to a large relative value. In scenarios involving measurements such as Floquet codes, this condensation can be performed by measuring the creation operator of the excitation, which, in turn, adds it to the stabilizer group of the instantaneous state while removing any previous instantaneous stabilizers that anticommute with it.  

We claim that for all the Floquet codes discussed in this work\footnote{For the 3D Floquet fTC, we state the rewinding schedule without relying on a condensation picture, even though we do not see any obstruction in constructing a parent stabilizer code using the above-mentioned definition. This is because our construction of 3D Floquet fTC generalizes the construction of the 2D Floquet TC to a trivalent lattice in three dimensions.}, a parent stabilizer code can be constructed by taking the stabilizers of the associated check group and adding in a maximal set of commuting local Pauli terms, that are products of ISG checks around closed loops. For the 2D Floquet TC, previous work has shown that the color code can be interpreted as a parent stabilizer code~\cite{Kesselring2022condensation} and it can be derived using this constructive definition. In Appendix~\ref{sec:Floquetcolorcode}, in which we construct the Floquet color code, we explicitly construct the parent stabilizer code by taking the stabilizers of the check group and closed loops of checks measured in the three rounds. For the 3D Floquet TC and the X-cube Floquet code, the construction involves coupling layers of 2D Floquet TCs. Even though the layers are coupled, we can form a maximal set of local commuting terms to consist of products of checks around closed loops within 2D layers; the resulting parent stabilizer code for both 3D Floquet TC and the X-cube Floquet code is simply stacks of color codes, as one might intuitively expect for the coupled layer construction. This is described explicitly in Sec.~\ref{sec:condensation_3DFloquet_TC}. 

We now discuss how the checks evolve under a sequence of measurements.
The evolution of an element of an ISG depends on whether it commutes with the checks in the subsequent round of measurements. In the 2D Floquet TC, any element in the current ISG that commutes with all of the checks measured in the subsequent round is a (static) stabilizer of the check group of the Floquet code; there are no products of checks, except stabilizers of the check group, in the current ISG that survive into the subsequent round as stabilizers. 
In the 3D Floquet TC, this is not the case. 
There are checks whose products are not stabilizers of the full check group and still survive as stabilizers into the next round because they commute with all of the checks of that next round. These checks are associated with condensation of nontrivial excitation across orthogonal layers in the ISG of stack of 2D toric codes. In general, it is the checks that condense an excitation of the ISG and not just that of the parent stabilizer code, whose products we expect to survive into subsequent rounds. Note that from here on, we will refer to these checks as the condensation checks\footnote{even though all checks can be interpreted as performing condensations in a parent stabilizer code} below, and they are said to evolve into bigger condensation checks under subsequent rounds of measurements. In subsequent rounds of the 3D Floquet code, this growth of condensation checks can keep happening until \textit{all} of them become non-local stabilizers. In order to obtain a desired ISG, we utilize rewinding to avoid this situation, i.e., we reverse the evolution before \textit{all} condensation checks evolve into non-local stabilizers. 
There is a physical explanation of the evolution of condensation checks in terms of the topological data of the parent stabilizer code. We provide this explanation for the 3D Floquet TC in Sec.~\ref{sec:condensation_3DFloquet_TC}.

}

\section{2D Floquet TC on a planar layout}
\label{sec:2DFloquetreviewandrewinding}
\subsection{Review: 2D Floquet TC with a rewinding schedule}
\label{sec:square-octagon_RGBRBGschedule}
We now review the 2D Floquet TC, which is essential to our 3D construction. 
Following Ref.~\cite{Paetznick2023Performance}, the 2D Floquet TC can be defined on a two-dimensional square-octagon lattice with periodic boundary conditions and a qubit at each vertex. We color the edges of the lattice red, green, and blue, as shown in Fig.~\ref{fig:square-octagonsubsystemcode}. The red, green, and blue colors determine the 2-qubit Pauli check operators $XX$, $YY$, or $ZZ$, respectively, associated with an edge.
We refer to the checks on the red, green, and blue edges as R-, G-, and B-checks, respectively. The stabilizers of the check group are generated by three types of operators, supported on either a square or an octagon, as shown in Fig.~\ref{fig:square-octagonsubsystemcode}(a). We refer to these as the square stabilizers and the octagon stabilizers.

\begin{figure*}[t]
    \centering
    \includegraphics[scale=0.068]{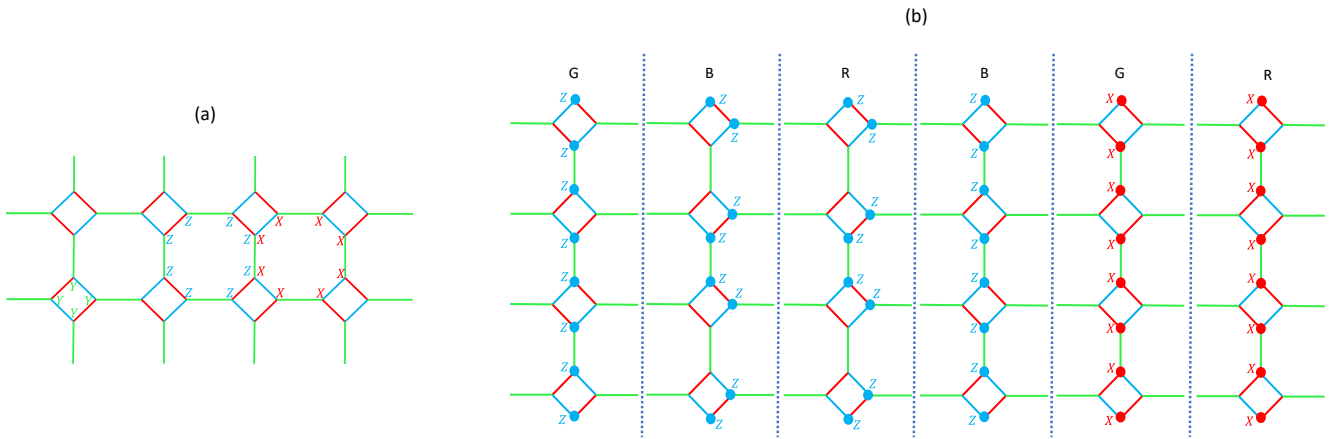}
    \caption{
    (a) The check operators of the Floquet code consist of two-body Pauli check operators  $XX$, $YY$, and $ZZ$ for the red, green, and blue edges, respectively. The stabilizers of the check group are generated by the products of check operators along closed loops, as shown; the product of check operators around a square plaquette consists of Pauli operators $Y$s, while the products of check operators around octagon plaquettes consist of Pauli operators $X$ or $Z$ depending on the sublattice. (b) Evolution of a representation of a logical operator in the GBRBGR schedule on the square-octagon lattice.
    The labels G, B, and R at the top indicate the rounds of the Floquet code.
    In each round, the representations commute with both the checks of the current and of the next round. At the end of the cycle of 6 rounds, the representation is back to the starting representation and hence, the automorphism is trivial.}
    \label{fig:square-octagonsubsystemcode}
\end{figure*}

\begin{figure*}
    \centering
    \includegraphics[scale=0.14]{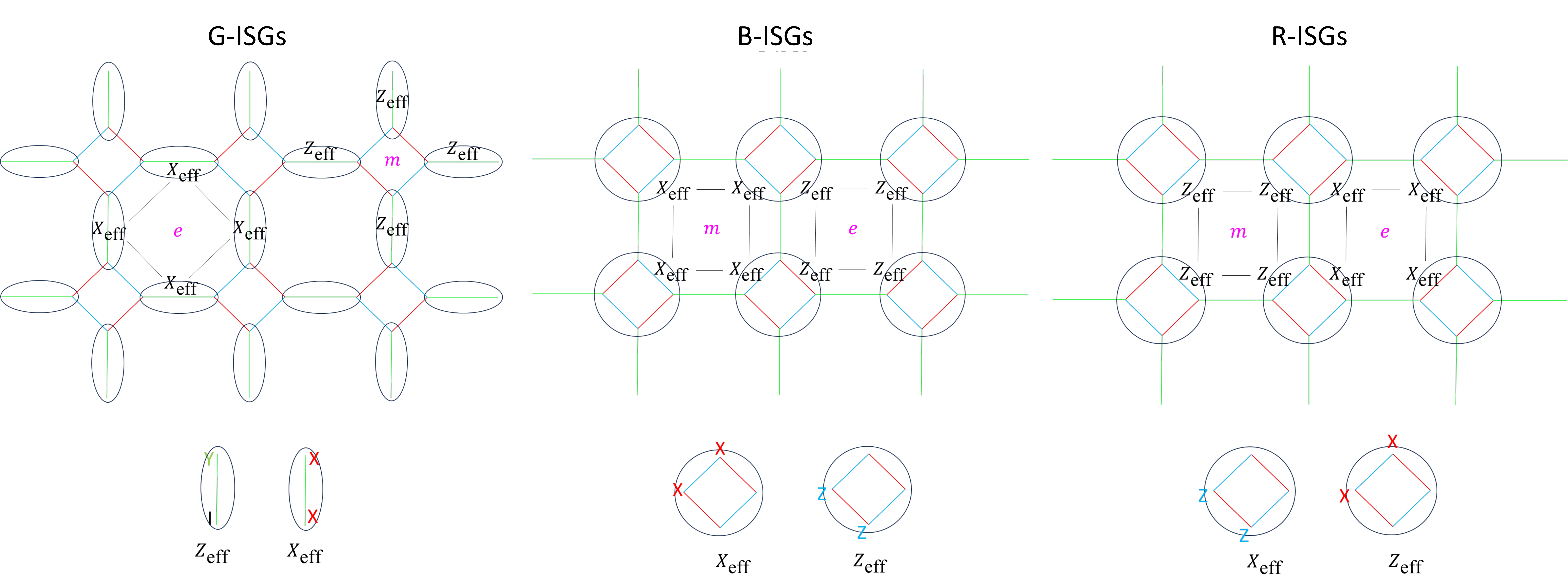}
    \caption{Bulk stabilizers of the effective TC ISGs in the 2D Floquet code. In the G-ISGs, the effective qubits (shown using ellipses) live on the green edges, and we get the usual square lattice TC stabilizers. In the B- and R-rounds, the effective qubits (shown using circles) live on the square plaquettes, and we get ISGs that are 2D TCs on rotated square lattices.}
    \label{fig:effective_toric_codes_2DFloquet}
\end{figure*}

\begin{figure*}
    \centering
    \includegraphics[scale=0.10]{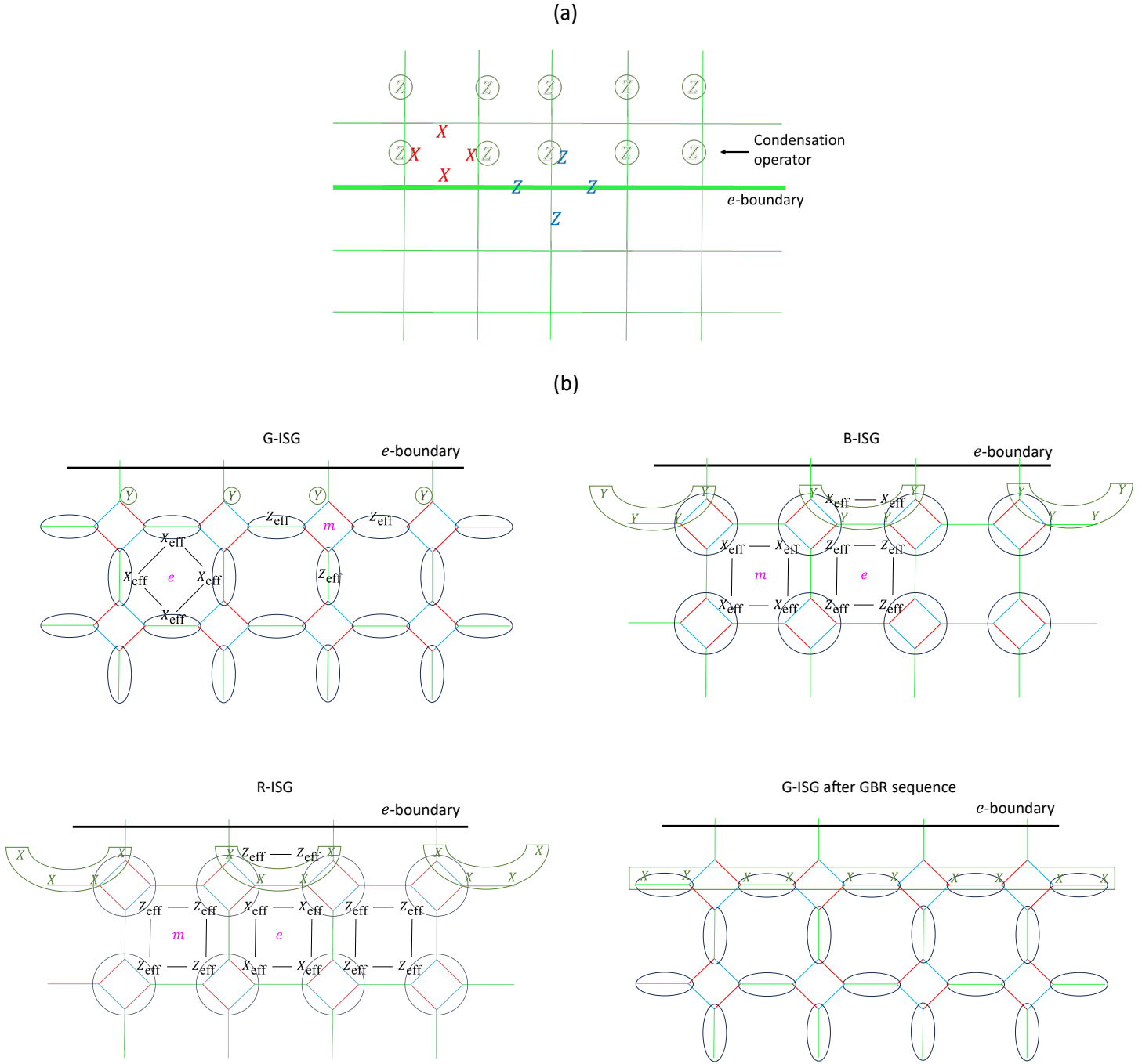}
    \caption{{The issue with the boundary construction (truncation through green checks) using the 3-round GBR schedule is expressed in terms of the evolution of condensation checks. (a) $e$-boundary of 2D TC, obtained by adding single-qubit condensation checks, the Pauli $Z$ operators (shown inside circles), on the qubits sitting at the vertical edges in the upper half-plane. This leads to condensation of $e$-charges in the upper half plane, forming an $e$-boundary of the TC as shown. (b) The evolution of condensation checks is at the boundary of the 2D Floquet code. The microscopic condensation checks are shown in green text. In the first G-ISG, the single-qubit Pauli $Y$ operators act (on the effective qubits) as the Pauli $Z$ (condensation) operators shown in (a). In the B-ISG, the product of the single-qubit $Y$ operators and the green check, as shown, is left invariant as a stabilizer, and hence, this is the evolved condensation check. In the R-ISG, the product of blue checks and the prior evolved condensation check is left invariant as a stabilizer. If we follow this GBR sequence with a G-round, as one would in a 3-round schedule, the condensation check evolves into a non-local stabilizer after multiplication with red checks. This non-local operator is a logical operator on a planar layout of the G-ISG with two $e$-boundaries and two $m$-boundaries. Thus, the logical information is measured out in the 3-round schedule. If the schedule is rewound, then the size of the condensation checks can be kept constant, and logical information preserved.}}
    \label{fig:2DFloquetboundaryGBRschedule}
\end{figure*}

\begin{figure*}
    \centering
    \includegraphics[scale=0.11]{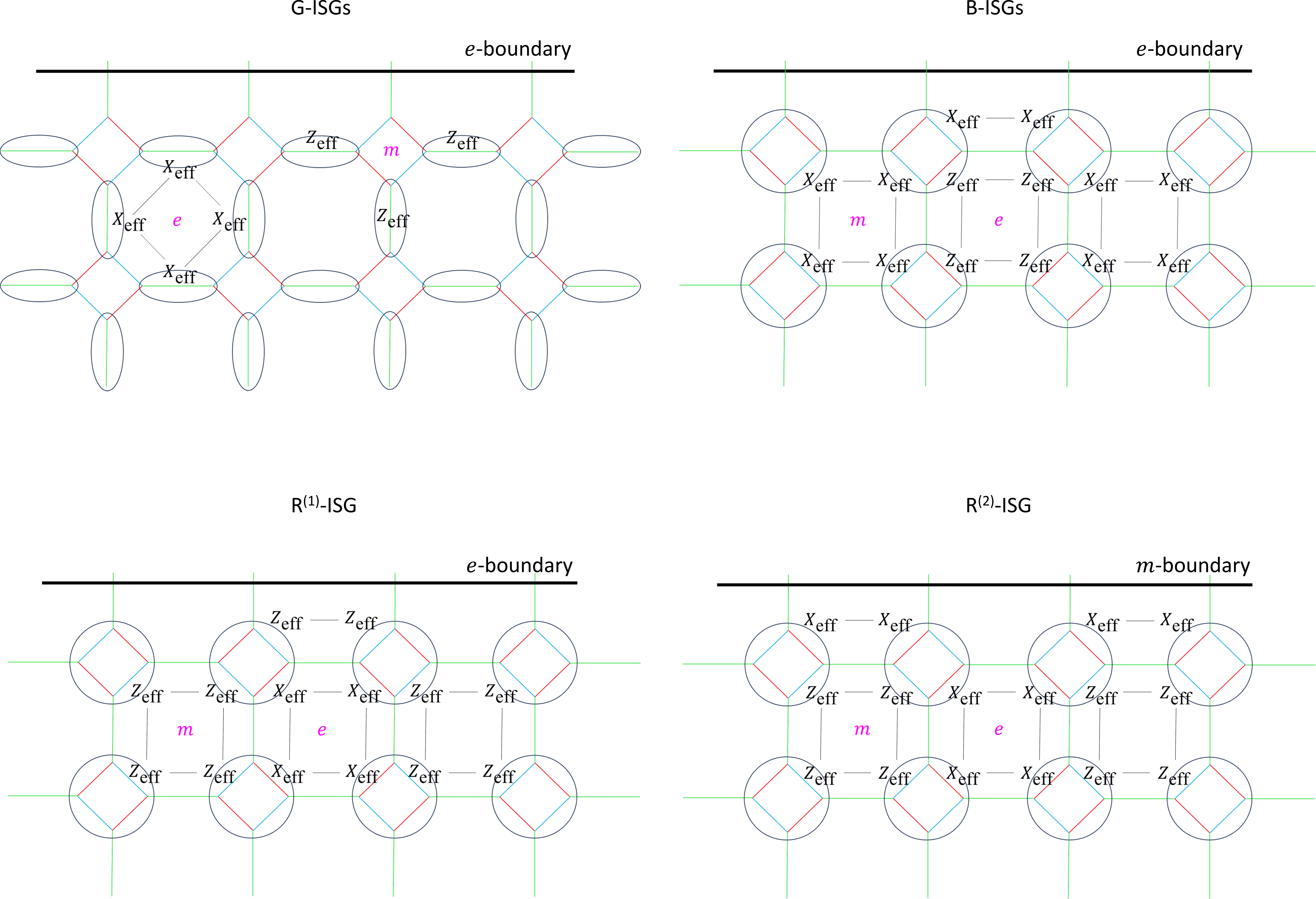}
    \caption{ Truncation through the green checks in the ISGs in the 2D Floquet TC on the square octagon lattice with the GBRBGR schedule. For the G-ISGs, the effective qubits (illustrated using ellipses) live on the green edges, and for the B- and R-ISGs, the effective qubits (illustrated using circles) live on the square plaquettes. The bulk and boundary stabilizers are written in terms of the effective Pauli operators. Top: The truncation through green checks condenses the $e$-anyons (convention chosen is shown) in the G- and B-ISGs.  
    Bottom: The truncation through green checks in the two R-ISGs. For these two R-ISGs, the boundary undergoes an automorphism in its type due to the bulk automorphism.   $\text{R}^{(1)}$-ISG indicates the ISG in the first R-round, and $\text{R}^{(2)}$-ISG indicates the ISG in the second R-round in the sequence. The truncation through green checks describes an $e$ ($m$) boundary in the $\text{R}^{(1)}$-ISG ($\text{R}^{(2)}$-ISG) respectively for the convention chosen.}
    \label{fig:boundaries_2D_Floquet_code}
\end{figure*}

To initialize the code, we measure the checks RBGR in sequence. Subsequently, in each period, we measure the sequence GBRBGR~\footnote{It can be equivalently written as RGBRBG due to periodicity, and then the correspondence to the 012021 notation mentioned earlier is clear.}. As it is essential in the construction of the 3D Floquet TC, we note that this schedule is rewinding, so the logical automorphism is trivial under the full period. We also note that this is the schedule used in Refs.~\cite{Gidney2022benchmarkingplanar} and \cite{Paetznick2023Performance} to define the 2D Floquet TC on a system with a boundary.

The ISGs obtained upon measuring the R-, G-, and B-checks are referred to as the R-, G-, and B-ISGs, respectively. These ISGs are generated by the stabilizers of the check group and the check operators measured in the round. The G-ISG, for example, is generated by the square and octagon stabilizers as well as the two-qubit $YY$ stabilizers on the green edges.

The G-ISG is precisely the 2D TC on the square lattice concatenated with a 2-qubit repetition code on the green edges. More specifically, the 2-qubit repetition code on each green edge is defined by a $YY$ stabilizer and logical operators 
\begin{align} \label{eq:GISGrepetition}
    X_{\text{eff}}&=XX\equiv ZZ, & Z_{\text{eff}}&=YI\equiv IY.
\end{align}
That is, in the subspace where all the green checks $YY$ are satisfied, we may define an effective qubit on each edge according to the above equation. In this subspace, the octagon stabilizers can be recast as products of $X_{\text{eff}}$'s on four effective qubits, allowing them to be interpreted as the vertex terms of the square lattice 2D TC. Similarly, the square stabilizers can be written as a product of $Z_{\text{eff}}$'s on four effective qubits and hence correspond to the plaquette terms of the 2D TC.
We define the violations of vertex and plaquette stabilizers to be $e$ anyons and $m$ anyons, respectively.
The $e$ anyons are created by the application of $Z_{\text{eff}}$ on the green edges, and the $m$ anyons are created by the application of $X_{\text{eff}}$ on the green edges. 

The R- and B-ISGs are precisely the 2D TC on a rotated lattice up to concatenation with a 4-qubit code. Since the B- and R-ISGs are symmetric, we make this explicit for the B-ISG.
Each square of the lattice consists of four qubits and three instantaneous stabilizers, i.e., one $YYYY$ stabilizer and an additional two $ZZ$ stabilizers from the measurements of the B-checks. Due to these stabilizers, an effective qubit can be defined on each square with the following effective (logical) operators,
\begin{align}
    X_{\text{eff}}&=ZIZI, & Z_{\text{eff}}&=XXII,
\end{align}
where the first two and last two qubits come from the two blue edges on the square, respectively. The above operators commute with the three instantaneous stabilizers on the square. In other words, each square supports a [[4,1,2]] code. 
In the logical subspace of the [[4,1,2]], the octagon stabilizers reduce to a product $X_{\text{eff}}^{\otimes 4}$ or $Z_{\text{eff}}^{\otimes 4}$ depending on the sublattice, as shown in Fig.~\ref{fig:effective_toric_codes_2DFloquet}. Thus, on the effective qubits defined by the [[4,1,2]] code, we have a 2D rotated TC.

{

We note that on a torus, we could also just use the 3-round sequence GBR instead of the 6-round rewinding sequence GBRBGR. However, the 3-round sequence does not work for 2D Floquet TC on a planar layout. Below, we discuss the 3-round sequence does not work for the planar layout, using the evolution of condensation checks at the boundary. 
\subsection{Rewinding Boundary for 2D Flouqet TC: evolution of condensation checks}
Let us consider how to create a boundary of the 2D TC using condensation checks. In order to create the $e$-boundary, we first note that the $e$-excitations (violations of vertex $X$ stabilizers) are created by single-qubit Pauli $Z$ operators. To define the boundary, we add these single-qubit Pauli $Z$ stabilizers at the boundary to the stabilizer group. Due to this, the $X$ stabilizers at the boundary are no longer in the stabilizer group, and the four-qubit $Z$ stabilizer generators can be broken down into two-qubit Pauli $Z$ stabilizers. Similarly, we can define the $m$-boundary. 

Now, for the 2D Floquet TC, each ISG is a TC on an effective lattice, and for one of the ISGs, we have the effective terms (Pauli $Z$ operators on the effective qubits) at the boundary as defined above. Under subsequent rounds, the condensation checks at the boundary evolve into larger-weight operators. As long as these boundary condensation checks are constant weight, they condense either the $e$ or $m$ bulk excitation at the boundary. In particular, we consider a truncation through green checks for the 3-round schedule. In this case, after a full cycle GBR, if we measure G-checks again, the condensation checks would evolve into a non-local operator that is a logical operator of the ISG. The evolution of the boundary condensation checks in the 3-round RGB schedule is shown in Fig.~\ref{fig:2DFloquetboundaryGBRschedule}. To not measure the logical information, instead of following up with G after the GBR part of the schedule, a rewinding schedule GBRBGR is used instead. This ensures that after the R-round, we reverse back to the B-round and G-round with constant weight condensation checks.  
In Fig.~\ref{fig:boundaries_2D_Floquet_code}, we illustrate the boundary for the case of the rewinding schedule GBRBGR. In this case, no logical information is measured, and we get the desired ISGs.

}


\section{3D Floquet TC}

\label{sec:3DbosonicFloquet}

We now present the 3D Floquet TC whose ISGs are FDLQC-equivalent to the usual 3D TC (or two copies). The construction is inspired by the coupled layer construction of the 3D TC presented in Ref.~\cite{ma2017}. 
More specifically, our strategy for building the 3D Floquet TC is to start with layers of 2D Floquet codes and add measurements that implement the condensation procedure of Ref.~\cite{ma2017} at the level of the ISGs. We use 2D Floquet codes on square-octagon lattices, which are stacked along the $x$-, $y$-, and $z$-axes, as our building blocks. 

Before going into the detailed construction of the 3D Floquet TC, we review the coupled layer construction of the 3D TC and the construction of the 2D Floquet TC with a rewinding schedule on the square octagon lattice. Readers who are familiar with the details of the coupled layer construction of 3D TC and the 2D Floquet TC are welcome to skip directly to Sec.~\ref{subsec:rewinding_schedule_3D_Floquet_bTC}.

\begin{figure}
    \centering
 \includegraphics[scale=0.145]{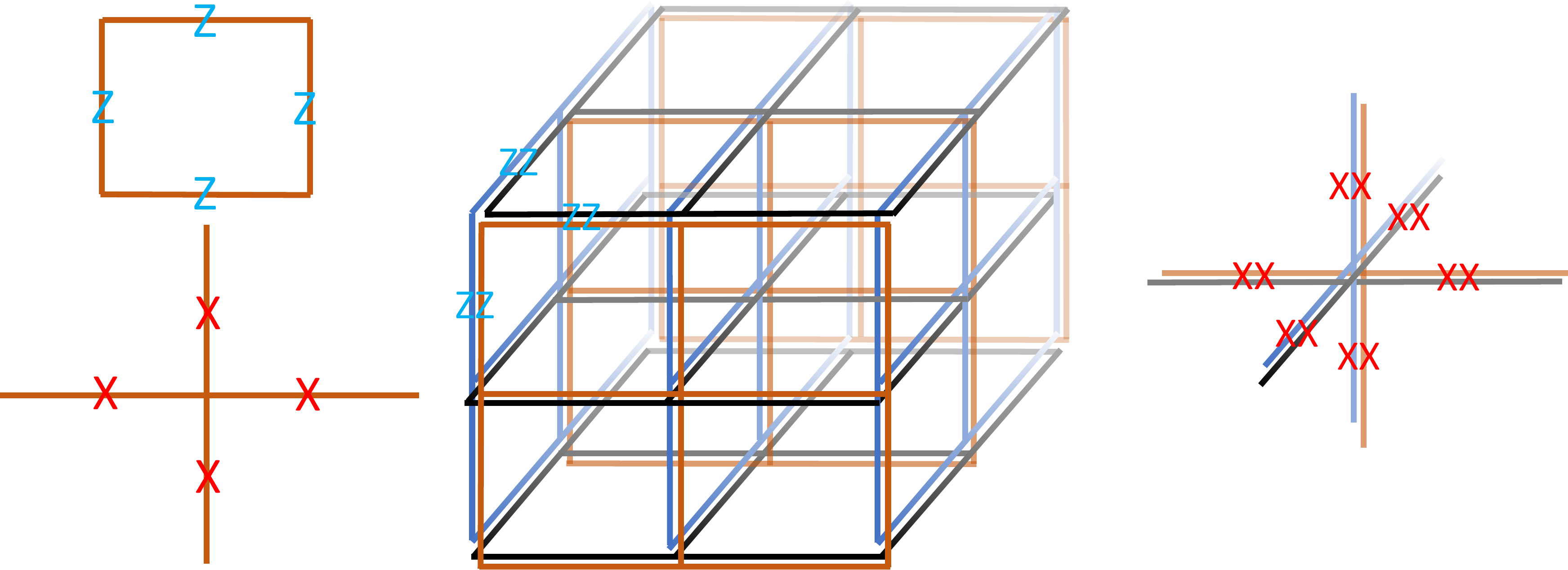}   
    \caption{Constructing three-dimensional topological codes from stacks of 2D TCs. A choice of vertex and plaquette stabilizers in each 2D TC layer of the stacks is shown on the left. Putting the three orthogonal stacks of 2D TC stacks together results in a lattice where each edge is now a composite edge that consists of two edges (qubits) from the 2D TC layers. Measuring $ZZ$ on the pair of qubits on each composite edge results in effective 3D TC stabilizers with an effective qubit (due to the $ZZ$ stabilizer) on each composite edge; some $ZZ$ stabilizers on the composite edges are shown. The resulting $X$-stabilizer is shown on the right. Measuring $XX$ instead of $ZZ$ on the pair of qubits on each composite edge results in the X-cube model stabilizers on the effective qubits.}
    \label{fig:foliation_3DTC}
\end{figure}

\subsection{Review: coupled layer construction}
\label{subsec:coupledlayerreview}

The coupled layer construction of the 3D TC in Ref.~\cite{ma2017} starts with stacks of 2D TCs, along the $x$-, $y$-, and $z$-axes, as shown in Fig.~\ref{fig:foliation_3DTC}(a). The 2D TCs here are defined on square lattices with qubits on the edges. 
Taken together, the layers of 2D TCs define a cubic lattice with two qubits on each edge. The edges parallel to the $z$-axis, for example, have one qubit from a 2D TC in a $yz$-plane and another from a 2D TC in an $xz$-plane. 

The 2D layers are then coupled together by forcing an interlayer $ZZ$ operator at each edge to be a stabilizer.\footnote{We remark that if the 2D TC layers are coupled together by an interlayer $XX$ stabilizer at each edge, then we obtain an effective X-cube model~\cite{ma2017}. This is because the $Z$-type stabilizers that remain in the stabilizer group are the products of 2D TC plaquette stabilizers around a cube, corresponding to the cube term of the X-cube model. This is used to build the X-cube Floquet code in Ref.~\cite{XcubeFloquet2022}.
} This requires removing the 2D TC stabilizers that fail to commute with the $ZZ$ operators. Note that the vertex terms of the 2D TCs do not commute with the interlayer $ZZ$ checks, but the product of three vertex stabilizers, one from each intersecting plane, does commute. Therefore, this product of $X$-type stabilizers remains in the stabilizer group.

Operationally, the $ZZ$ stabilizers define a single effective qubit at each edge. The effective stabilizer group is then equivalent to the usual 3D TC up to concatenation with two-qubit repetition codes; the product of three vertex stabilizers becomes the vertex stabilizer on the cubic lattice and the plaquette interlayer $ZZ$ checks, the logical operators of the 3D TC can be represented by $e$ string operators along non-contractible paths and membranes built from stacks of $m$ string operators. 

Intuitively, the interlayer $ZZ$ operators create pairs of $e$ anyons. By adding the $ZZ$ operators to the stabilizer group, we have condensed pairs of $e$ anyons from intersecting 2D TCs. Along the $z$-axis, for example, the $ZZ$ operator creates a pair of $e$ anyons with one from the $yz$-plane and the other from the $xz$-plane. Heuristically, the condensation of interlayer pairs of $e$ anyons implies that the $e$ anyons can transfer without any energy cost between layers at an intersection while the $m$ anyons in individual layers become confined.

\begin{figure*}
    \centering
 \includegraphics[scale=0.17]{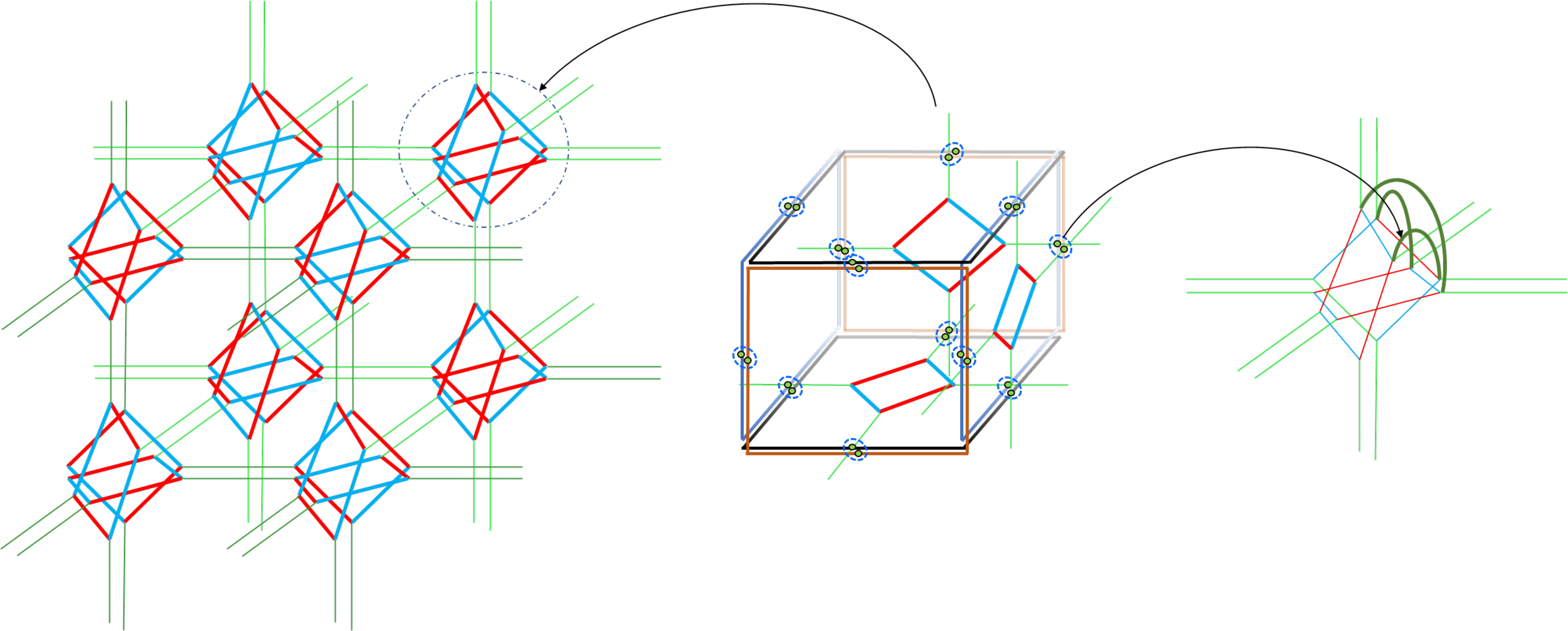}   
    \caption{Stacks of square-octagon lattice (see Fig.~\ref{fig:square-octagonsubsystemcode}) along three orthogonal directions to prepare the resource stacks of 2D TC layers before reaching the 3D TC instantaneous stabilizer group (ISG) of the 3D Floquet TC. For the 3D TC ISG in G-round (see table~\ref{tab:3DTCschedule}), the plaquette operators on the octahedron map to the plaquette operators on the cubic lattice in the manner as shown. In the middle, we show how the plaquette stabilizers on the circled ``octahedron'' in the ISG of green check measurements are mapped from the plaquette stabilizers in the stacks of the square lattice 2D TC layers. On the right are shown the condensation checks that couple the stacks of square-octagon models such that the resulting ISG in G-round is the usual representation of the 3D TC on the cubic lattice in terms of the effective qubit Pauli operators.
    }
    \label{fig:foliation_3DTC2}
\end{figure*}

\subsection{6-round rewinding schedule}
\label{subsec:rewinding_schedule_3D_Floquet_bTC}

To build the 3D Floquet TC, we start with layers of 2D Floquet TCs on square-octagon lattices, which are stacked along the $x$-, $y$-, and $z$-axes. The stacks of square-octagon lattices define a 3D truncated cubic lattice with two qubits at each vertex; see also Ref.~\cite{XcubeFloquet2022}. For clarity, we resolve the vertices in Fig.~\ref{fig:foliation_3DTC2} to show the connectivity of each square-octagon foliation. 

The check group of the 3D Floquet TC is generated by the check operators of the 2D Floquet TCs and 2-qubit interlayer $YY$ check operators that couple the layers together as in Fig.~\ref{fig:foliation_3DTC2}. There are three interlayer $YY$ checks for each octahedron of the truncated cubic lattice. These correspond to the three ways of pairing the $xy$-, $yz$-, and $xz$-planes. The stabilizers of the check group are generated by the square stabilizers of the 2D Floquet TCs and the product of three octagon stabilizers sharing a truncated cube, as shown in Fig.~\ref{fig:static_stabilizers_3D_Floquet}.

\begin{figure}
    \centering
  \includegraphics[scale=0.12]{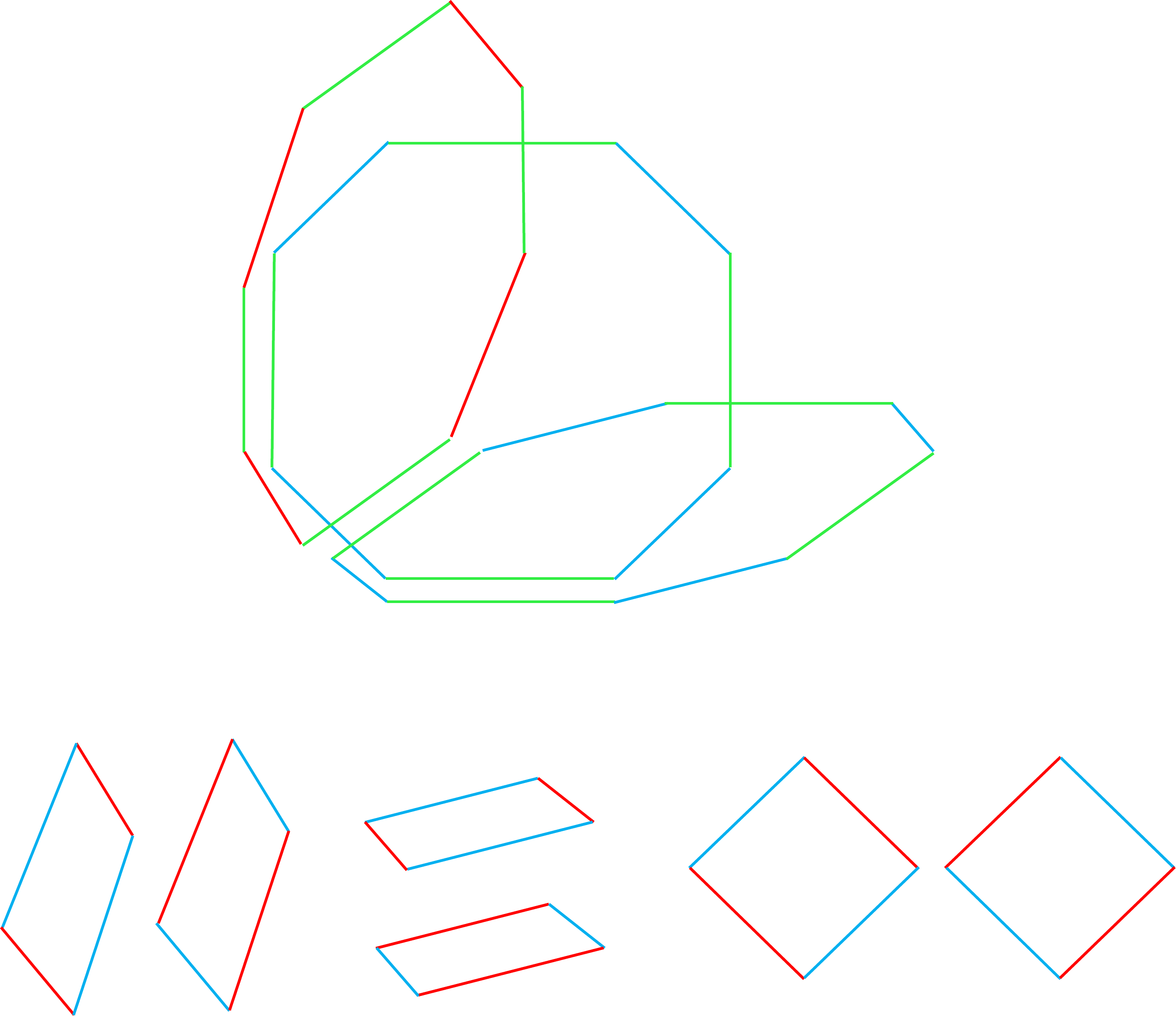}
    \caption{Stabilizers of the check group of the 3D Floquet TC. They form a subgroup of the stabilizers of the square-octagon layers such that they commute with the condensation checks. Top: in each cube, the product of three octagon stabilizers from orthogonal foliations, as shown, commutes with the condensation checks. Note that the octahedrons can have four different configurations of edges in the unit cell shown in Fig.~\ref{fig:foliation_3DTC2}, so all of the stabilizers that are products of three octagons on a cube do not look exactly as the one shown. Bottom: the square stabilizer terms from the square-octagon layers commute with the condensation checks. 
    }
    \label{fig:static_stabilizers_3D_Floquet}
\end{figure}

\begin{table}[]
    \centering
    \begin{tabular}{|c|l|c|}
    \hline
Round   & Measurement  & ISG\\
\hline
0& G (Green $YY$+ interlayer $YY$) & 3D TC\\
1& B (Blue $ZZ$) & 3D TC $\times$ 3D TC\\
2& R (Red $XX$) & 3D TC $\times$ 3D TC\\
3& B (Blue $ZZ$) & 3D TC $\times$ 3D TC\\
4& G (Green $YY$+ interlayer $YY$) & 3D TC \\
5& R (Red $XX$) & 3D TC $\times$ 3D TC\\
\hline  
\end{tabular}
\caption{The GBRBGR schedule of measurements for the 3D Floquet TC. The measured checks and the instantaneous stabilizer groups (ISGs) in each round are written. We note that in the ISG with two copies of 3D TC, the two copies are in an entangled logical state due to nonlocal stabilizers of the form $\overline{Z}_{1,i}\overline{Z}_{2,i}$ where $\overline{Z}_{1,i}$ and $\overline{Z}_{2,i}$ are logical string operators of the two 3D TCs along nontrivial cycles $i$ of the associated 3D tori.}
\label{tab:3DTCschedule}
\end{table}

\begin{figure*}
    \centering
  \includegraphics[scale=0.13]{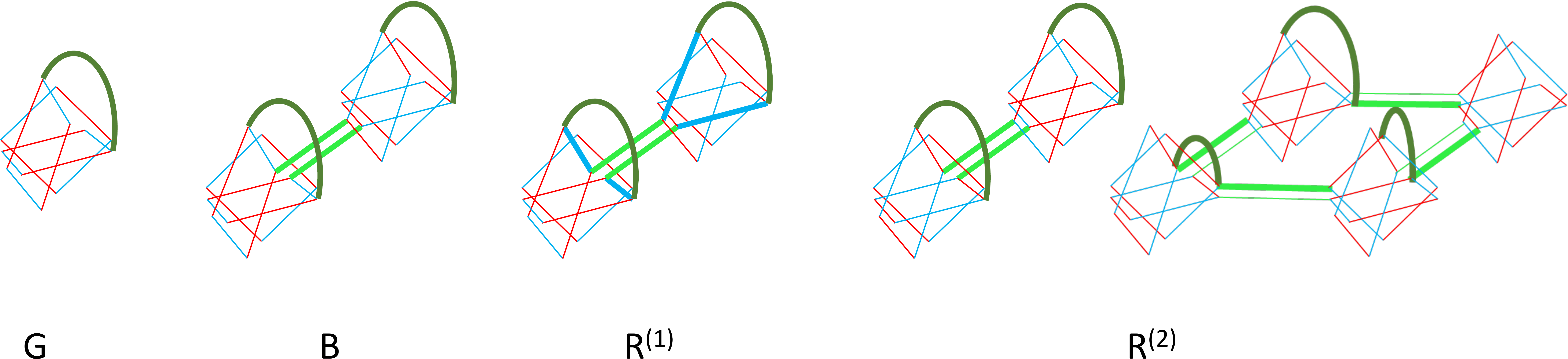}
    \caption{The evolution of one condensation check in the 3D Floquet TC with schedule GB$\text{R}^{(1)}$BG$\text{R}^{(2)}$ where we have added the superscripts to the R-rounds to specify their position. The different rounds are labeled as shown. There is no difference in the evolved condensation checks in the ISGs of the two B-rounds (and of the two G-rounds) due to rewinding and hence, we show only one B-round (and one G-round). In G-rounds, the condensation check is a two-qubit operator. In B-rounds, the interlayer term that survives is a product of green edges (thickened) and two original two-qubit condensation checks. Blue checks are multiplied by the representation in the B-round to get the representation in the $\text{R}^{(1)}$-round. In $\text{R}^{(2)}$-round we get a representation of the evolved condensation check similar to the B-round but across different octahedrons. We get another interlayer term in the $\text{R}^{(2)}$-round, which is a product of green checks and 2-qubit condensation checks of the G-round.}\label{fig:evolution_condensation_op_3DTC}
\end{figure*}
\begin{figure}
    \centering
    \includegraphics[scale=0.1,angle=0.3]{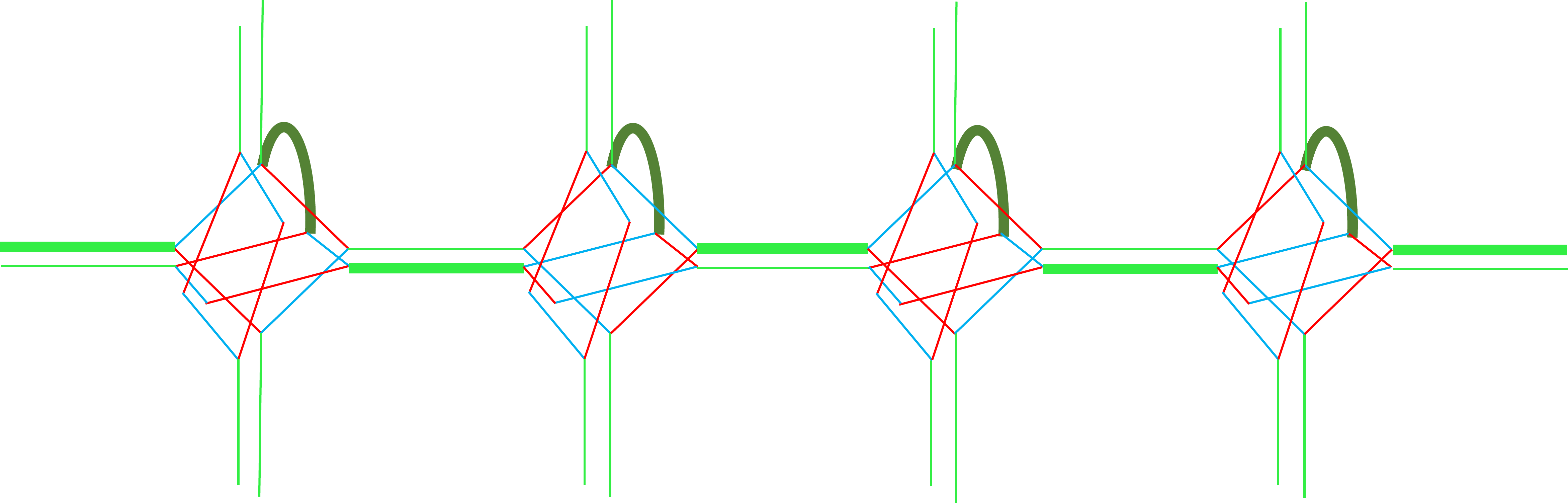}
    \caption{A nonlocal stabilizer made of Pauli $Y$ operators in the B-ISG. It is a product of the 2-qubit condensation checks and green checks that are shown with thick lines. Two out of the three two-qubit condensation checks on each octahedron evolve into nonlocal stabilizers along different nontrivial cycles. The nonlocal stabilizer is a product of two $e$ logical string operators from the G-ISG. These string operators are no longer stabilizer equivalent in B-ISG as they are logical string operators of two different copies of 3D TCs.}
    \label{fig:nonlocalstab}
\end{figure}

\begin{figure*}
    \centering
    \includegraphics[scale=0.18]{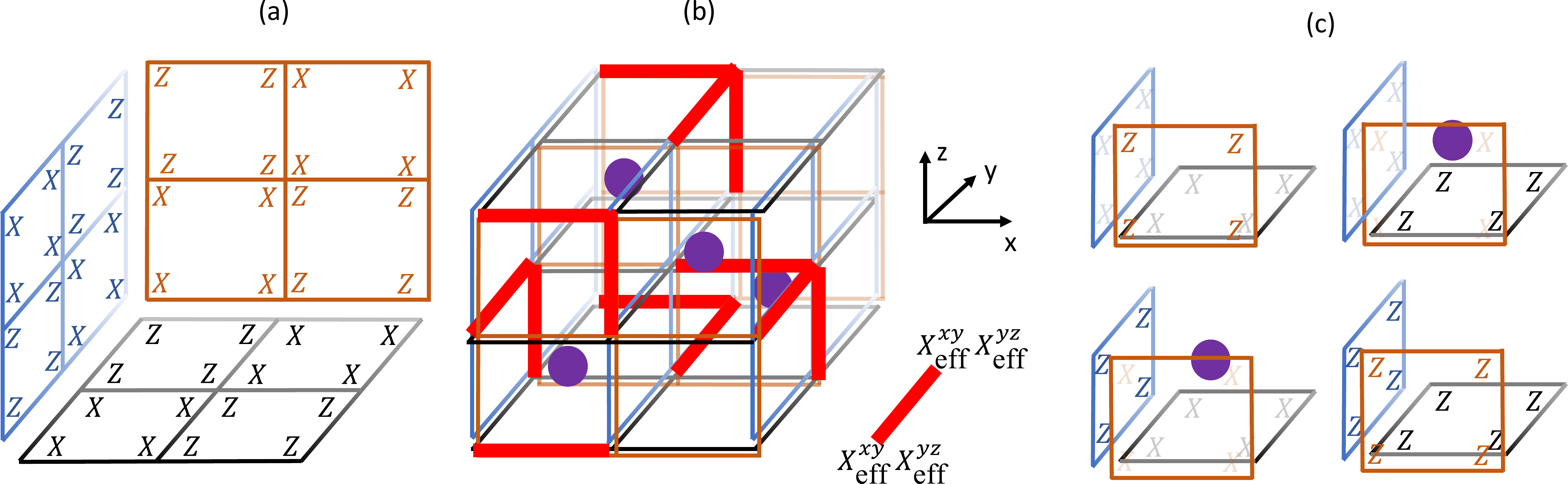}
    \caption{
    Coupled layer construction in a 3-foliated stack of 2D rotated TC layers to describe the B-ISG of 3D Floquet TC. (a) In the B-round or R-round of 2D Floquet TC on the square-octagon lattice, we get an effective TC on a rotated square lattice (check the effective description of the square-octagon layers in the B-round of the 2D Floquet code in Fig.~\ref{fig:effective_toric_codes_2DFloquet}). To describe the B-ISG for 3D Floquet TC, consider the 3-foliated (stacks of 2D planes along three directions) stack of 2D rotated TCs along with the evolved condensation checks in B-ISG; a 2D rotated TC layer from each foliation is shown in (a). We omit the subscript $\text{eff}$ from the Pauli operators in both (a) and (c) for simplicity. (b) The unit cell of the 3-foliated stack of 2D rotated TCs. Each coarse-grained vertex, where the three foliations meet, has 3 qubits. The evolved condensation checks are illustrated using thick red edges. Each thick red edge denotes a 4-qubit operator $X_{\text{eff}}^{\otimes 4}$; the thick red edge along $y$ direction denotes a 4-qubit Pauli $X_{\text{eff}}^{\otimes 4}$ operator on the qubits that belong to foliations $xy$ and $yz$, as shown on the right. Due to these evolved condensation checks, only certain products of $Z$-stabilizers from 2D rotated TC layers survive as stabilizers (shown in (c)), and depending on the cube, these products correspond to the $e_1$ and $e_2$-charges of the two 3D TCs in the B-ISG. The cubes corresponding to $e_1$ ($e_2$)-charges are (not) highlighted using purple dots, which form a checkerboard pattern.
    (c) On each cube, we consider a product of $Z$ stabilizers of 2D rotated TC layers that commute with the condensation checks. For each of the cubes in the front layer of the unit cell in (b), the products of $Z$-plaquette terms of 2D rotated TCs that 
    survive as stabilizers in the B-ISG are shown. The $X$-stabilizers from the 2D rotated TC layers are left transparent just to clarify how the plaquette terms appear on a cube.}    
    \label{fig:sublattices_BISG}
\end{figure*}

\begin{figure*}
    \centering
    \includegraphics[scale=0.1,angle=0.7]{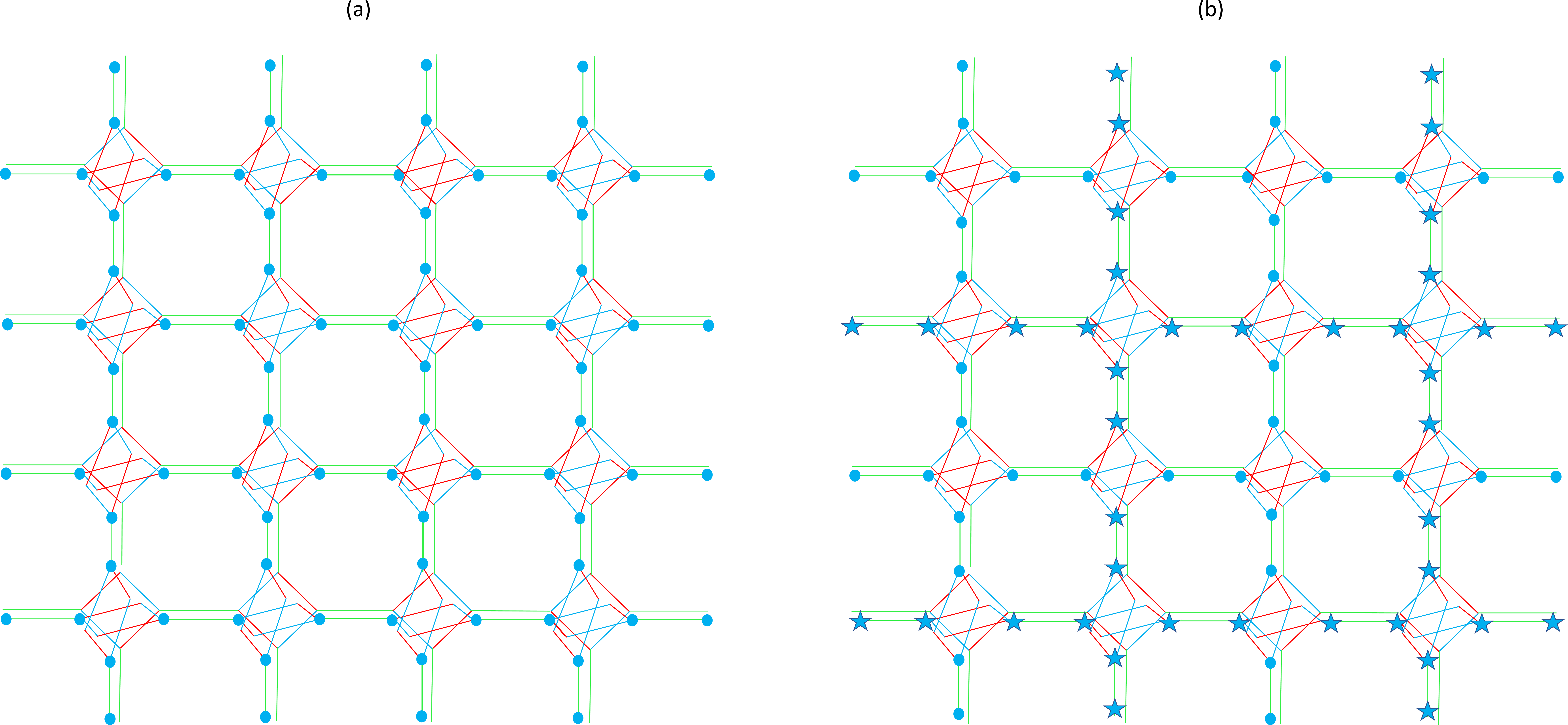}
    \caption{(a) A logical membrane operator of the 3D TC ISG in the G-rounds of the 3D Floquet TC; the membrane is a product of $m$-logical strings (made of Pauli $Z$ operators shown as blue dots) of the 2D TCs in the planes orthogonal to the membrane. (b) In the B-rounds, the membrane operator splits into two Pauli $Z$ membrane operators corresponding to two copies of 3D TCs that form the B-ISG. One membrane consists of Pauli $Z$ operators acting on qubits indicated by the blue dots, while the other acts on those indicated by the blue stars.}
\label{fig:membrane_3DFloquetTC}
\end{figure*}

\begin{figure*}
    \centering
    \includegraphics[scale=0.22]{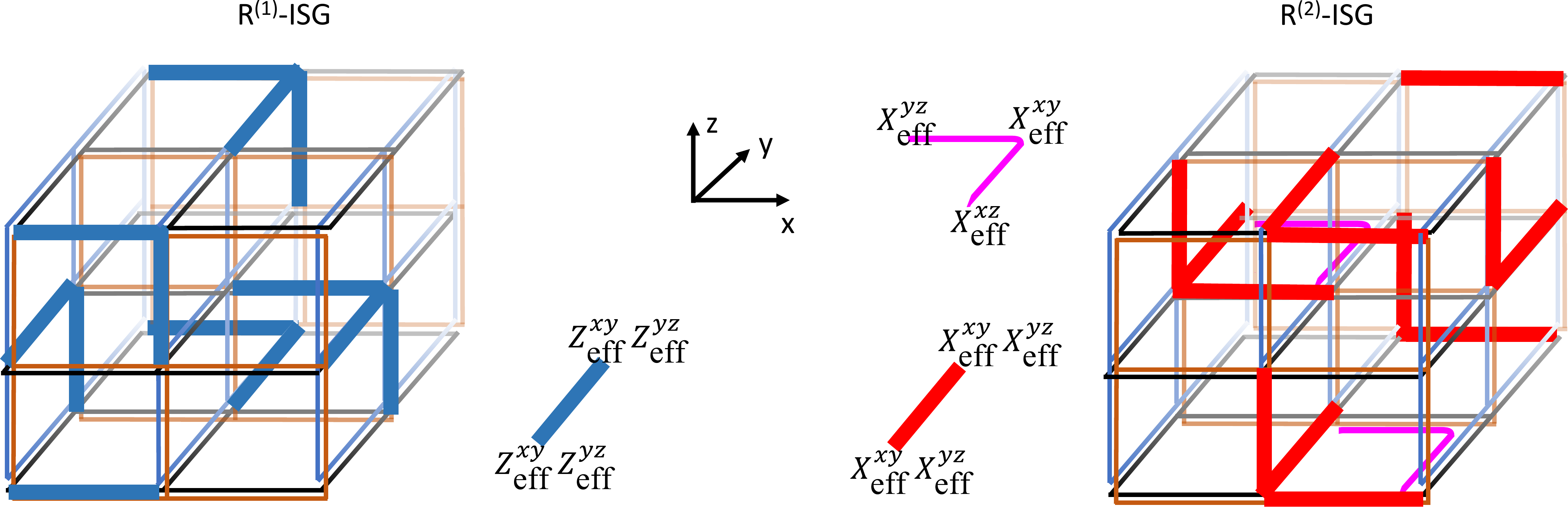}
    \caption{Configuration of condensation checks in the two R-ISGs is shown in the 3-foliated stack of rotated lattice TC layers (check the effective description of the square-octagon layers in the R-round of the 2D Floquet code in Fig.~\ref{fig:effective_toric_codes_2DFloquet}). In the $\text{R}^{(1)}$-ISG ($\text{R}^{(2)}$-ISG), the evolved condensation checks are  $Z_{\text{eff}}^{\otimes 4}$ ($X_{\text{eff}}^{\otimes 4}$ in red and $X_{\text{eff}}^{\otimes 3}$ in pink) operators; representations for the 4-qubit condensation checks along the $y$-direction and for the 3-qubit condensation check are shown; here the superscripts $xy$ and $yz$ in Pauli operators $X^{xy}_{\text{eff}}$ and $X^{yz}_{\text{eff}}$, denote the foliations the qubits belong to. }
\label{fig:gapped_boundaries_3D_Floquet_bTC}
\end{figure*}


To initialize the 3D Floquet TC, we first initialize the stack of 2D Floquet codes with the four-round measurement sequence RBGR. At this point, the ISG is equivalent to a stack of 2D TCs up to concatenation with two-qubit repetition codes. We then measure the G-checks and the interlayer $YY$ checks simultaneously (round 0 listed in Table~\ref{tab:3DTCschedule}). In terms of the effective Pauli operators $X_\text{eff}$ and $Z_\text{eff}$ in the G-ISG of the 2D Floquet TC [Eq.~\eqref{eq:GISGrepetition}], the interlayer $YY$ checks are precisely the $Z_\text{eff}Z_\text{eff}$ terms of the coupled layer construction as described in Sec.~\ref{subsec:coupledlayerreview} and Fig.~\ref{fig:foliation_3DTC}. Hence, the interlayer $YY$ checks create pairs of $e$ anyons on intersecting layers. Therefore, the interlayer $YY$ checks condense pairs of $e$ anyons, and after measuring the $YY$ checks, the ISG is the 3D TC, up to concatenation with 2-qubit repetition codes on the green edges. 

We continue with a periodic measurement schedule of the checks, according to the sequence GBRBGR, where the G round implicitly includes the interlayer $YY$ checks. We note that the interlayer $YY$ checks fail to commute with the subsequent B- and R-checks, so they are removed from the B- and R-ISGs. However, products of the interlayer $YY$ checks and possibly checks measured in the preceding rounds survive as stabilizers. We refer to these new instantaneous stabilizers as the evolved condensation checks, and these are shown in shown in Fig.~\ref{fig:evolution_condensation_op_3DTC}. On each octahedron of the truncated cubic lattice, only one of the three interlayer $YY$ checks (depending on the octahedron) evolves to a constant-weight stabilizer.
The other two evolve into nonlocal stabilizers. An example of a non-local evolved condensation check is shown in Fig.~\ref{fig:nonlocalstab}.


The generators of each of the ISGs thus consist of (i) the checks measured in that round, (ii) the condensation checks or evolved condensation checks, (iii) some octagon stabilizers (which octagon terms are in ISG depends on the precise round), and (iv) the stabilizers of the check group. 
We list the topological order of the ISGs in each round of the schedule in Table~\ref{tab:3DTCschedule}. 
Explicit circuits to map the ISGs to the canonical form of the 3D TC (or two copies of it) are given in the supplementary \texttt{Mathematica} file. The counting of logical qubits is described in Appendix~\ref{sec:counting}. 

We note that the rewinding GBRBGR schedule ensures that, in each round, there is an extensive number of constant-weight evolved condensation checks. This is not the case for the 3-round schedule RGB, in which all the condensation checks evolve into non-local stabilizers, leading to an ISG of a stack of 2D TCs with some logical operators fixed as non-local stabilizers. Intuitively, rewinding the schedule prevents the possibility that all the evolved condensation checks are nonlocal.
We also note that the automorphism of logical operators is trivial, which follows from the rewinding property discussed in Sec.~\ref{subsec:trivial_automorph_rewinding}.

\subsubsection{Splitting into two copies of 3D TC}

Interestingly, the B-ISGs and the R-ISGs are FDLQC-equivalent to two copies of the 3D TC, up to nonlocal stabilizers. Here, we elaborate on the mechanism by which the single 3D TC splits into two copies of the 3D TC via measurements. 
In short, the constant-weight evolved condensation checks can be interpreted as short string operators that create pairs of $e$ anyons on the next-nearest neighbor octagons of the 2D square-octagon lattices. The configuration of evolved condensation checks is such that the ISG is FDLQC-equivalent to two copies of the 3D TC with a constraint on the logical subspace given by the condensation checks that evolve into nonlocal stabilizers. 

The fact that two copies of 3D TC arise in the B-ISG is best understood in the effective picture of TC layers. In Sec.~\ref{sec:square-octagon_RGBRBGschedule}, we showed that in the B- and R-ISGs of the 2D Floquet TC, we get TCs on rotated lattices with an effective qubit on each square plaquette of the square-octagon lattice. The B-ISG of the 3D Floquet TC can hence be understood as a coupled layer construction starting from rotated 2D TC layers stacked along three orthogonal directions as shown in Fig.~\ref{fig:sublattices_BISG}. 

In the effective description of 2D rotated TC layers, the evolved condensation checks act as 4-qubit Pauli $X$ operators, which create pairs of $e$-anyons across intersecting layers, with the violations of $Z$-stabilizers in the 2D rotated TCs corresponding to the $e$-anyons. The evolved condensation checks are illustrated on the effective lattice using thick red edges in Fig~\ref{fig:sublattices_BISG}(b). The fact that these evolved condensation checks belong to the stabilizer group implies that only certain
products of $Z$-stabilizers of the 2D layers, as shown in  Fig.~\ref{fig:sublattices_BISG}(c), survive as stabilizers after condensation.
The violations of these products of $Z$-stabilizers correspond to the $e$-charges of the two 3D TCs. 

Given the structure of the evolved condensation checks, the $e$-charges on nearest-neighboring cubes, in fact, belong to inequivalent superselection sectors. Hence, we label them as $e_1$ and $e_2$ corresponding to the two copies of the 3D TC. This illustrates that the two 3D TCs ``live'' on two sublattices of the full cubic lattice. This splitting is captured by the membrane logical operators of the two ISGs illustrated in Fig.~\ref{fig:membrane_3DFloquetTC}.

We now consider the two R-ISGs appearing in the GBRBGR schedule. We label these two consecutive R-ISGs as the $\text{R}^{(1)}$-ISG and the $\text{R}^{(2)}$-ISG. 
The key difference in the $\text{R}^{(1)}$-ISG, compared to the preceding B-ISG, is that we now have 4-qubit Pauli $Z_{\text{eff}}$ evolved condensation checks in the effective description of 2D rotated TC layers. The effective 2D rotated TC $X_{\text{eff}}$ ($Z_{\text{eff}}$) stabilizers are also now changed to $Z_{\text{eff}}$ ($X_{\text{eff}}$) stabilizers, since the [[4,1,2]] codes on the squares now have $XX$ checks as stabilizers; see Fig.~\ref{fig:effective_toric_codes_2DFloquet} for our convention of $X_{\text{eff}}$ and $Z_{\text{eff}}$ in the R-ISGs of the 2D Floquet TC. Thus, this R-ISG is equivalent to the B-ISG up to a basis change, and we again get two copies of 3D TC (up to nonlocal stabilizers). The full configuration of evolved condensation checks in the unit cell is shown in Fig.~\ref{fig:gapped_boundaries_3D_Floquet_bTC}(b). The nonlocal stabilizers, in this case, are the same as those of the B-ISG in Fig.~\ref{fig:nonlocalstab}, but with Pauli $Y$ operators replaced by Pauli $X$ operators. These can be obtained by multiplying the nonlocal stabilizers of the B-ISG by the B-checks.

In the $\text{R}^{(2)}$-ISG, the evolved condensation checks are 4-qubit $X_{\text{eff}}^{\otimes 4}$ and 3-qubit $X_{\text{eff}}^{\otimes 3}$; the microscopic representations are shown in Fig.~\ref{fig:evolution_condensation_op_3DTC}(d) and the effective representations are shown in Fig.~\ref{fig:gapped_boundaries_3D_Floquet_bTC} along with the configuration of evolved condensation checks in the unit cell. We again get two copies of 3D TCs (up to nonlocal stabilizers) as the ISG. The nonlocal stabilizers, in this case, are similar to those of B-ISG but shifted in space.

{\subsubsection{Evolution of condensation checks using topological data of the parent stabilizer code}
\label{sec:condensation_3DFloquet_TC}
As mentioned in Sec.~\ref{subsec:condensation}, the reason for the growth of condensation checks is that after their measurements, the next round of checks may only commute with certain products of the condensation checks. It is these products of condensation checks that survive as stabilizers into the next round. Rewinding helps to ensure that not all condensation checks evolve into non-local stabilizers. In the case of the 3D Floquet TC, if all condensation checks become non-local, we obtain an ISG that is FDQC-equivalent to a 3-foliated stack of 2D TCs. Hence, rewinding helps to avoid this ISG, and instead, we obtain 3D TC(s) as ISGs. We now discuss how the evolution of condensation checks can be described using the braiding and fusion data of the excitations in the parent stabilizer code of the 3D Floquet code, i.e., stacks of 2D color codes. 

We first state why a stack of color codes is the parent stabilizer code for the 3D Floquet TC. Invoking our construction from Sec.~\ref{subsec:condensation}, we can consider products of checks around closed loops that commute with each other and the stabilizers of the check group shown in Fig.~\ref{fig:static_stabilizers_3D_Floquet}. The product of green checks around each octagon, products of blue checks around each octagon, products of red checks around each octagon, and products of red checks around each square form closed-loop stabilizers that, along with the stabilizers of the full check group of Fig.~\ref{fig:static_stabilizers_3D_Floquet}, form the stabilizer generators of a parent stabilizer code. This code is FDQC-equivalent to a 3-foliated stack of 2D color codes. This is also what one may intuitively expect given that the parent stabilizer code for the 2D Floquet code is the color code~\cite{Kesselring2022condensation} and we use a coupled layer construction involving a stack of 2D Floquet codes to get the 3D Floquet code. 

Using the notation of Ref.~\cite{Kesselring2022condensation} for the anyons of the color code, the measurements of the 2D Floquet code correspond to condensing the anyons $rX$, $gY$ and $bZ$ in order. After each condensation, we obtain 2D TC ISGs and transition between them without loss of logical information. The deconfined and confined anyons for the steps when $gY$ and $bZ$ anyons are condensed are shown at the top in Fig.~\ref{fig:condensation_3D_Floquet_TC}. 

\begin{figure}[t]
    \centering
    \includegraphics[scale=0.13]{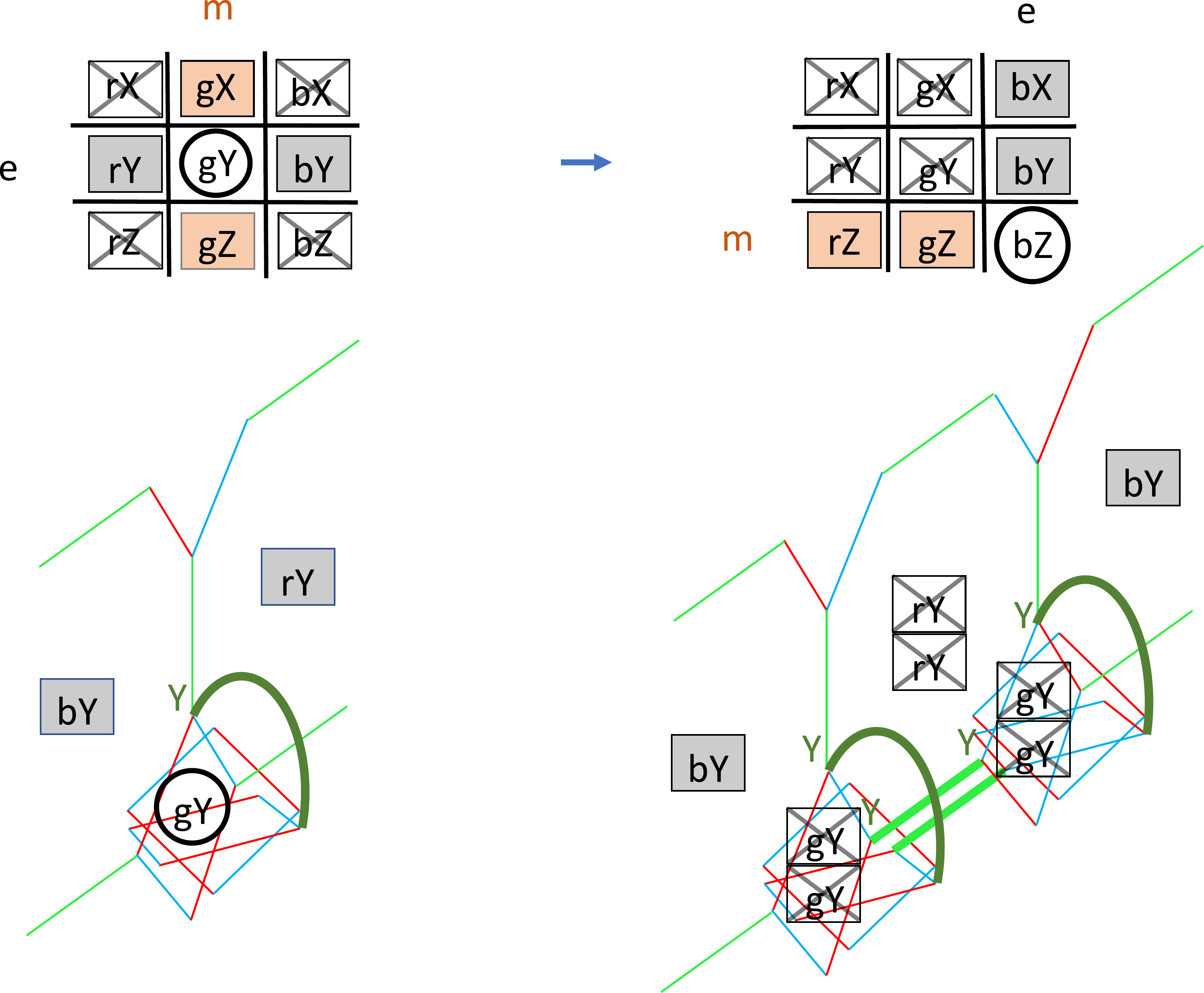}
    \caption{{Top: Condensation of $gY$ (left) and $bZ$ (right) anyons in the 2D color code, respectively. When $gY$ is condensed, meaning $gY$ is equivalent to vacuum $\mathbf{1}$, the crossed-out anyons are confined and 
    we obtain a 2D TC; we label the deconfined anyons, $rY\equiv bY$ as the $e$-anyon and $gX\equiv gZ$ as the $m$-anyon of 2D TC. Similarly, when $bZ$ is condensed, the crossed-out anyons, shown in the right table, are confined, and we label $bX\equiv bY$ and $rZ\equiv gZ$ as the $e$- and $m$-anyons of the resulting 2D toric code. Bottom: In the left part, the two-qubit Pauli $Y$ condensation check is shown using a dark green arc connecting orthogonal 2D color code layers of the parent stabilizer code. These are applied in the G-round along with intralayer green checks, which condense $gY$ anyons in color code layers. A single Pauli $Y$ operator in the dark green arc acts on a parent color code as shown, resulting in the creation of $gY$, $bY$ and $rY$ anyons. Since $gY$ is condensed before, $gY\equiv \mathbf{1}$ and $bY\equiv rY$ correspond to the $e$-anyon, we are left with a pair of $e$-anyons in the color code layer. Applying the two-qubit Pauli $Y$ operator on the dark green arc results in such pairs for both of the orthogonal layers. On the right, we show that when $bZ$ is condensed in the B-round following the G-round, the single $Y$'s now create confined excitations $rY$ and $gY$, and these confined excitations need to be canceled out. This is possible by multiplying the products of two arc-shaped two-qubit Pauli $Y$ operators with the green checks shown with thick lines. Hence, such a product survives as an evolved condensation check in the B-round.}}
    \label{fig:condensation_3D_Floquet_TC}
\end{figure}

In the 3D Floquet code, if we ignore the interlayer measurements that condense pairs of $e$-anyons, only the $rX$-$gY$-$bZ$ sequence of condensations would be performed in each color code layer of the parent code. To be consistent with our Pauli notation used for the 3D Floquet code, consider that we do the interlayer pair condensation across orthogonal color code layers along with the $gY$ condensation. 
In the round of $gY$ condensation, it is $gX\equiv gZ$ and $rY\equiv bZ$ that form the $e$ and $m$ anyons of the 2D TC ISG, respectively. This is illustrated in the table on the top left of Fig.~\ref{fig:condensation_3D_Floquet_TC}. The condensation operator is a two-qubit Pauli $YY$ operator, as shown in Fig.~\ref{fig:evolution_condensation_op_3DTC}. In the parent code of a stack of 2D color codes, the Pauli $Y$ operators act on orthogonal color code layers. Application of a single $Y$ operator in one color code layer creates $rY$, $gY$, and $bY$ anyons. However, because $gY$ is already condensed, meaning it is equivalent to vacuum $\mathbf{1}$, $gY\equiv \mathbf{1}$, we are left with $rY$ and $bY$, which are equivalent and correspond to the $e$-anyons as mentioned. This is illustrated in the figure at the bottom left of Fig.~\ref{fig:condensation_3D_Floquet_TC}. Thus, the two-qubit Pauli $Y$ condensation operator creates two pairs of $e$ anyons, one pair in each of the orthogonal color code layers. 

In the following round, $bZ$ is condensed, and thus, $rY$ and $gY$ are confined. Thus, we can no longer use the equivalences $gY\equiv 1$ and $rY\equiv bY$. Moreover, $bY\equiv bX$ and $rZ\equiv gZ$ are the deconfined anyons, and we choose to label $bY\equiv bX$ as the $e$-anyon. This is illustrated in the table on the top right of Fig.~\ref{fig:condensation_3D_Floquet_TC}. The confined excitations $rY$ and $gY$ must be canceled to find an evolved condensation check. This is possible if we take the 
product of two pair condensation checks (the two-qubit Pauli $Y$ operator across orthogonal layers) and the condensation operator for $gY$ (the green checks). This product cancels out the confined excitations $gY$ and $rY$ and creates two pairs of $bY$ on orthogonal color code layers. This is illustrated in the figure at the bottom right of Fig.~\ref{fig:condensation_3D_Floquet_TC}. Thus, again, we have a pair of $e$-anyons that are condensed across orthogonal layers. The requirement of the cancelation of the confined excitations serves as an alternate physical explanation for the evolution of condensation checks into bigger condensation checks. Moreover, there are two inequivalent flavors of such composites corresponding to the two sublattices and hence, the B-ISG is FDQC-equivalent to two copies of 3D TC, as described microscopically earlier. This is up to the caveat that a non-local stabilizer fixes the state of the two copies of 3D TC such that only one logical qubit remains. A similar description in terms of condensation holds for the R-ISGs.   

}

\subsection{Boundary construction}
\label{subsec:3dFbTC_boundaries}

\begin{figure}
    \centering \includegraphics[scale=0.1]{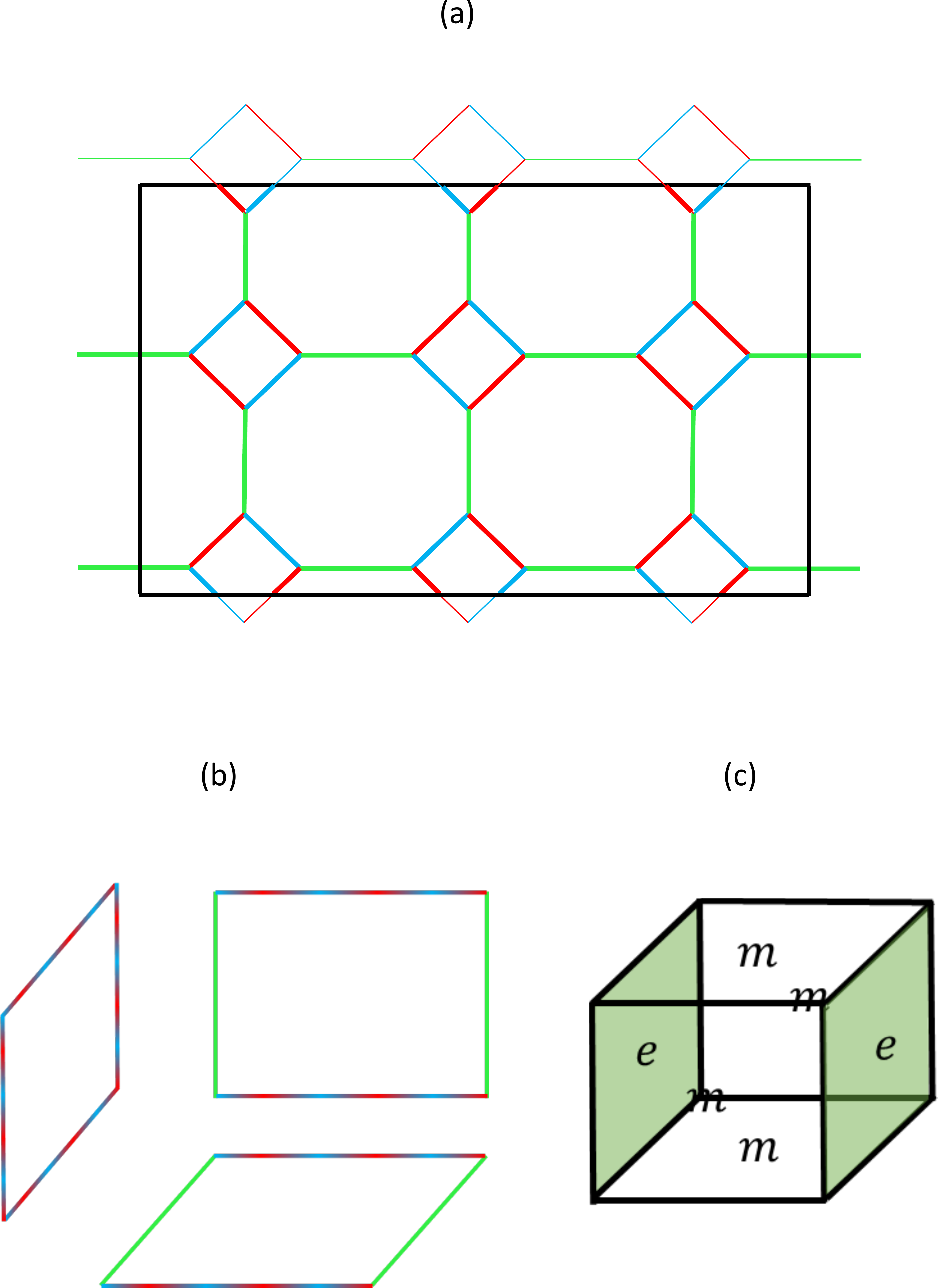}
    \caption{(a) Truncation of a layer of the square-octagon lattice in the 2D Floquet TC. (b) The boundary truncations of the square-octagon layers that form the lattice for the 3D Floquet TC. (c) Boundaries for a planar configuration of the 3D Floquet TC: we get $e$-condensing boundaries on the left and right while the $m$-condensing boundaries on the other four sides. Note that in the ISGs that are FDLQC-equivalent to two copies of 3D TC, the boundary labeled $e$ should be interpreted as having two sublattices, one condensing $e_1$ and the other condensing $e_2$; the same holds for the boundary labeled $m$. }
    \label{fig:truncation_layers}
\end{figure}

We now consider a boundary construction for the 3D Floquet TC. Since the 3D Floquet TC is based on a coupled layer construction, we start by reviewing the boundaries for the 2D Floquet TC on the square-octagon lattice~\cite{Gidney2022benchmarkingplanar,Paetznick2023Performance}. We consider, in particular, the truncation of the square-octagon lattice shown in Fig.~\ref{fig:truncation_layers}(a). Here, any check that is truncated is included in the check group as a single-qubit check.
In the first R-ISG of the schedule GBRBGR, we get $m$-boundaries (i.e., smooth boundaries)
where the truncation cuts through the blue and red edges, and $e$-boundaries (i.e., rough boundaries) where the truncation cuts through the green edges. After half a period of the rewinding schedule, the boundaries switch types. 

We now use this construction of boundaries for the 2D Floquet code to determine the boundaries of the 3D Floquet TC. The truncation for the three foliations of square-octagon layers is specified in Fig.~\ref{fig:truncation_layers}(b). 
For one of the foliations, the truncation goes through only the blue and red edges while for the other two, the truncation goes through the red and blue edges on the top and bottom and green edges on the left and right sides. 
The resulting 3D Floquet TC is such that, for each ISG, we have charge-condensing boundaries on the left and right and the loop-condensing boundaries on the remaining four sides, as in Fig.~\ref{fig:truncation_layers}(c). 

We note that the particular choice to truncate the blue and red edges on the lower halves of the square plaquettes ensures that in the 3D Floquet TC lattice, the two-qubit condensation checks in the G-ISG are never truncated. 
This choice can nonetheless lead to truncated evolved condensation checks. 
The truncated evolved condensation checks survive as stabilizers along the boundary with truncated green edges, while they do not survive along the boundaries with truncated blue and red edges since they do not commute with the single-qubit blue and red checks. 

The result of these truncations is that, for each ISG, the charges (possibly two types $e_1$ and $e_2$, depending on the round) are condensed on the right and left boundaries, while the loop-like excitations are condensed on the remaining four sides. 

It is straightforward to understand this result from the perspective of the 2D TCs on the effective lattice. In the G-ISG, the $e$-charge is associated with the product of octagonal plaquettes, which also correspond to $e$-anyons of the layers. Since the truncation through green checks creates an $e$-condensing boundary for the 2D G-ISGs, the $e$-charge of the 3D code also condenses at that boundary. Similarly, the loop excitations are condensed at the truncation through the blue and red edges. 

In both of the B-ISGs, we have the effective picture of the 2D rotated TCs, and the evolved condensation checks in the bulk are of the form $X_{\text{eff}}^{\otimes 4}$, which implies that the $e$-charges are associated with products of $Z_{\text{eff}}$-stabilizers. For the B-ISG of the 2D Floquet code, the truncation through green checks condenses the violations of the $Z_{\text{eff}}$-stabilizers supported on the octagonal plaquettes consisting of red and green edges. This is illustrated in Fig.~\ref{fig:boundaries_2D_Floquet_code}(c). Thus, for the B-ISG of the 3D Floquet code, at the truncation through green edges, we condense the violations of these $Z_{\text{eff}}$-stabilizers, corresponding to point-like excitations, and at the truncation through blue and red edges, we condense the violations of the $X_{\text{eff}}$-stabilizers corresponding to loop-like excitations. 

After the B-round, we have the $\text{R}^{(1)}$-ISG. For both of the R-ISGs, we can again use the effective picture of rotated toric layers along three foliations and the action of evolved condensation checks. For the 2D Floquet TC, the truncation through green edges gives an $e$- ($m$-) boundary in the $\text{R}^{(1)}$-ISG ($\text{R}^{(2)}$-ISG). This is illustrated in Fig.~\ref{fig:boundaries_2D_Floquet_code}, where the stabilizers at the boundary are shown in terms of the effective Pauli operators. The evolved condensation checks in the $\text{R}^{(1)}$-ISG and $\text{R}^{(2)}$-ISG are given by $Z_{\text{eff}}^{\otimes 4}$ and $X_{\text{eff}}^{\otimes 4}$, respectively, and are shown in Fig.~\ref{fig:gapped_boundaries_3D_Floquet_bTC}. 
In the $\text{R}^{(1)}$-ISG, since the evolved condensation checks are given by $Z_{\text{eff}}^{\otimes 4}$, the stabilizer products corresponding to $e$-charges ($m$-loops) are given by the $X_{\text{eff}}$ ($Z_{\text{eff}}$) stabilizers. In the $\text{R}^{(1)}$-ISG of the 2D Floquet code, the truncation through green edges condenses $X_{\text{eff}}$ stabilizers; see Fig.~\ref{fig:boundaries_2D_Floquet_code}. Hence, in the $\text{R}^{(1)}$-ISG of the 3D Floquet TC, the truncation through green edges condenses the point-like excitations. Similarly, the truncation through the blue and green edges condenses the loop-like excitations of the 3D TCs.   

Similarly, in the $\text{R}^{(2)}$-ISG, the evolved condensation checks are given by $X_{\text{eff}}^{\otimes 4}$ and $X_{\text{eff}}^{\otimes 3}$ as shown in Fig.~\ref{fig:gapped_boundaries_3D_Floquet_bTC} and the stabilizer products corresponding to $e$-charges ($m$-loops) are given by the $Z_{\text{eff}}$ ($X_{\text{eff}}$) stabilizers. As shown in Fig.~\ref{fig:boundaries_2D_Floquet_code}, the truncation through green edges condenses the excitations of the $Z_{\text{eff}}$-stabilizers in the $\text{R}^{(2)}$-ISG. Hence, in the $\text{R}^{(2)}$-ISG of the 3D Floquet TC, the truncation through green edges again condenses the point-like excitations, and the truncation through the blue and green edges condenses the loop-like excitations of the 3D TCs. 
To conclude, even though the 2D Floquet code layers undergo a boundary transformation from $e$-type to $m$-type, the 3D Floquet TC boundaries of given types condense the same type of excitations in each ISG respectively. 

\subsection{Transversal non-Clifford gate}

One key computational advantage of the 3D TC is that it allows for an implementation of the transversal logical non-Clifford gate~\cite{Vasmer2019transversal}. Such an advantage is retained for our 3D Floquet TC. This is because the 3D TC in the G-ISG is precisely the conventional cubic lattice 3D TC up to concatenation with a 2-qubit repetition code. We can stack the 3D Floquet TC with two copies of the 3D TC on the checkerboard lattice to yield the transversal $CCZ$ gate as an on-site symmetry of the stabilizer group~\cite{Vasmer2019transversal}. More specifically, on a system with both the 3D Floquet TC and two 3D checkerboard lattice TCs, we can perform the transversal logical $CCZ$ gate in the round of the Floquet cycle in the G-ISG. Since the logical information is preserved under subsequent measurements for both the 3D Floquet TC and the 3D checkerboard lattice TCs, one can wait to do the $CCZ$ gate transversally in the G-ISG and then proceed with subsequent rounds.

\begin{figure*}
    \centering
    \includegraphics[scale=0.1]{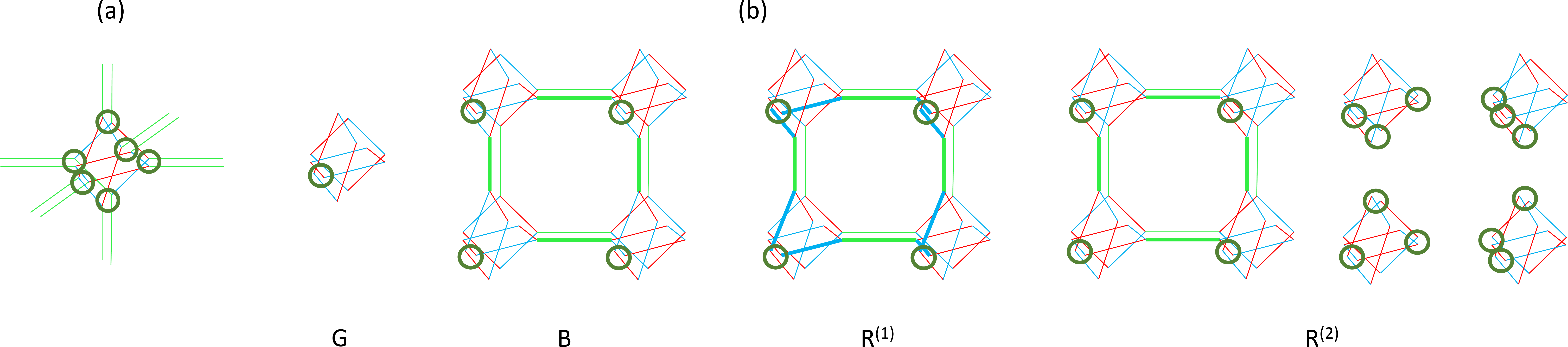}
    \caption{(a) Two-qubit condensation checks (green round) in the X-cube Floquet code. (b) The evolution of one condensation check in the Floquet code with schedule GB$\text{R}^{(1)}$BG$\text{R}^{(2)}$. The different rounds are labeled. There is no difference in the ISGs in the two B-rounds (and in the two G-rounds) due to rewinding and hence only one B-round (G-round) is shown. In the $\text{R}^{(2)}$-ISG, besides products of 2-qubit condensation checks and green checks (shown on the left in $\text{R}^{(2)}$ subfigure), we also have products of 2-qubit condensation checks around every octahedron (shown on the right in $\text{R}^{(2)}$ subfigure) that survive as stabilizers.}
    \label{fig:Xcubeevolution}
\end{figure*}

The checkerboard lattice surface code (planar variant) can be obtained as instantaneous stabilizer codes of the 3D subsystem TC, which uses measurements of three-qubit checks.~\cite{kubica2022}. Hence, one can stack the planar variant of our 3D Floquet code with two copies of the 3D subsystem TC to do the non-Clifford $CCZ$ gate in the G-round of the 3D Floquet TC.

\section{Rewinding X-cube Floquet code}
\label{sec:Xcube}

The X-cube Floquet code of Ref.~\cite{XcubeFloquet2022} has ISGs that are for some rounds, stacks of 2D TC, and for other rounds, FDLQC-equivalent to the X-cube model or another fracton model. {This Floquet code is again based on a coupled layer construction and uses stacks of 2D Floquet TCs on the square-octagon lattices, as used for our 3D Floquet TC in the previous section. Hence, the microscopic lattice is the same as given in Fig.~\ref{fig:foliation_3DTC2}, and the effective underlying lattice for the B-ISG is the same as given in Fig.~\ref{fig:sublattices_BISG}(a). However, the sequence of measurements is different and hence, the stabilizers supported on those lattices are different.} 

In this section, we propose a rewinding schedule for the X-cube model such that each ISG is a fracton model, and unlike Ref.~\cite{XcubeFloquet2022}, we do not go to an intermediate ISG of stacks of TC. We consider a rewinding schedule of the form GB$\text{R}^{(1)}$BG$\text{R}^{(2)}$ where $1$ and $2$ label the position of the two R-rounds in the sequence. The on-site condensation checks, as shown in Fig.~\ref{fig:Xcubeevolution}(a), are measured along with the green checks in the G-round. The rewinding schedule ensures that in each subsequent round, not all condensation checks evolve into nonlocal stabilizers. 
The evolution of the condensation checks under the rewinding schedule is shown in Fig.~\ref{fig:Xcubeevolution}. Note that, if we had used the schedule GBR, then the R-round would have been followed by the G-round, and all the condensation checks would have evolved into non-local stabilizers, resulting in an ISG of a subspace of stacks of 2D TC. This is because after a full cycle GBRG where no extra condensation checks are measured, the measurements of the local checks in the BRG part of the sequence (without the 2-qubit condensation checks) result in an ISG of 3-foliated stack of 2D TCs while the 2-qubit condensation checks measured in the first G-round evolve into logical operators of the stacks; this leads to an ISG that is a subspace of the stack of 2D TCs. 

The ISGs in the rewinding X-cube Floquet code are FDLQC-equivalent to those listed in Table~\ref{tab:XCschedule}. 
Explicit FDLQCs to map the ISGs to these models are given in the supplementary \texttt{Mathematica} files. The G-ISG is exactly the canonical X-cube model concatenated with 4-qubit repetition codes on the composite green edges. The B-ISG is FDLQC-equivalent to a product of decoupled models (up to non-local stabilizers), including the X-cube model, 3D TC, and a 3-foliated stack of 2D TC.  
Similar to the 3D Floquet TC, the ISG in the $\text{R}^{(1)}$-round is related to that of the B-rounds by a basis change and hence is FDLQC-equivalent to the same models. The ISG in the $\text{R}^{(2)}$-round, which follows immediately after the G-round, has different evolved condensation checks as shown in Fig.~\ref{fig:Xcubeevolution}. The $\text{R}^{(2)}$-ISG is FDLQC-equivalent to a product of the X-cube model and a 3-foliated stack of 2D TCs. We note that such splitting of topological order also happens in the (non-rewinding) X-cube Floquet of Ref.~\cite{XcubeFloquet2022}. However, it was not identified in that work.

\begin{figure}
    \centering
    \includegraphics[scale=0.1]{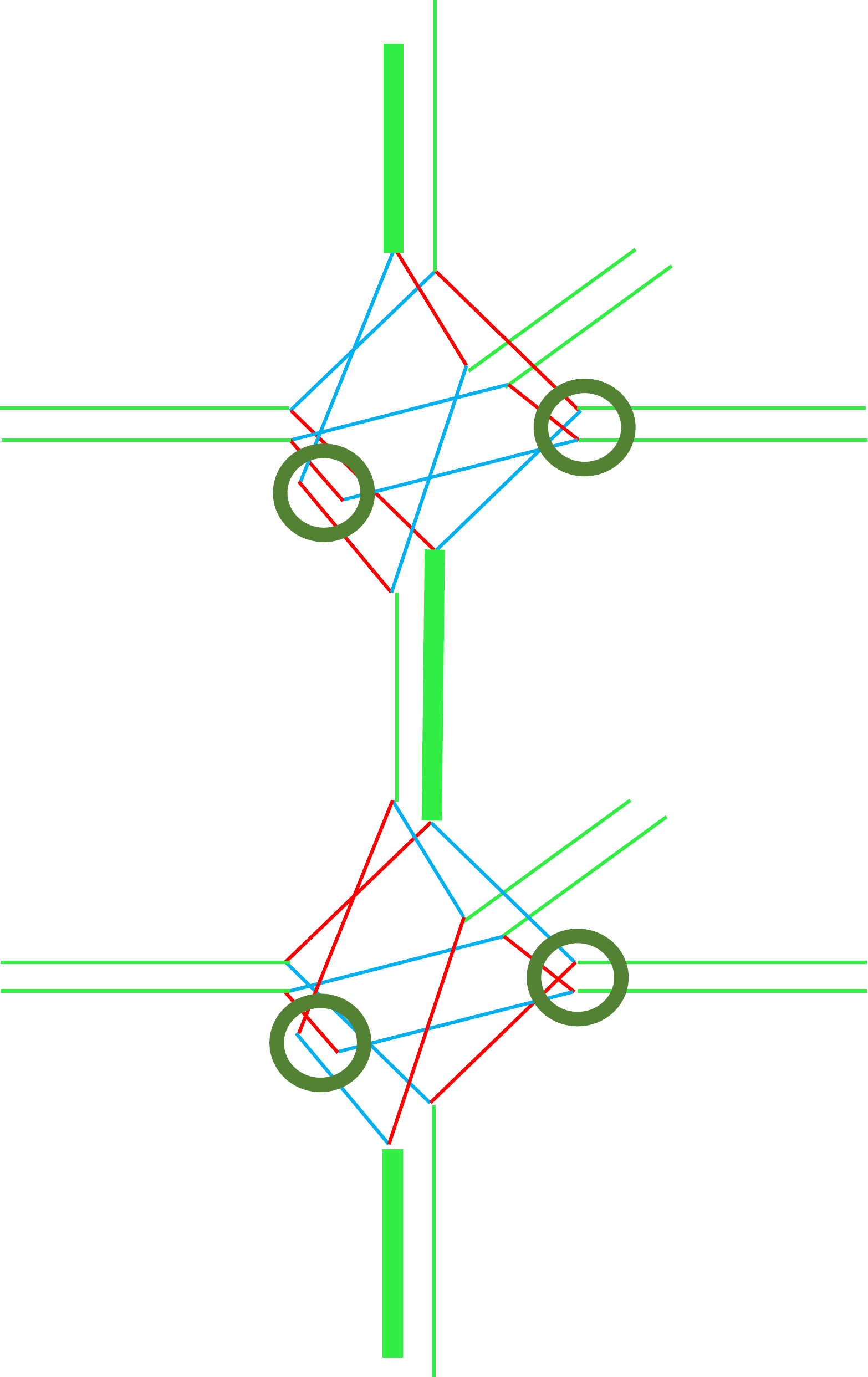}
    \caption{An example of a non-local stabilizer (only a cut-out part is illustrated) in the B-ISG of the X-cube Floquet code. It is a product of the green checks shown using thick lines and 2-qubit condensation checks marked using thick circles.}
    \label{fig:Xcube_nonlocal_stabilizer}
\end{figure}

\begin{table}[]
    \centering
    \begin{tabular}{|c|l|c|}
    \hline
Round   & Measurement  & ISG\\
\hline
0& G (Green $YY$+ interlayer $YY$) & XC\\
1& B (Blue $ZZ$) & XC $\times$ 3D TC $\times$ stacks\\
2& R (Red $XX$) & XC $\times$ 3D TC $\times$ stacks\\
3& B (Blue $ZZ$) & XC $\times$ 3D TC $\times$ stacks\\
4& G (Green $YY$+ interlayer $YY$) & XC \\
5& R (Red $XX$) & XC $\times$ stacks\\
\hline  
\end{tabular}
\caption{The ISGs in the (rewinding) GBRBGR schedule of measurements for the X-cube Floquet code. The measured checks and the instantaneous stabilizer groups (ISGs) in each round are written in the Measurement column. In the ISG column, XC denotes the X-cube model, 3D TC denotes 3D TC, and ``stacks'' denote a 3-foliated stack of 2D TCs.  For even system sizes $L=2n$, the 3-foliated stack consists of $n$ layers of 2D TC along each of the three orthogonal lattice directions.}
\label{tab:XCschedule}
\end{table}

We now discuss the counting of logical qubits in the rewinding X-cube Floquet code. 
As mentioned, the G-ISG is exactly the canonical X-cube model concatenated with 4-qubit repetition codes on the composite green edges. Hence, we have $6L-3$ logical qubits in the G-ISG~\cite{vijay2016}. To count the number of logical qubits in the B-ISGs and $\text{R}^{(1/2)}$-ISG, we consider even system sizes $L=2n$ for simplicity. The FDLQC-equivalent model, as stated in Table~\ref{tab:XCschedule} implies a total of $6L$ logical qubits as we get $6n-3$ logical qubits from the X-cube model, 3 logical qubits from the 3D TC, and $6n$ logical qubits from the 3-foliated stack of 2D TCs. The B-ISG has three independent non-local stabilizers; an example is shown in Fig.~\ref{fig:Xcube_nonlocal_stabilizer}. Due to these three non-local stabilizers, the number of logical qubits the B-ISG is $6L-3$ which in turn implies that the $6L-3$ qubits from the G-ISG are all preserved. For an effective description of the B-ISG in terms of rotated 2D TC stack and the counting of logical qubits using that, see Appendix~\ref{sec:eff_description_and_counting_BISG_XC}. The $\text{R}^{(1)}$-ISG is the same as the B-ISG up to a basis change, and hence, we have $6L-3$ logical qubits. Due to rewinding, the second B-ISG is exactly the same as the first B-ISG. The $\text{R}^{(2)}$-ISG is FDLQC-equivalent to the X-cube model and a 3-foliated stack of 2D TCs; hence, the number of logical qubits is $6L-3$. Overall, $6L-3$ logical qubits are preserved in the rewinding X-cube Floquet code.

\begin{figure}[tb]
\centering
    \includegraphics[scale=0.2]{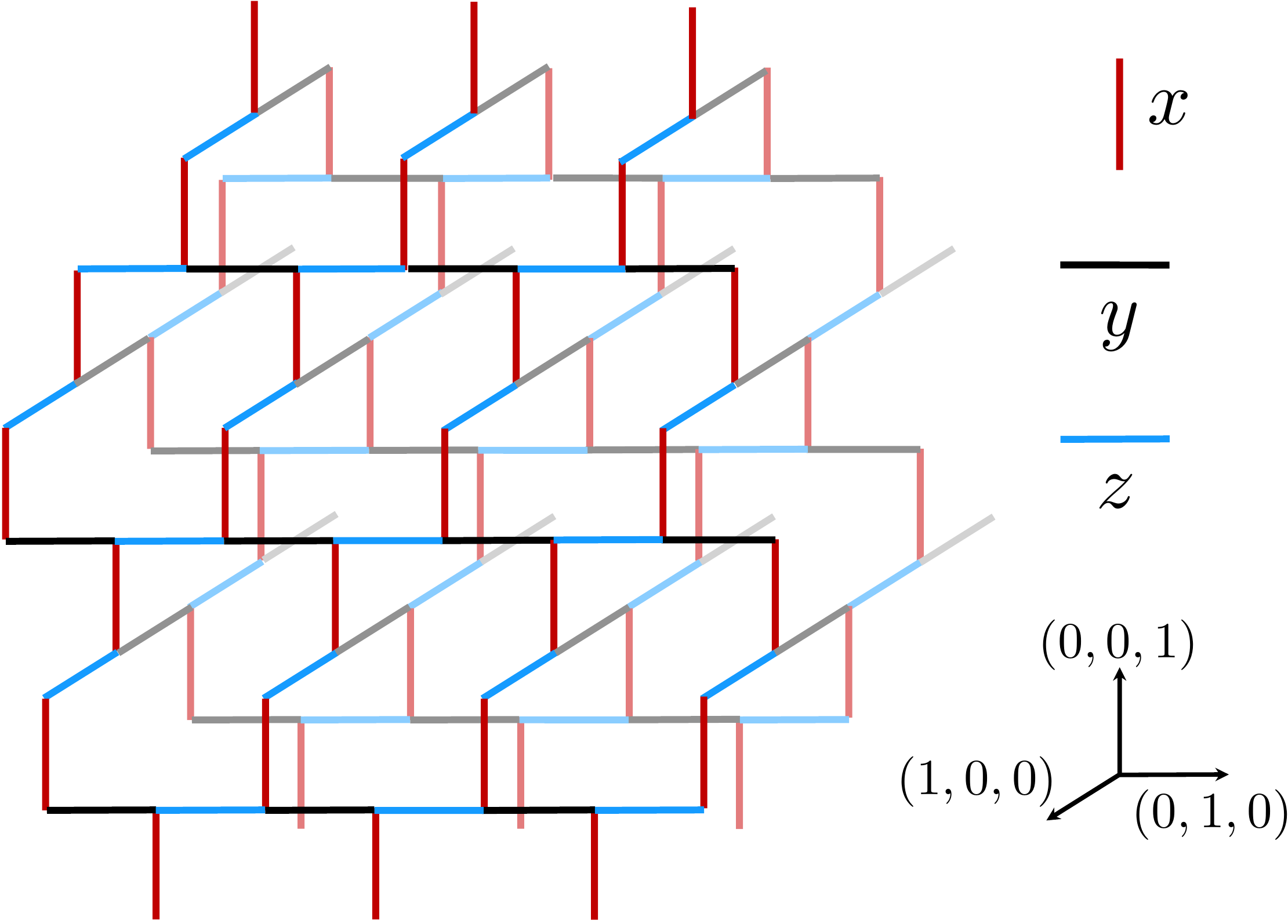}
    \caption{
    The 3D fTC is defined on a trivalent lattice. The edges are labeled by $x$ (red), $y$ (black), and $z$ (blue) corresponding to XX, YY and ZZ checks.}
    \label{fig:3DFloquetlattice}
\end{figure}

\begin{figure*}[tb]
    \centering
   \subfloat[]{ \includegraphics[width=.14\textwidth]{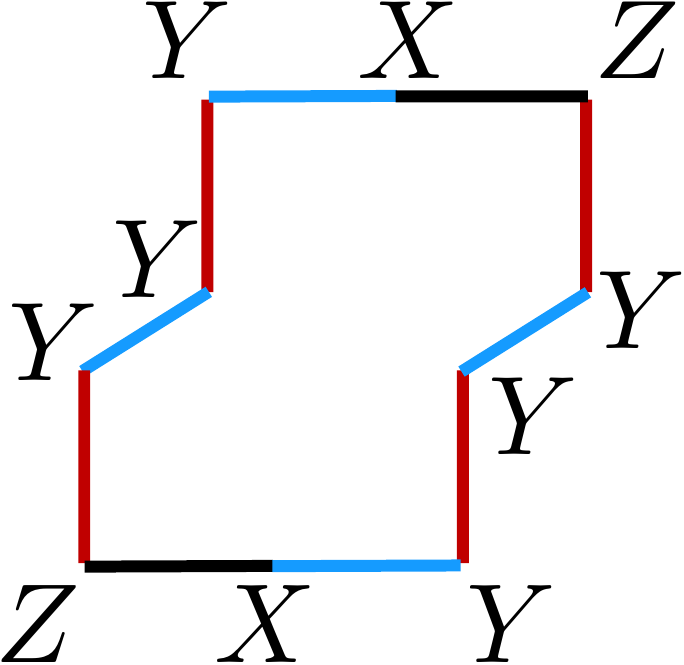}}\qquad
    \subfloat[]{ \includegraphics[width=.14\textwidth]{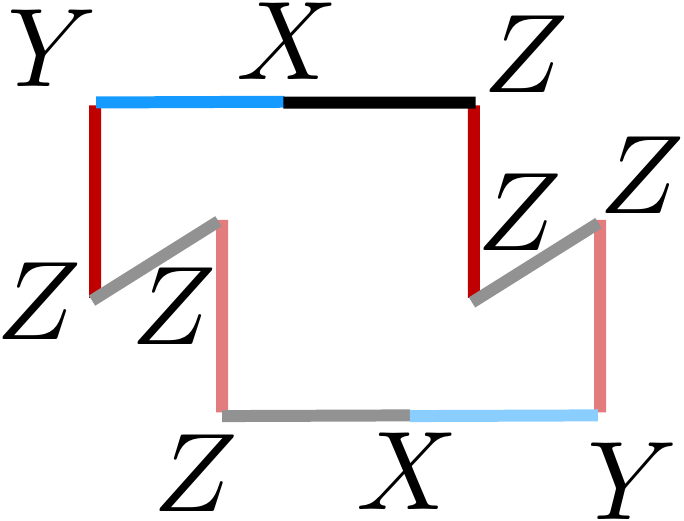}}\qquad
     \subfloat[]{ \includegraphics[width=.12\textwidth]{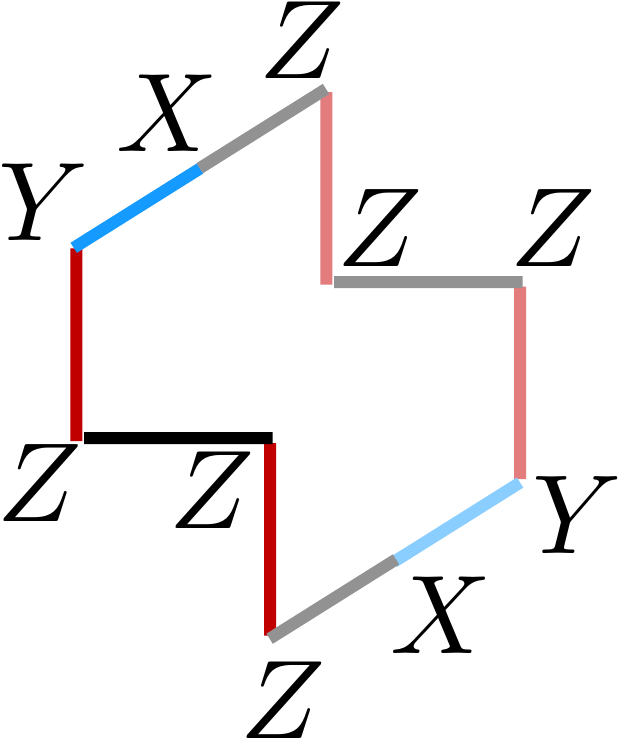}}\qquad
      \subfloat[]{ \includegraphics[width=.115\textwidth]{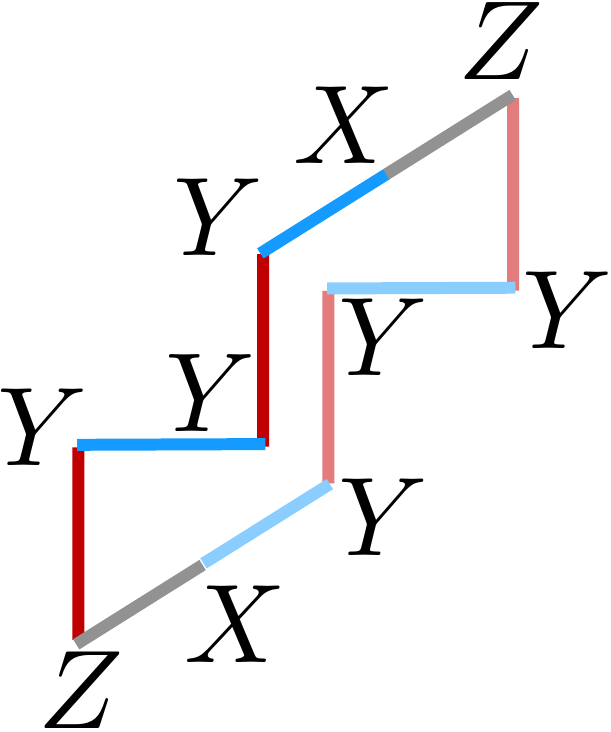}}\qquad
      \subfloat[]{ \includegraphics[width=.18\textwidth]{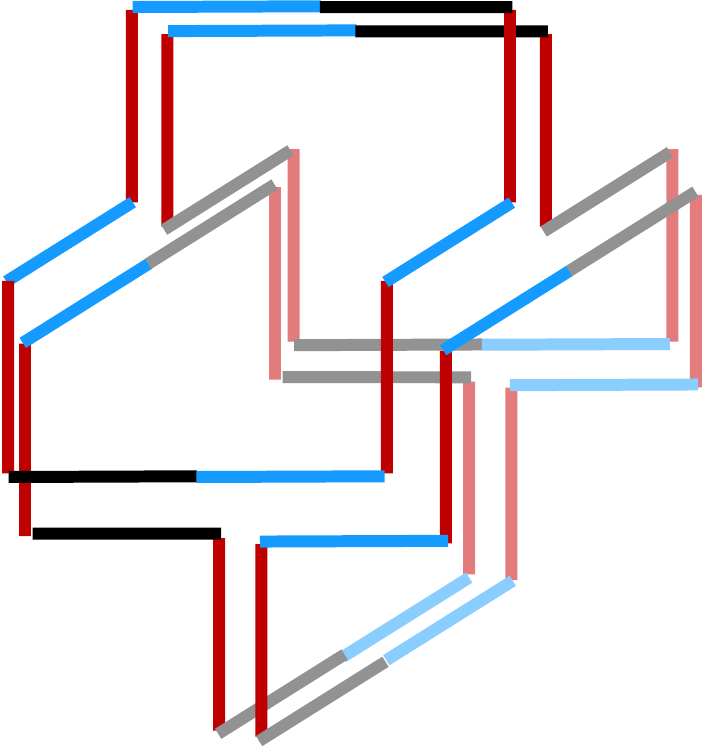}}
    \caption{There are four types of local generators of the stabilizer group: the (a) front, (b) back, (c) right, and (d) left armchair stabilizers, which are equivalent to a product of checks around the plaquette. (e) The four armchair stabilizers satisfy a local relation.}
    \label{fig:armchairorientations}
\end{figure*}

\begin{figure*}[]
    \centering
   \subfloat[]{ \includegraphics[scale=0.058]{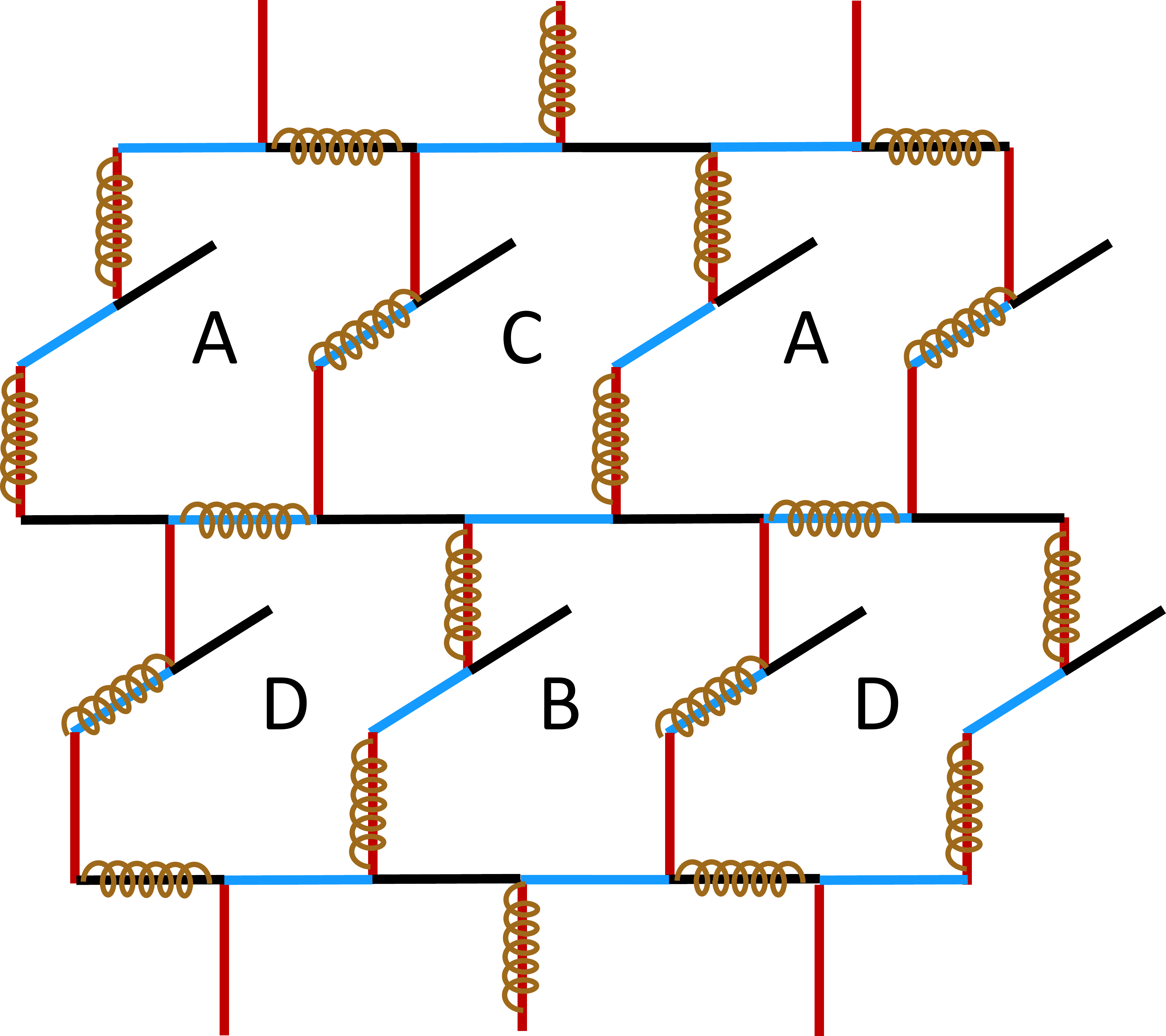}}\quad
    \subfloat[]{ \includegraphics[scale=0.058]{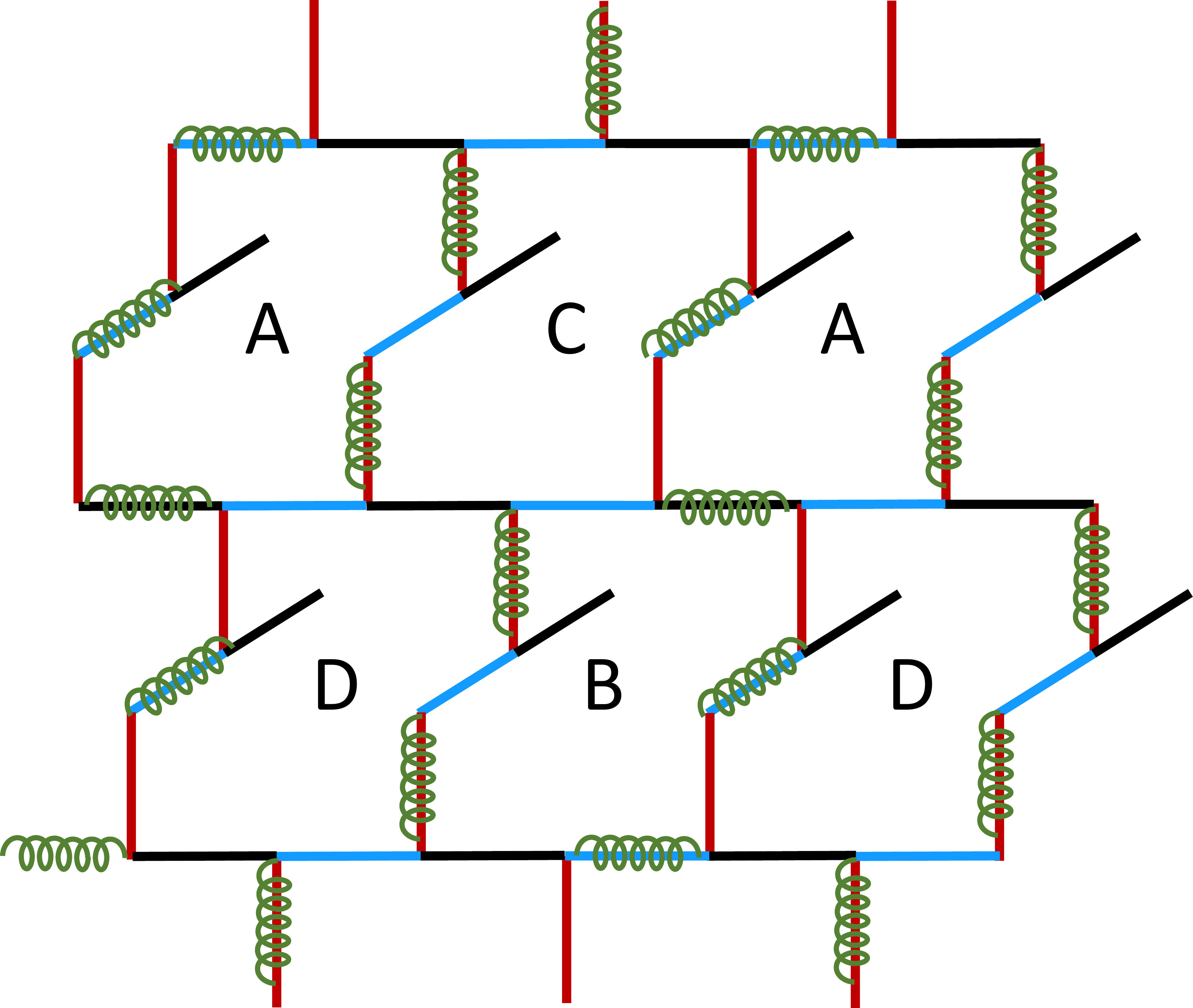}}
    \vspace{4mm}
     \subfloat[]{ \includegraphics[scale=0.058]{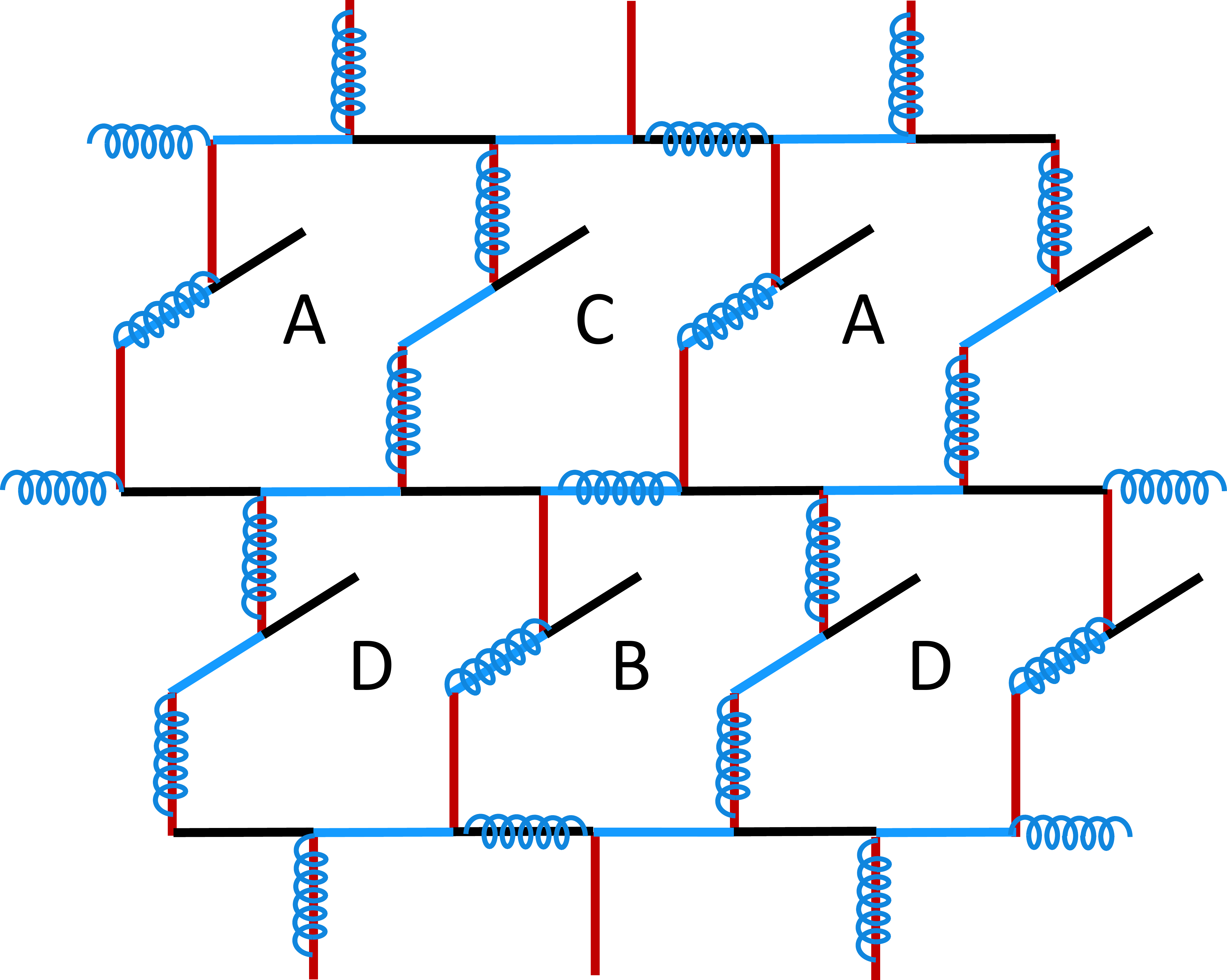}}\quad
      \subfloat[]{ \includegraphics[scale=0.058]{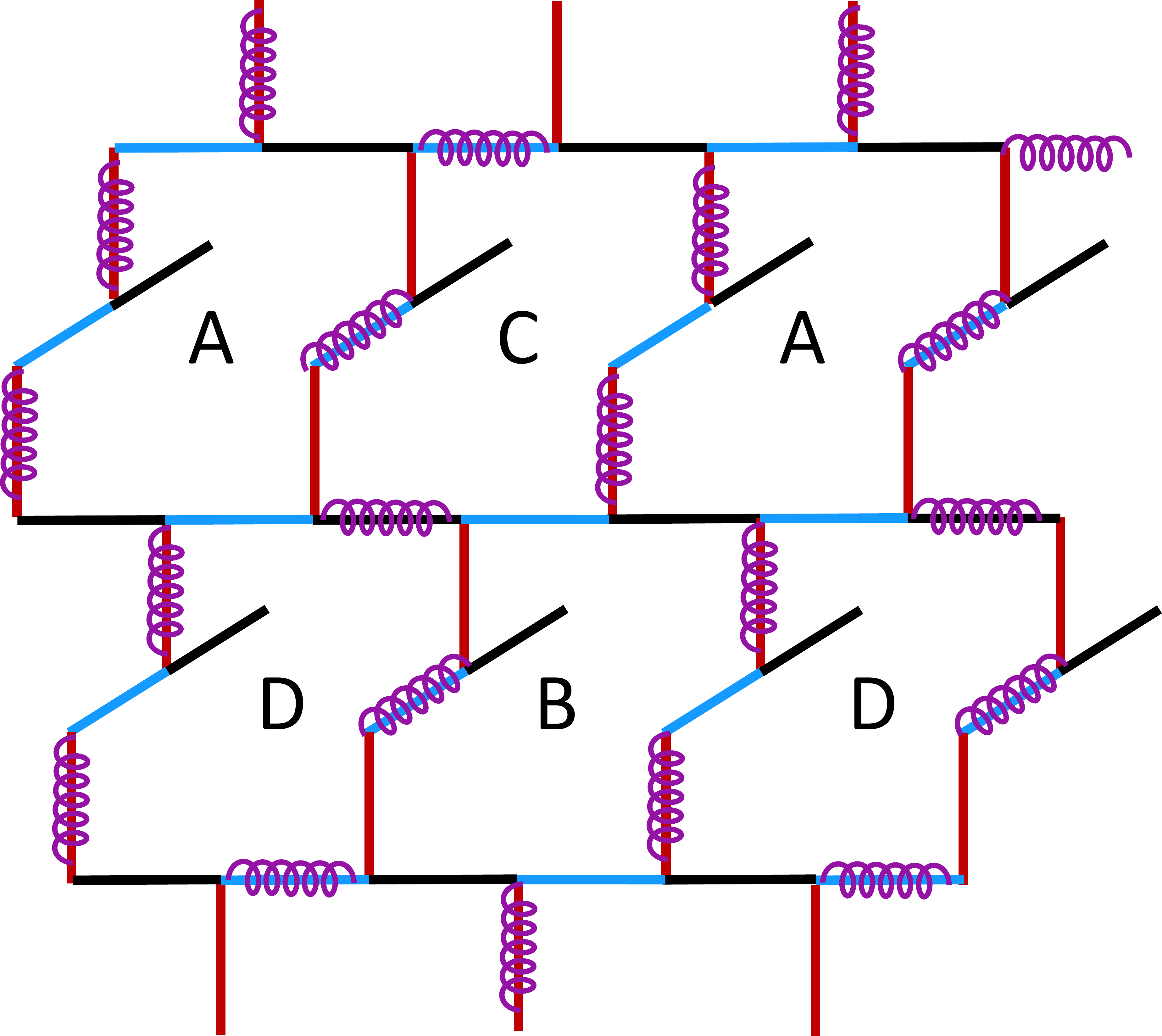}}
    \caption{(a)-(d) The front-facing armchair stabilizers can be inferred from five rounds of measurements. The measurements in rounds 0-3 are shown in brown, green, blue, and pink, respectively. The final round is a repetition of round 0. The measurements for the back-facing and left-facing armchairs are defined similarly. The labels A, B, C, and D shown for the front-facing armchair stabilizers are decided according to the pairs of rounds of checks used to infer them i.e., 01, 12, 23, and 30 respectively as described in the text; see also Fig.~\ref{fig:ABCDlabels}.}
    \label{fig:3Dschedule}
\end{figure*}

\begin{figure}[tb]
\centering
    \includegraphics[scale=0.25]{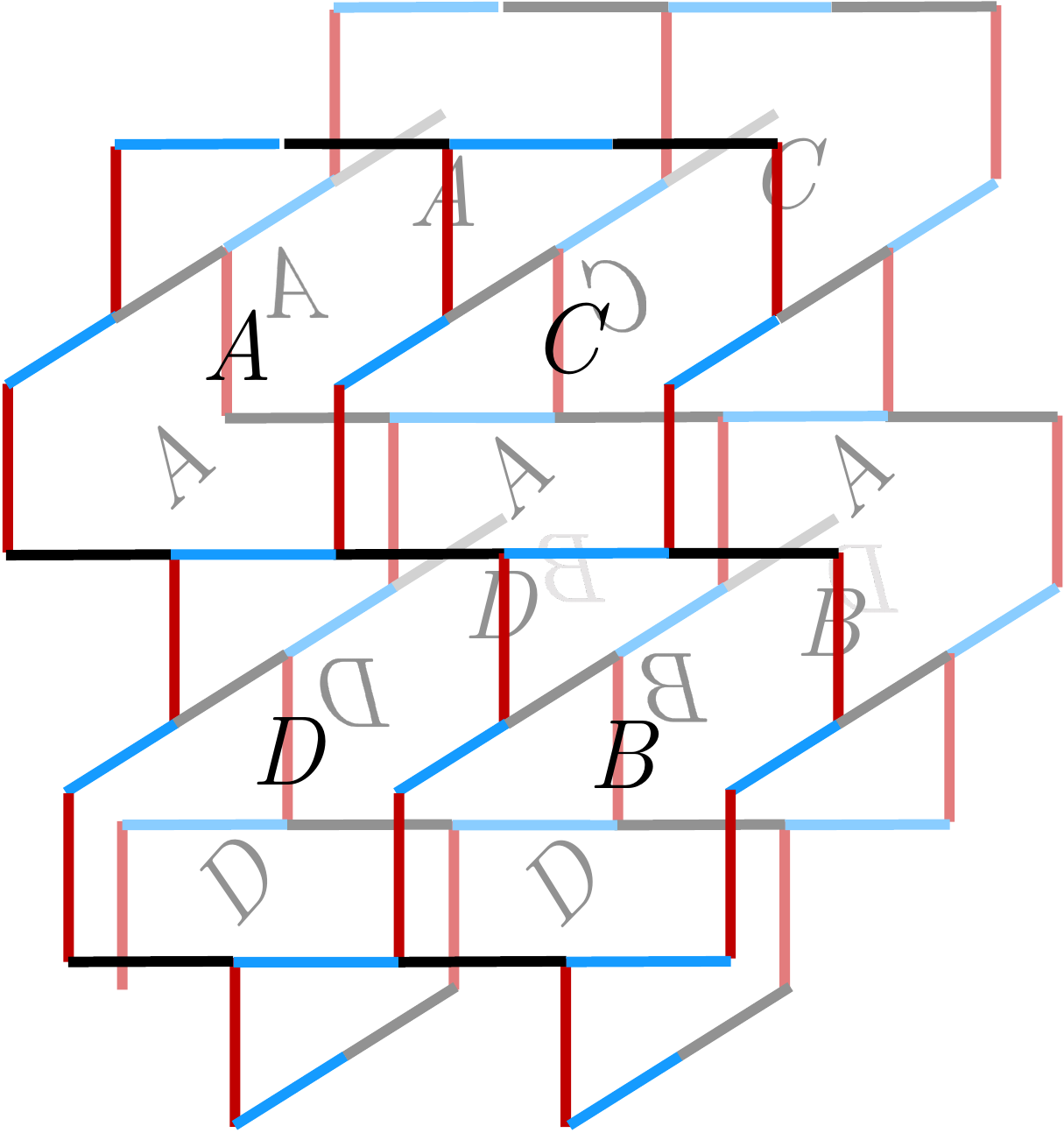}
    \caption{
    There are four labels for the armchairs: A, B, C, and D. The labels for front-facing armchairs are written with the usual orientation of $A$, $B$, $C$, $D$, while those for the right- and back-facing armchairs are tilted and reflected. The labels for the $B$ and $C$ right-facing armchairs do not appear in this portion of the lattice.}
    \label{fig:ABCDlabels}
\end{figure}

\begin{figure*}[tb]
    \centering
   \subfloat[]{ \includegraphics[width=.26\textwidth]{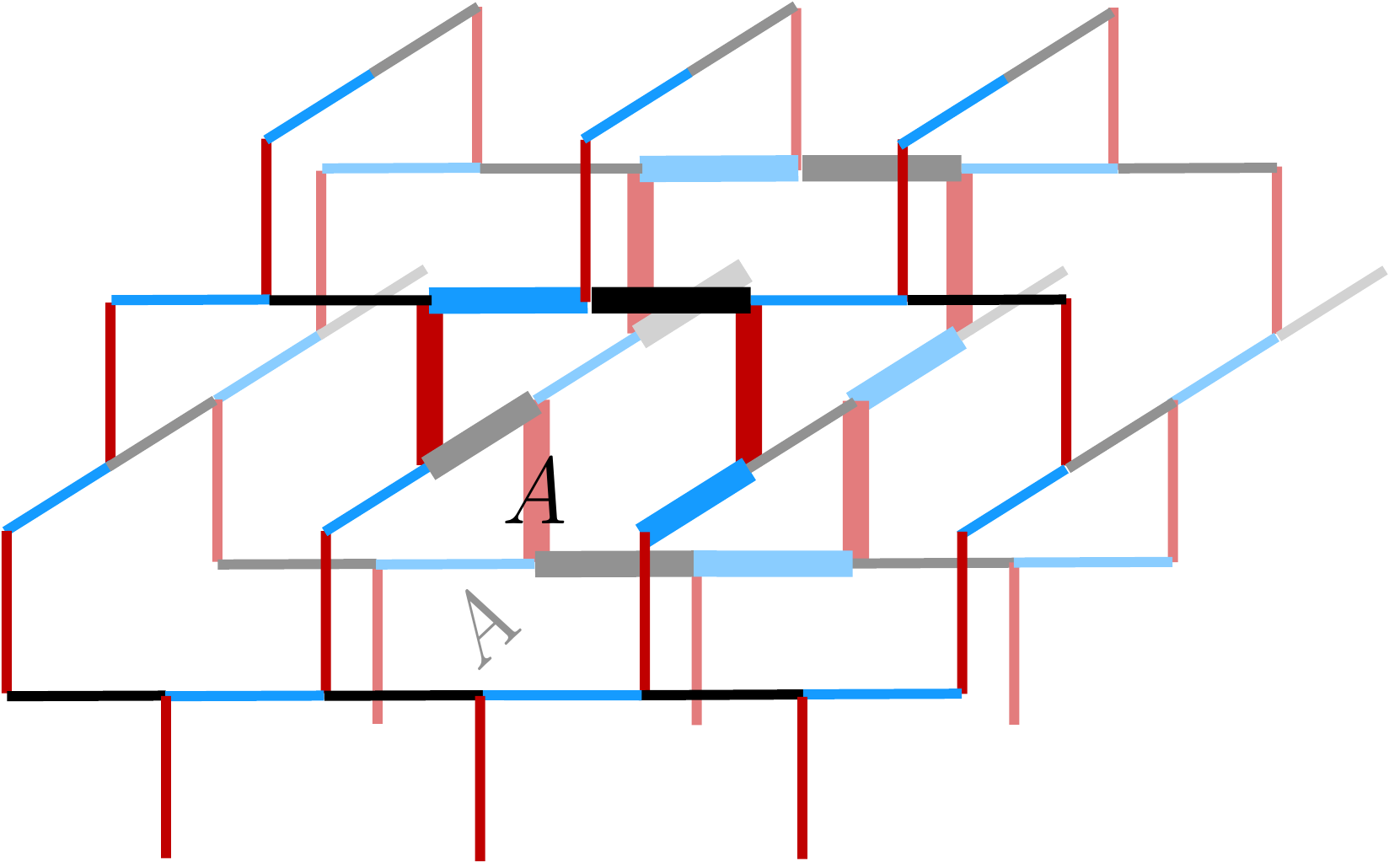}} \qquad
    \subfloat[]{ \includegraphics[width=.26\textwidth]{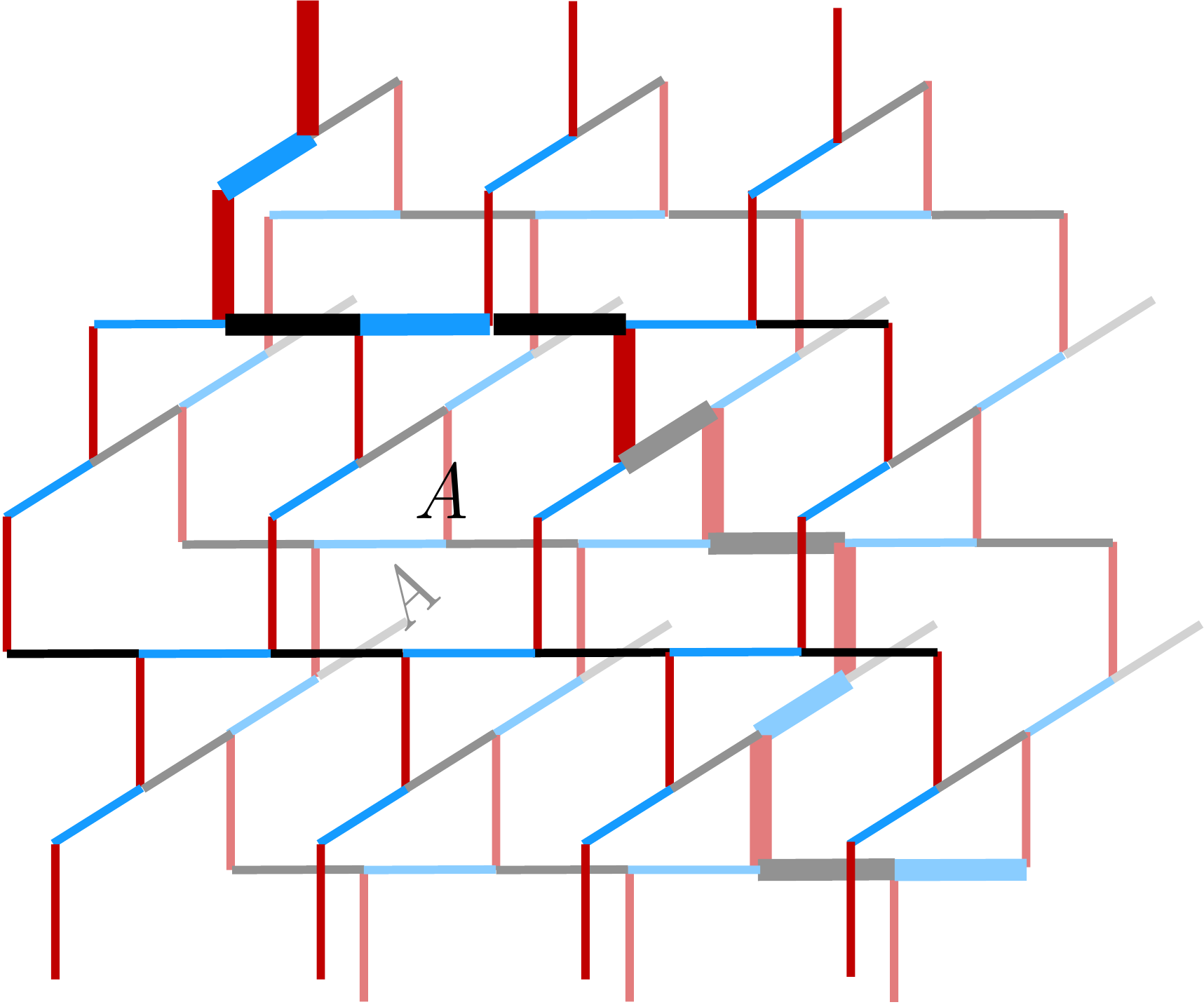}} \qquad
     \subfloat[]{ \includegraphics[width=.26\textwidth]{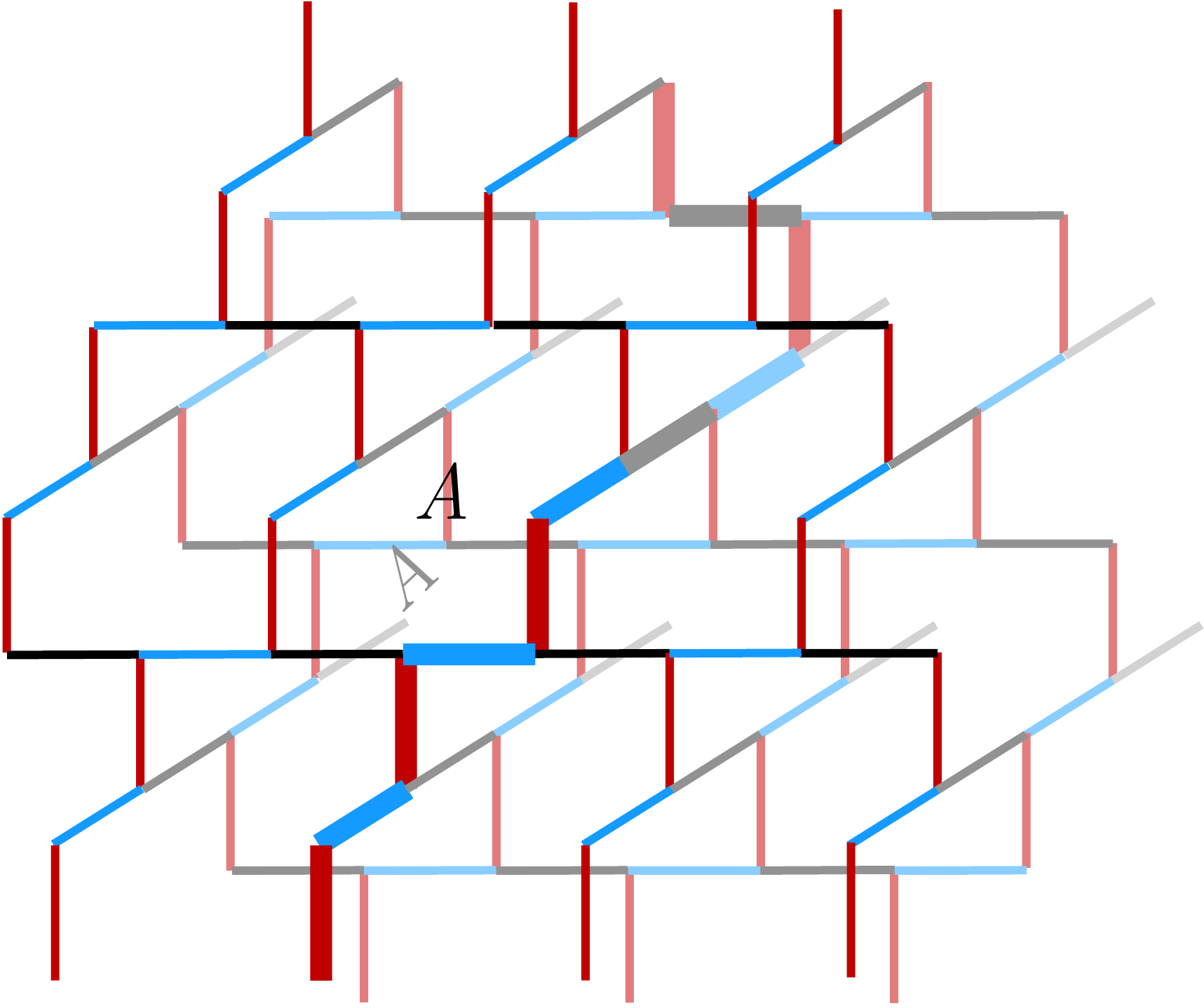}}
    \caption{Logical operators are inadvertently measured using the FBR schedule when transitioning between inferring the armchair stabilizers with different orientations. The logical operators measured in the transitions (a) F to B, (b) B to R, and (c) R to F are shown. We note that the two specified $A$ plaquettes are sufficient to determine the labeling of the other plaquettes.}
    \label{fig:innerlogicalsmeasured}
\end{figure*}

\begin{figure*}[tb]
    \centering
   \subfloat[]{ \includegraphics[width=.3\textwidth]{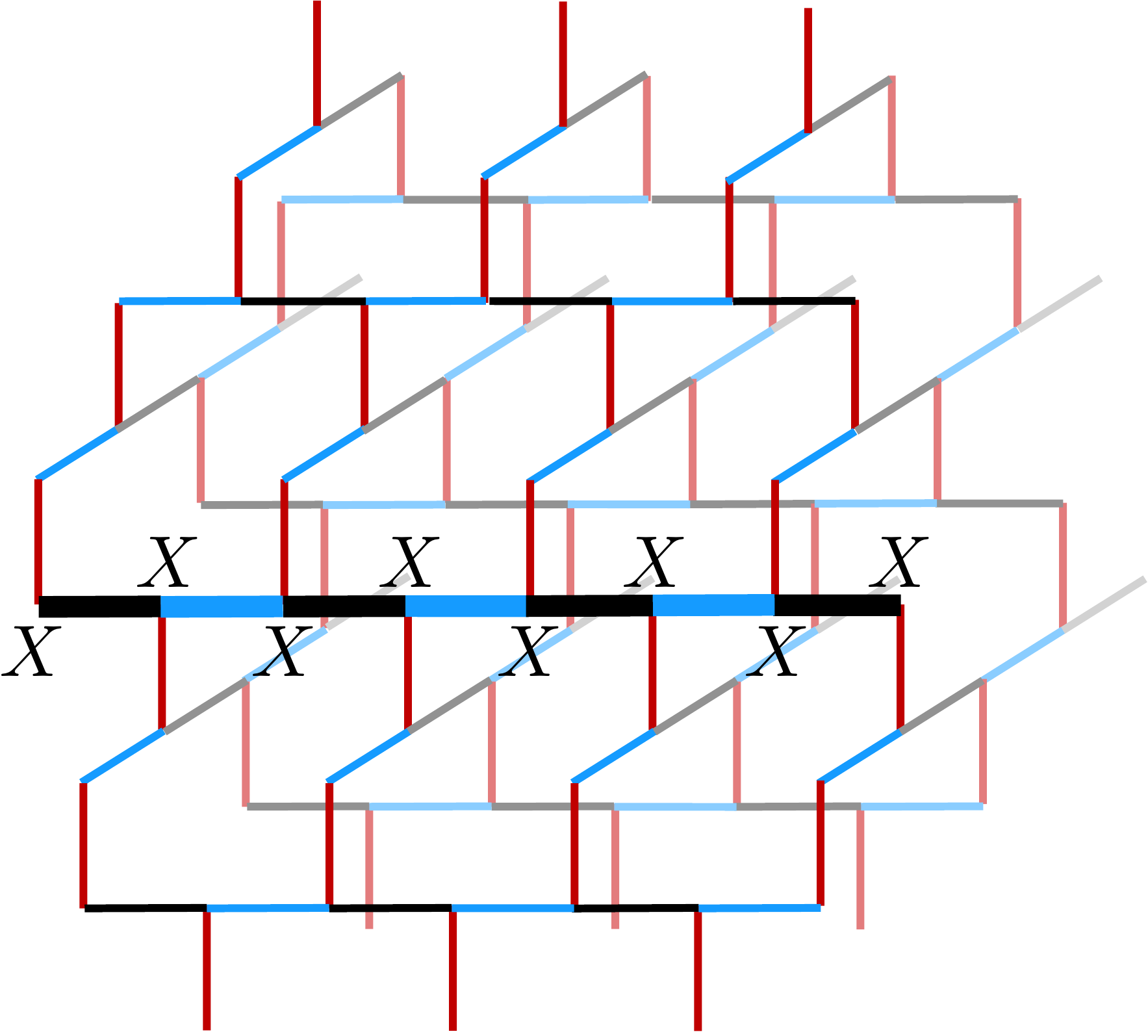}} \qquad
    \subfloat[]{ \includegraphics[width=.3\textwidth]{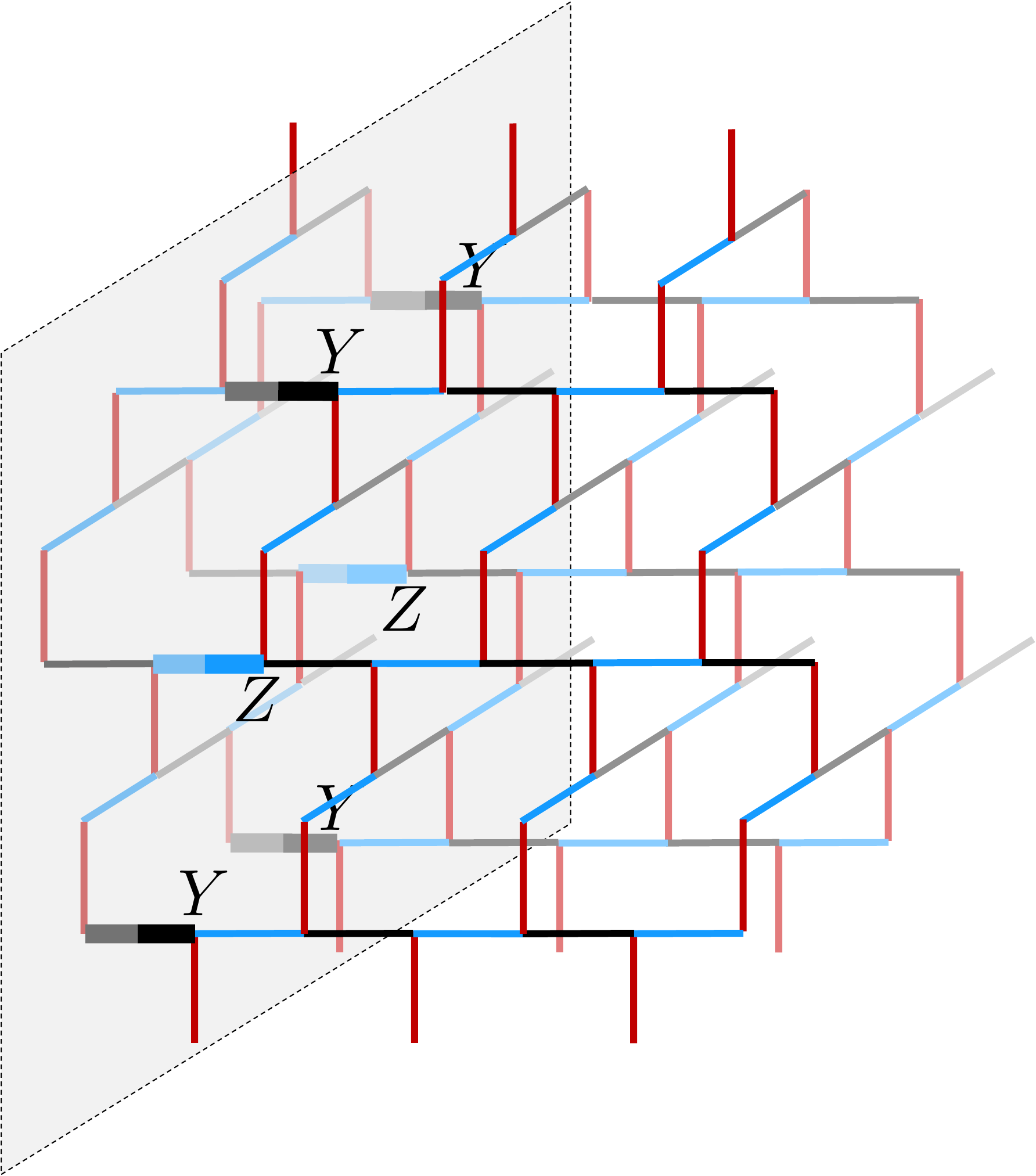}} 
    \caption{The Floquet 3D fTC encodes one logical qubit. (a) The $Z$ logical operator is a product of 2-body check operators along the $(0,1,0)$ direction. (b) The $X$ logical operator can be represented by an operator supported on a membrane that is orthogonal to the string of the logical $Z$ operator.}
    \label{fig:3Dlogicals}
\end{figure*}

\section{3D Floquet fermionic toric code}
\label{sec:3DfermionicFloquet}

In this section, we present a Floquet code that has ISGs that are FDLQC-equivalent to the 3D fermionic TC (fTC) ~\cite{Levin2003Fermions,Yuan2019bosonization3D}. The construction is based on the 3D generalization of Kitaev's honeycomb model introduced in Ref.~\cite{mandal2009}, which has a fixed point gapped Hamiltonian of the (static) 3D fTC. {Even though the 3D fTC encodes three logical qubits,} our Floquet code with 3D fTC ISGs only encodes a single logical qubit on a system with periodic boundary conditions due to inadvertently measuring a subset of the logical operators. Again, we utilize a rewinding schedule to avoid measuring the logical operators for the remaining logical qubit.

\subsection{16-round measurement schedule} \label{sec: 3D Floquet}

Our first example is defined on the trivalent lattice in Fig.~\ref{fig:3DFloquetlattice} with periodic boundary conditions.\footnote{We note that any 2D lattice can be coarse-grained into a square lattice with a constant number of vertices per unit cell. Since the square lattice is 4-valent, the model can be generalized to a 3D diamond lattice by replacing the sites of the diamond lattice with the unit cell of the square lattice model. The lattice in Fig.~\ref{fig:3DFloquetlattice} can be constructed in this way starting from the honeycomb lattice.} 
We place a qubit at each vertex and label the edges with $x$, $y$, and $z$, as in Fig.~\ref{fig:3DFloquetlattice}. The 2-body check operators on the edges $x$, $y$, and $z$ are $XX$, $YY$, and $ZZ$, respectively.

Similar to the 2D Floquet TC in Section~\ref{sec:square-octagon_RGBRBGschedule}, the stabilizers of the check group are generated by products of the 2-body checks along closed paths, which include stabilizers supported on non-contractible paths. The local generators of the stabilizer group are 10-body products of check operators around a plaquette. Given the shape of these plaquettes, we refer to these stabilizers as the `armchair' stabilizers. 
There are four possible orientations for the armchairs: front-, back-, left-, or right-facing [see Fig.~\ref{fig:armchairorientations}(a)-(d)]. For each 3-cell, there exists a local relation between four armchair stabilizers, with one facing in each direction as pictured in Fig.~\ref{fig:armchairorientations}(e). Thus, we only need to infer the measurement outcomes of three out of the four orientations of armchair stabilizers.

Our measurement schedule is designed to extract the syndrome for one orientation of armchair stabilizers at a time. We use a separate set of five rounds of measurements to infer the front-, back-, and right-facing armchair stabilizers. We do not need to infer the measurement of the left-facing armchairs, given the local relation in Fig.~\ref{fig:armchairorientations}(e). The five rounds of measurements are shown in Fig.~\ref{fig:3Dschedule} and are labeled as 0, 1, 2, 3, and 0, where the 0 round is repeated at the end of the cycle. If we label the armchairs of a unit cell as $A$, $B$, $C$, and $D$, as in Fig.~\ref{fig:ABCDlabels}, we see that after applying consecutive $01$, $12$, $23$, and $30$ checks, the $A$, $B$, $C$, and $D$ armchair stabilizers are inferred, respectively. Therefore, the above schedule measures all the armchair stabilizers for a given orientation, as desired. We proceed similarly for the back- and right-facing armchair stabilizers.

Naively, a potential measurement schedule is to periodically measure the sequence FBR, where F, B, and R stand for the 5-round measurement schedules for extracting the front-, back-, and right-facing armchair syndromes. This schedule, however, does not exhibit any dynamically generated logical qubits. This is because, in transitioning between the orientations, e.g., F to B, we inadvertently measure the stabilizers supported along non-contractible paths around the torus. Specifically, in transitioning from F to B, B to R, and R to F, we measure the logical operators supported on non-contractible paths along the $(1,0,0)$-, $(0,1,1)$-, and $(1,0,1)$-directions, respectively (see Fig.~\ref{fig:innerlogicalsmeasured}). As such, the FBR schedule gives a Floquet code that does not encode any qubits.

To rectify this, we rewind the schedule at the level of the armchair orientations, i.e., we periodically measure the sequence FBFR. This sequence avoids the transition from B to R, implying that we do not inadvertently measure the stabilizer supported on a non-contractible path along the $(0,1,1)$-direction. We note that the second 5-round sequence F can be replaced with a single round of 0 measurements for the front-facing armchair. In summary, our schedule consists of repeating the following sequence of 16 rounds of measurements:
\begin{align}
    (0,1,2,3,0)_\text{F}(0,1,2,3,0)_\text{B}0_\text{F}(0,1,2,3,0)_\text{R},
\end{align}
where the subscripts denote the orientations of the armchairs. This modification to the FBR schedule ensures that we retain a single dynamically generated logical qubit.

After each round of measurements, the ISG is FDLQC-equivalent to the 3D fTC. 
To verify this, we use entanglement renormalization group~\cite{Haah2014ERG,Dua2020ER} to construct an explicit circuit to map the ISGs to the canonical form of the 3D fTC, as shown in the supplementary \texttt{Mathematica} file. 
We note that, at a high level, the stabilizers of the check group generate a so-called anomalous 2-form symmetry~\cite{Yuan2019bosonization3D}, i.e., the symmetry operators (the stabilizers) are supported on loops and the point-like excitation at the endpoint of a truncated string operator have nontrivial (in this case fermionic) exchange statistics. This implies that each ISG has an anomalous 2-form symmetry, which is sufficient to guarantee that the ISGs necessarily have an emergent fermion.

The 3D fTC has three logical qubits on a system with periodic boundary conditions. However, given that our schedule inadvertently infers the measurement of the stabilizers supported on non-contractible paths along the $(1,0,0)$- and $(1,0,1)$-directions, our Floquet 3D fTC encodes a single logical qubit. The logical Pauli $Z$ operator can be represented by a product of 2-body check operators along the $(0,1,0)$-direction, as in Fig.~\ref{fig:3Dlogicals}(a). This can be interpreted as the nonlocal stabilizer of the check group that remains unmeasured throughout the measurement schedule. The logical Pauli $X$ operator can be represented by an operator supported on a membrane perpendicular to the $(0,1,0)$-direction [Fig.~\ref{fig:3Dlogicals}(b)]. 
We find that the sequence $(0,1,2,3,0)_\text{F}$ implements the trivial automorphism (and similarly for the B and R sequences). Thus, due to the rewinding nature of the FBFR schedule, the full 16-round period undergoes a trivial automorphism.

\section{Discussion}
\label{sec:discussion}
In this paper, we constructed examples of 3D topological Floquet codes with ISGs that are FDLQC-equivalent to the 3D TC, X-cube model, and 3D fermionic TC. We also constructed the 2D Floquet color code in Appendix~\ref{sec:Floquetcolorcode}. To construct the 3D Floquet codes, we utilized rewinding measurement schedules {and their description in terms of the evolution of condensation checks. As mentioned in Sec.~\ref{sec:intro}, we expect the principles behind our explicit microscopic examples to aid in the construction of a wider class of Floquet codes, including those with quantum LDPC codes as ISGs.}
Below, we comment on aspects of our Floquet codes that could lead to directions for future work.

\subsubsection{Local reversibility and possible fault-tolerance}
\label{subsubsec:localrev}
In Ref.~\cite{aasen2023measurement}, it was suggested that the local reversibility of a topological Floquet code could imply the existence of a non-zero threshold. We consider the implications of this conjecture for our Floquet codes. To do this, we first state some definitions from Ref.~\cite{aasen2023measurement} for completeness: If every consecutive pair of ISGs in a Floquet code form a locally reversible pair, then the code is said to be locally reversible. To define a locally reversible pair of ISGs, we first define a locally generated ISG. An ISG is --locally generated above a subgroup (LGAS)-- if there is a set of local operators that, along with the subgroup elements, form a generating set of the ISG. The ISGs in our Floquet codes are all locally generated above a subgroup, with the subgroup being the intersection of two ISGs. Now, two ISGs, let's say 0-ISG and 1-ISG, form a locally reversible pair if, (a) for both of them, there is a choice of local generating sets above the intersection (LGAI) of the two ISGs i.e., we have 0-LGAI and 1-LGAI respectively with the intersection subgroup 0-ISG$\cap$1-ISG, and if (b) one generator from 
0-LGAI (1-LGAI) anticommutes with exactly one generator from 1-LGAI (0-LGAI). Such anticommuting pairs are referred to as conjugate pairs. In Appendix~\ref{sec:Floquetcolorcode}, we show how local reversibility holds for our Floquet color code construction. Hence, based on the above-mentioned conjecture, we expect it to have a non-zero fault-tolerant threshold.

We expect the schedules for the 3D Floquet TC and 3D Floquet fTC to be also locally reversible. There are non-local stabilizers in the ISGs but they exist as stabilizers in the intersection of any consecutive pair of ISGs and do not preclude the existence of locally conjugate generating sets. Here, we discuss only the reversibility (instead of local reversibility) of ISGs in the 3D Floquet TC. We can consider the pair of ISGs, 3D TC in the G-round, and a subspace of two copies of 3D TCs in the B-round. Any string logical operator of G-ISG 3D TC is indeed also a logical string operator of the ISG of two copies of 3D TCs up to nonlocal stabilizers. Hence, the two ISGs share their full set of logical operators and hence, as defined in Ref.~\cite{aasen2023measurement}, this is a sufficient criterion for reversibility.

\subsubsection{Bifurcation of topological order under measurements}

In the 3D Floquet TC, we observed that as the condensation checks evolved into 4-qubit operators, we obtained the ISG of two copies of the 3D TC, up to nonlocal stabilizers. Starting from stacks of TC, if we are allowed measurement of higher-but-constant weight condensation checks, we can obtain a higher constant number of copies of 3D TC. It would be interesting to see if there is a Floquet code arising from coupled layers, such that the condensation checks always evolve by a \textit{constant} factor so that topological order self-bifurcates under every round of measurements~\footnote{Here, we distinguish between the splitting of topological order in our Floquet code examples and the bifurcation of topological order because the former did not happen recursively under the associated dynamics.}. A similar question holds for the bifurcation of the Floquet codes of fracton models under measurements as in the X-cube Floquet code; we observed a splitting of the X-cube topological order into the X-cube model and TCs.  
Bifurcation is already known to occur for fracton topological order under entanglement renormalization group~\cite{Haah2014ERG,Shirley2019ERG,Dua2020ER}. However, topological orders are conventional fixed points under the entanglement renormalization group. Thus, it is curious if, under specialized measurement sequences, conventional topological order such as the 3D TC can exhibit bifurcation. 
On the other hand, it would also be interesting if there is a fundamental obstruction to having such measurement dynamics.

\subsubsection{3D Floquet TC on a fractal}

In Sec.~\ref{subsec:3dFbTC_boundaries}, we discussed the planar realization of the 3D Floquet TC. 
It is known that in the 3D TC, one can punch holes with smooth boundaries to build a fractal lattice code embedded in 3D with fractal Hausdorff dimension $2<D_H<3$~\cite{zhufractal2022,dua2022quantum}. Starting with the G-ISG, we can, in principle, punch holes with smooth boundaries by truncating the checks across the red and blue edges. In subsequent rounds, as discussed in Sec.~\ref{subsec:3dFbTC_boundaries}, these boundaries always remain loop-condensing boundaries, although they may become loop-condensing boundaries of two copies of the 3D surface code (3D TC with boundaries) depending on the round. Thus, we expect our construction with holes to yield a Floquet code with ISGs that are fractal surface codes embedded in 3D~\cite{dua2022quantum,zhufractal2022}. It would be interesting to explore such Floquet codes on fractal lattices further for both memory and computation. 

\subsubsection{Classification of measurement schedules}

One fundamental question motivated by our work is what schedules exist for a given check group and what automorphisms of the logical information they exhibit. For instance, in Appendix~\ref{sec:Floquetcolorcode}, we wrote three schedules for the Floquet color code, one which exhibited a $\mathbb{Z}_3$ automorphism of logical operators, a rewound version that exhibited a trivial automorphism, and another 6-round schedule that exhibited a trivial automorphism. It would be interesting to determine whether there is a notion of equivalence between the two schedules with a trivial automorphism, such as the notion of simple equivalence discussed in Ref.~\cite{aasen2023measurement}. It would also be interesting to have a recipe to construct the inequivalent Floquet codes, given a set of checks on a lattice. For the 3D Floquet fTC, we presented a schedule with 16 rounds that preserved one logical qubit. A more efficient schedule could exist, one that preserves all logical qubits. However, our attempts with different schedules and on different trivalent lattices were unsuccessful. Hence, we leave this as a future direction. 

One potential source of inspiration for new schedules, although not periodic, comes from Refs.~\cite{sriram2022topology} and \cite{lavasani2022monitored}, which
considered weighted random measurements of $XX$, $YY$, and $ZZ$ checks in the 2D Floquet TC on the honeycomb lattice. In particular, they found a regime in which the topological information is protected. One could consider such schedules for our Floquet code examples, especially for the 3D Floquet fTC. We were not able to preserve all three logical qubits, but this may be possible with an appropriate random measurement schedule; similar to the 2D case, we expect that there is a regime in which all three logical qubits are preserved under the dynamics.

\vspace{3mm}

\vspace{2mm}
\begin{center}
\small{\MakeUppercase{\textbf{Acknowledgement}}}
\end{center}
\vspace{2mm}
We thank the authors of Ref.~\cite{davydova2023quantum} and Ref.~\cite{townsendteague2023floquetifying} for informing us about their works. N.T., J.S., and T.D.E. acknowledge the 2023 Boulder Summer School on Non-Equilibrium Quantum Dynamics, where part of this work was completed. 
A.D. is supported by the Simons Foundation through the collaboration on Ultra-Quantum Matter (651438, AD) and the Institute for Quantum Information and Matter, an NSF Physics Frontiers Center (PHY-1733907). N.T. is supported by the Walter Burke Institute for Theoretical Physics at Caltech. J.S. is supported by DOE DE-SC0022102. The $\texttt{Mathematica}$ codes used in this paper are available on GitHub at https://github.com/dua-arpit/floquetcodes. 

\clearpage

\appendix
\vspace{5mm}

\section{2D Floquet color code} \label{sec:Floquetcolorcode}

\begin{figure*}
    \centering
    \includegraphics[scale=0.09]{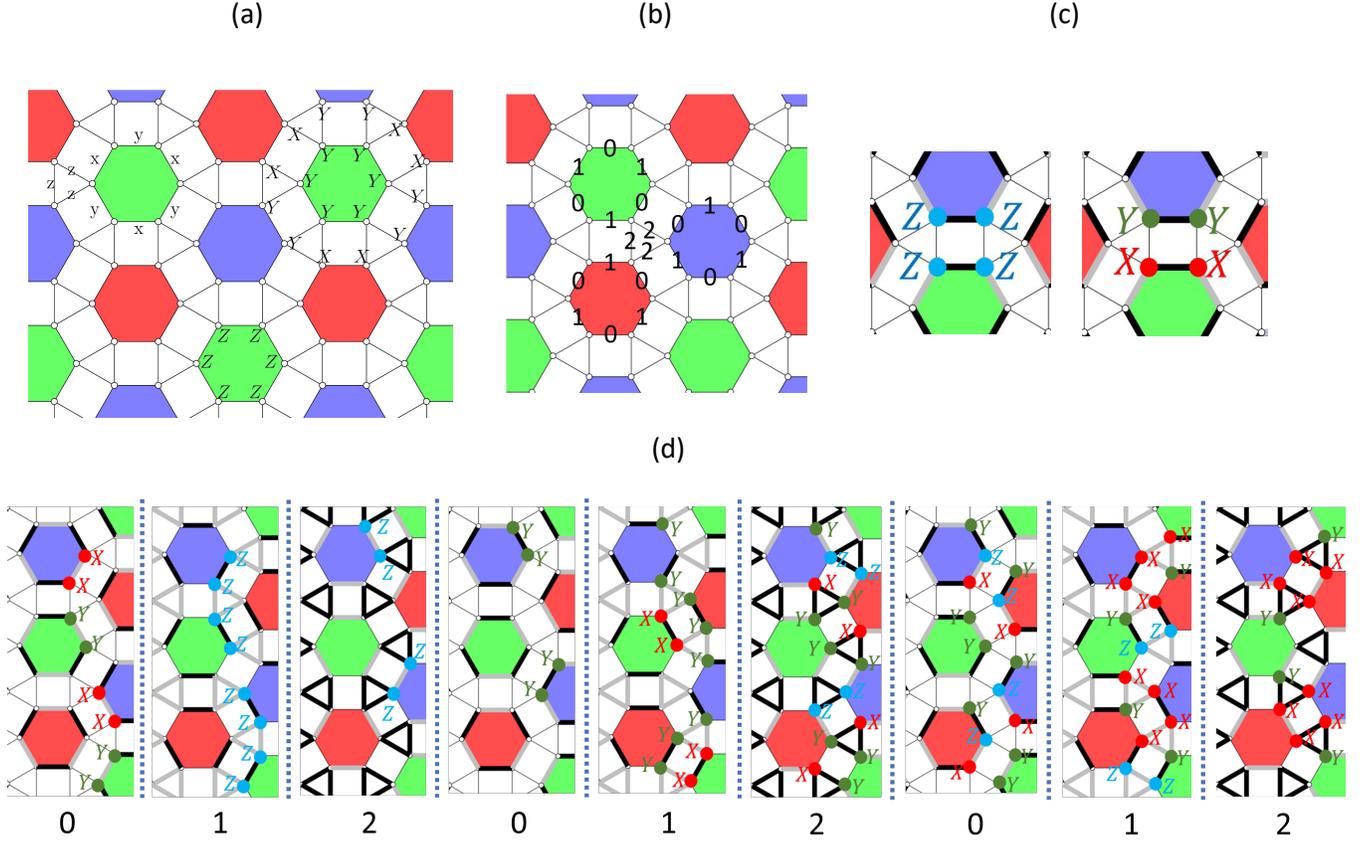}
    \caption{
    (a) The ruby lattice on which the Floquet color code is defined. The edges of the hexagon are labeled as $x$ and $y$, which correspond to the $XX$ and $YY$ check operators. All the edges connecting different colored hexagons are labeled as $z$ corresponding to a $ZZ$ check operator. The generators of the stabilizer group consist of two operators supported on every ``inflated'' hexagon. We show the two generators on two green inflated hexagons on the right. 
    (b) The 3-round schedule of measurements of Floquet color code; all the checks on the triangle are measured in 2-round but only some are shown for clarity. (c) Instantaneous square stabilizers in rounds 0 and 1 that form from products of checks in preceding rounds. These square stabilizers are significant in the effective measurement or inference of the stabilizers on the green inflated hexagon and in working out the automorphism of logical string operators. In (c) and (d), the checks of the current round are shown in black and the checks of the next round are shown in gray.
    (d)
    $\mathbb{Z}_3$ automorphism of dynamically generated logical operators of the Floquet color code.}
\label{fig:colorcode_schedule_automorphism}
\end{figure*}
In this appendix, we construct the 2D Floquet color code, whose ISGs are FDLQC-equivalent to the 2D color code. We then discuss the $\mathbb{Z}_3$ automorphism of logical operators and the construction of a rewinding schedule that trivializes the automorphism. 

The Floquet color code is based on the topological subsystem code of Ref.~\cite{Bombin2010Subsystem}, which is defined 
on the ruby lattice with a qubit at each vertex, as depicted in Fig.~\ref{fig:colorcode_schedule_automorphism}(a). The group of check operators is generated by two-qubit operators, specified by assigning a label $x$, $y$, or $z$ to each edge of the lattice. The corresponding check operators on the $x$-, $y$-, and $z$-edges are the two-qubit Pauli operators, $XX$, $YY$, and $ZZ$, respectively.\footnote{Through out this paper, we omit the tensor product between the Pauli operators for simplicity of notation. For example, we write $X\otimes X$ as $XX$.}  

The stabilizers of the check group are generated by two types of operators, both of which are supported on each ``inflated'' hexagon of the ruby lattice, as shown in Fig.~\ref{fig:colorcode_schedule_automorphism}(a). We call the generator that is a product of $Z$ operators around a hexagon, the hexagon stabilizer, and the generator that is a product of $X$ and $Y$ operators the inflated hexagon stabilizer.

\subsection{3-round measurement schedule}
The measurement schedule of the Floquet color code consists of three rounds, labeled 0, 1, and 2, with the checks of each round shown in Fig.~\ref{fig:colorcode_schedule_automorphism}(b). We note that this schedule is also given in Ref.~\cite{Sarvepalli2012} in the context of the topological subsystem code.\footnote{Although the schedule has appeared before in the literature, here, we view the code as a Floquet code. The important difference is that the Floquet code has dynamically generated logical qubits in addition to the logical qubits of the subsystem code.} This schedule ensures that the stabilizers of the check group are inferred once every cycle. 

In each round, the ISG is FDLQC-equivalent to the color code. We demonstrate this for round 2 by showing that the ISG is precisely the color code concatenated with the 3-qubit repetition code. 
In round 2, the measured checks consist of $ZZ$ operators on every edge of the triangles. Since there are three qubits and two independent checks per triangle, an effective qubit (3-qubit repetition code) lives on each triangle with effective Pauli operators
\begin{align}
  X_{\text{eff}} &= XYY \equiv XXX &   Z_{\text{eff}} = ZII \equiv ZZZ
  \label{eq:concatcolorcode}
\end{align}
where $\equiv$ is the equivalence up to the 2-qubit $ZZ$ checks. In terms of the effective Pauli operators, the instantaneous stabilizers reduce to those of the color code. Therefore, the 2-round ISG is the color code up to concatenation with 3-qubit repetition codes. For rounds 0 and 1, the FDLQCs that map the ISGs to the color code are given in the \texttt{Mathematica} file.

As an aside, the logical gates that can be implemented transversally for the (CSS) color code~\cite{Yoshida2015,kubica2015}, can still be implemented transversally in round 2 when the ISG is the color code up to concatenation with 3-qubit repetition codes. If we label the four logical qubits 1 through 4, then the transversal logical gates for a single copy of color code (in particular choice of basis) are $\overline{\text{CNOT}_{12}\text{CNOT}_{34}}$ and $\overline{\text{SWAP}_{12}\text{SWAP}_{34}}$.
To realize the former, we act with $S_{\text{eff}} = SII$ on one orientation of triangles (say, left-pointing) and $S^\dagger_{\text{eff}} = S^\dagger II$ on the other orientation (right-pointing) after the second round of measurements. Similarly, we can act with an effective Hadamard gate $H_{\text{eff}} = (X_{\text{eff}}+ Z_{\text{eff}})/\sqrt{2}$ on all triangles to realize the latter logical gate.

Since the logical information is preserved under the Floquet code dynamics, these gates can be done in 2-round and the processed logical information carries over to subsequent rounds. We note that a planar realization of the color code would allow for a transversal implementation of all logical Clifford gates~\cite{bombin2015gauge,kubica2015}. We leave the construction of a planar variant of our Floquet color code to future work.

\subsection{$\mathbb{Z}_3$ automorphism of logical operators}
Given that the ISGs are FDLQC-equivalent to the color code, the Floquet color code encodes four logical qubits on a torus. This implies that the Floquet code has two dynamically generated logical qubits since the topological subsystem code of Ref.~\cite{Bombin2010Subsystem} hosts only two logical qubits. We refer to the logical operators of the subsystem code as the static logical operators and the logical operators, which are neither static logical operators nor non-local stabilizers of the check group, as the dynamically generated logical operators~\footnote{The static subsystem code is characterized by the 3-fermion anyon theory~\cite{Bombin2009fermions,Bombin2012universal,Bombin2014structure, Ellison2022subsystem}; thus, the dynamically generated logical operators are associated with a second copy of the 3-fermion theory. The two 3-fermion theories together are equivalent to the color code.}. Indeed, since the static logical operators have representations that commute with all check operators, they do not evolve under the dynamics. As discussed below, there is a nontrivial automorphism of the dynamically generated logical operators after a full period. This automorphism happens through multiplication with dynamically generated logical operators that belong to the check group but do not commute with all the elements of the check group. This is in contrast to the 2D Floquet TC, where the automorphism happens due to multiplication with a dynamically generated logical operator that belongs to the center of the check group and has been referred to as the inner logical in Ref.~\cite{hhdynamic2021}.

For the 3-round Floquet cycle, the dynamically generated logical operators exhibit a $\mathbb{Z}_3$ automorphism, as shown in Fig.~\ref{fig:colorcode_schedule_automorphism}(d). The logical operator representation in each round $r$ commutes with the ISGs of rounds $r$ and $r+1$. The logical operator representation in each round $r$ is related to the one in the preceding round $r-1$ via multiplication with the ISG stabilizers of round $r$. Besides the checks measured in round $r$, we used the instantaneous square stabilizers shown in  Fig.~\ref{fig:colorcode_schedule_automorphism}(c) to work out the representations. 
Since there are no nonlocal stabilizer generators of the check group, the automorphism occurs due to multiplication with products of dynamically generated logical operators belonging to the check group. In other words, the product of the check operators that get multiplied after a full cycle is a logical of the ISG but not a static logical or a non-local stabilizer of the check group. Due to the $\mathbb{Z}_3$ automorphism that occurs every three rounds, it takes nine rounds for a trivial automorphism.

To see the automorphism explicitly, we can apply the mapping to effective qubits in round 2 (Eq.~\eqref{eq:concatcolorcode}) to the string operators shown in Fig.~\ref{fig:colorcode_schedule_automorphism}(d). In the first instance of round 2, we find that the logical consists of $Z_{\text{eff}} =ZII$ on each triangle connecting the red plaquettes. If we truncate such a string in the effective color code pictures, the $X$ and $Y$ red plaquettes of the effective color code will be violated. In the second instance of round 2, we have a string of $X_{\text{eff}} = XYY$ on each triangle connecting blue plaquettes. Thus, truncating such a string will violate the $Y$ and $Z$ blue plaquettes of the effective color code. Lastly, in the third instance of round 2, the strings are a product of $Y_{\text{eff}} =XXY$ connecting green plaquettes. Truncating these strings will violate the $X$ and $Z$ green plaquettes of the effective color code.

\subsection{6-round rewinding schedule}
Instead of trivializing the automorphism through 9 measurement rounds, we can use the 6-round rewinding schedule 012102, in which the $\mathbb{Z}_3$ automorphism action after 012 is rewound back. The evolution of logical operator representations is the same as shown in Fig.~\ref{fig:colorcode_schedule_automorphism}(c) for any three consecutive rounds with 0-, 1- and 2-checks. In principle, this rewinding schedule enables the construction of the Floquet color code on a triangular geometry, such as in Refs.~\cite{bombin2015gauge,kubica2015}. 
Below, we write another 6-round schedule for the Floquet color code, which exhibits a trivial automorphism. It would be interesting to see if that 6-round schedule is equivalent to the above rewinding schedule using some notion of equivalence, such as simple equivalence discussed in Ref.~\cite{aasen2023measurement}.

\subsection{Condensation picture}

\begin{figure}[]
    \centering
    \includegraphics[scale=0.1]{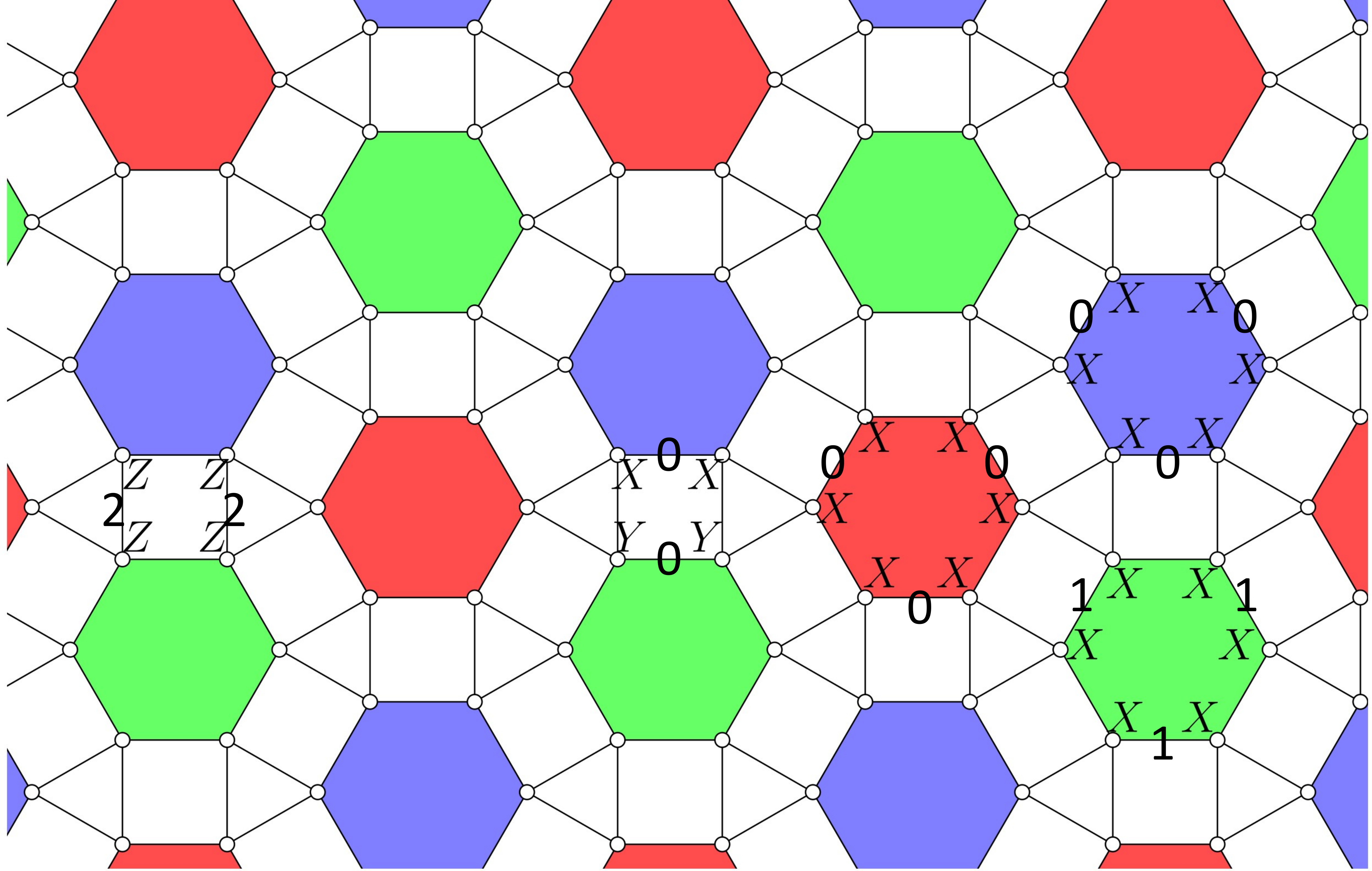}
    \caption{{Stabilizers in the parent stabilizer code for the Floquet color code. The stabilizers shown and the static stabilizers shown in Fig.~\ref{fig:colorcode_schedule_automorphism}(a) form the stabilizer generators of the parent code. 
    We show how these stabilizers are products of checks of rounds 0, 1 and 2 around closed loops. We have 6-body X stabilizers around the red and blue hexagons, respectively, formed from products of 0-checks. We have 6-body X stabilizers around the green hexagons made from products of 1-checks. We have 4-body stabilizers of two kinds on the squares between green and blue hexagons. One kind is a product of 0-checks, and the other one is a product of 2-checks, as shown. 
    The resulting parent stabilizer code is FDLQC-equivalent to two copies of the color code.}}
    \label{fig:parent_color_Floquet}
\end{figure}

\begin{figure}
    \centering
    \includegraphics[scale=0.085]{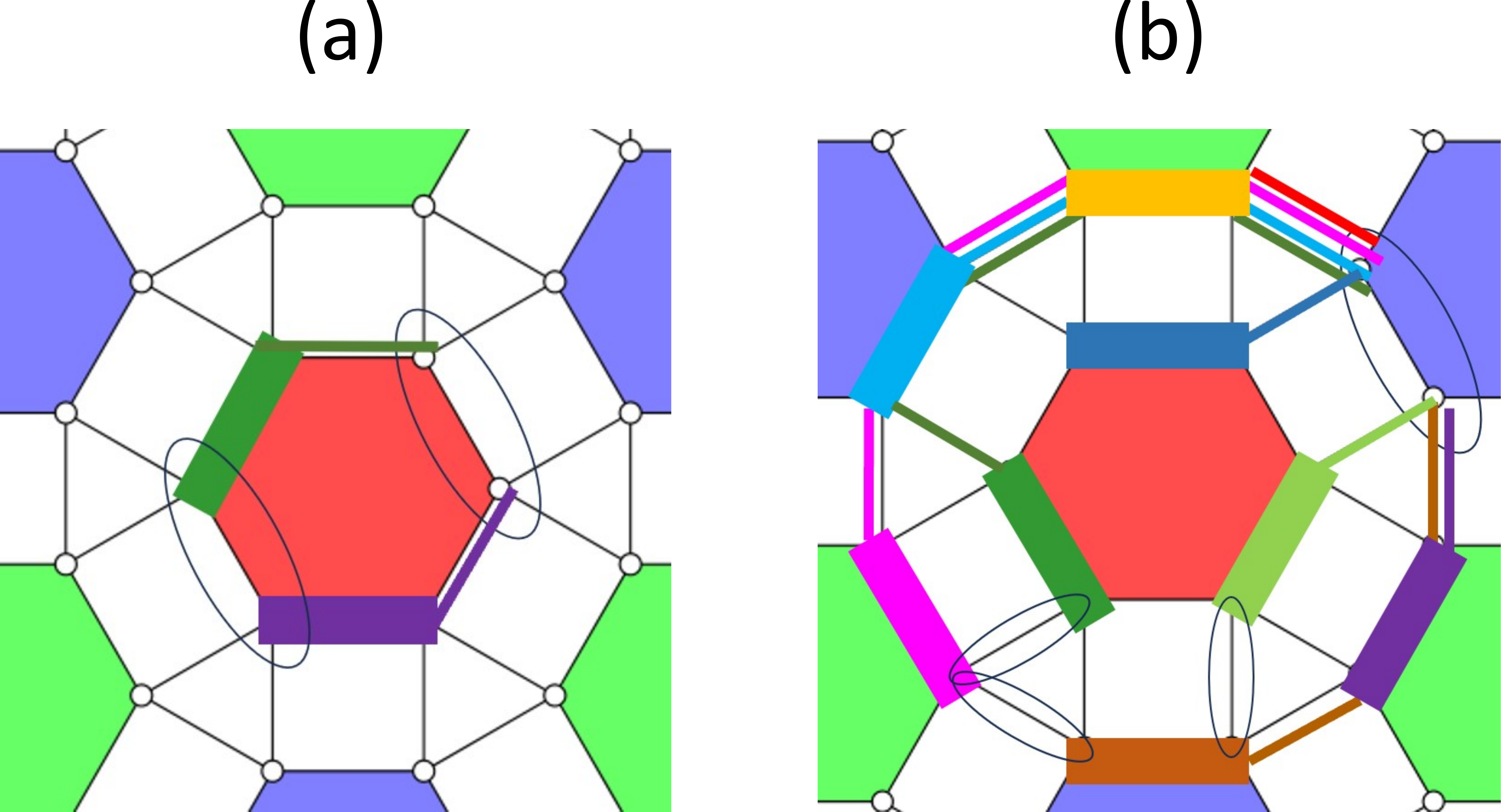}
    \caption{(a) Local reversibility of ISGs in rounds 0 and 1 of Floquet color code. The 0-checks on the red hexagon are marked using thick edges. Their conjugate partner 1-checks are marked using a thin edge of the same color as the 0-check. The circled checks are the checks that we do not need to consider to form the pairs of conjugate local operators \textit{above} the intersection of 0-ISG and 1-ISG; the intersection includes the product of 0-checks and the product of 1-checks around the hexagon. (b) Local reversibility of ISGs in rounds 1 and 2 of Floquet color code. The 1-checks on the red hexagon are marked using thick edges. Their conjugate partners are products of 2-checks that are marked using thin edges of the same color as the 1-check. We do not need to consider the circled checks in the pairs of conjugate local operators above the intersection of 1-ISG and 2-ISG; the intersection contains the product of all 1-checks on the red inflated hexagon and the inflated hexagon stabilizers of the check group.}
\label{fig:localreversibility}
\end{figure}

It is insightful to understand the Floquet codes in this work in terms of sequences of condensations in a parent stabilizer code~\cite{Kesselring2022condensation}. As mentioned in Sec.~\ref{subsec:condensation}, a natural generating set for the stabilizer group of the parent stabilizer group is the stabilizers of the check group and closed loops of checks from each round that commute to form a stabilizer group. For the Floquet color code, we write down a parent stabilizer code, as shown in Fig.~\ref{fig:parent_color_Floquet}. Since none of the individual ISG checks are included in the parent stabilizer code, the measurements of the check operators can then be interpreted as condensing an excitation of this parent stabilizer code, resulting in the ISGs of the Floquet code. 
The parent code shown in Fig.~\ref{fig:parent_color_Floquet} is FDLQC-equivalent to two copies of color code or four copies of TC. The exact circuit to map this code to four copies of the TC is given in the supplementary \texttt{Mathematica} files.}

\subsection{Local reversibility}

In Sec.~\ref{subsubsec:localrev}, we mentioned the definition of local reversibility used in Ref.~\cite{aasen2023measurement} and the conjecture that it could imply a non-zero fault-tolerant threshold. We now discuss the local reversibility of the Floquet color code. The reader is advised to check Sec.~\ref{subsubsec:localrev} for definitions. In the Floquet color code, the intersection 0-ISG$\cap$1-ISG contains the product of 0-checks and the product of 1-checks around each hexagon. Hence, we can remove one 0-check and 1-check from our local generating sets. 
The removed checks and conjugate pairs for 0- and 1-ISGs are illustrated in Fig.~\ref{fig:localreversibility}(a). 
The pairs of ISGs in rounds 1 and 2 as well as rounds 2 and 0 are also locally reversible. We now explain this for a pair of rounds 1 and 2. The intersection 1-ISG$\cap$2-ISG includes the product of 1-checks on the inner and outer boundary of the red-inflated hexagons and the stabilizers of the check group on each of the three inflated hexagons. Hence, we remove one 1-check and three 2-checks on the red-inflated hexagon from our consideration because we need to form the conjugate pairs only above the intersection. Then, the conjugate pairs are the products of 2-checks in a chain whose one end intersects with its conjugate partner 1-check in one qubit and the other end intersects the removed 1-check. The removed checks and conjugate pairs for 1- and 2-ISGs are illustrated in Fig.~\ref{fig:localreversibility}(b).

\subsection{Alternative 6-round Floquet color code schedule with trivial automorphism} 

\begin{figure*}
    \centering

    \sidesubfloat[]{\includegraphics[scale=0.13]{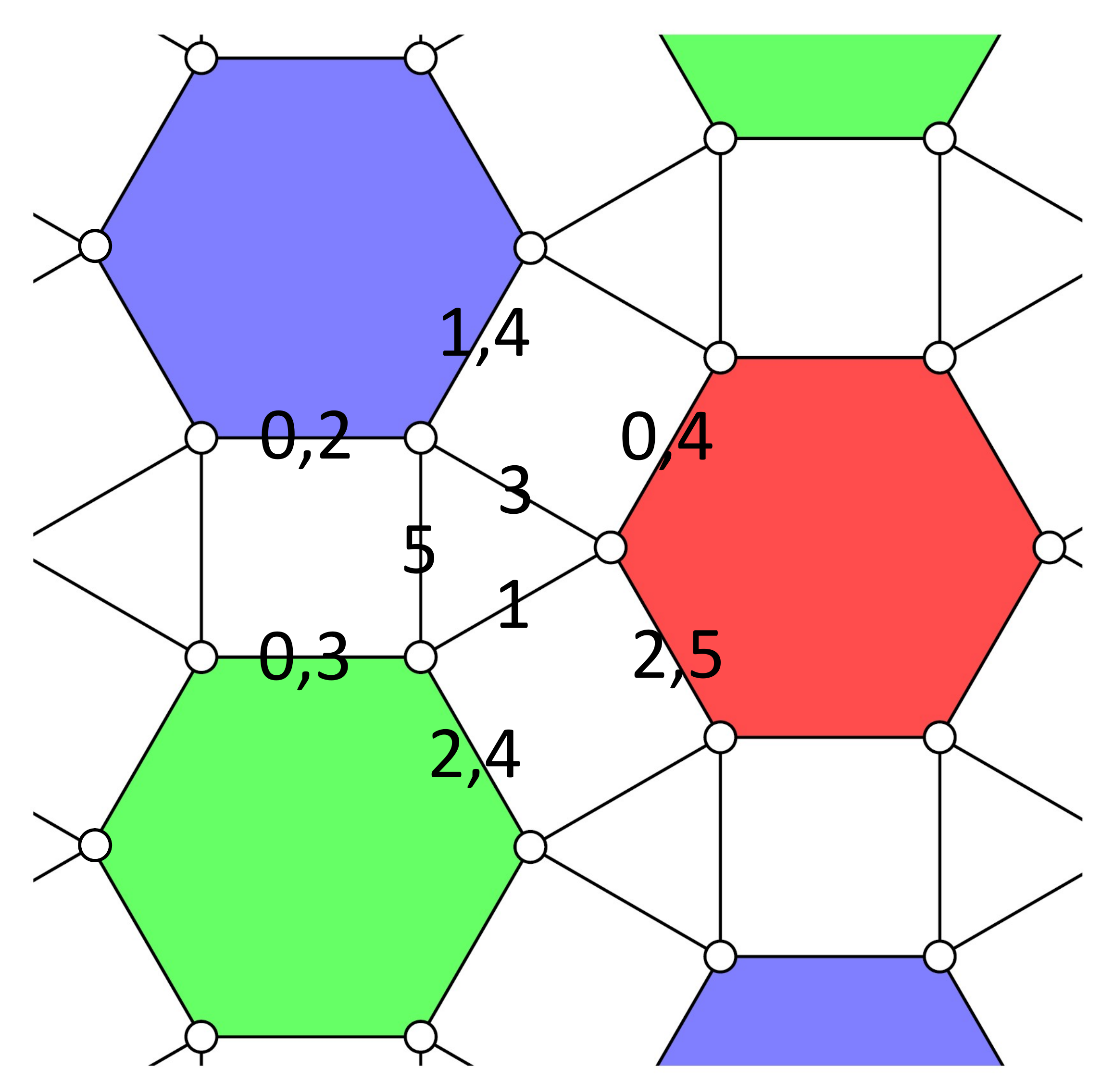}}
    
  \sidesubfloat[]{ \includegraphics[scale=0.45]{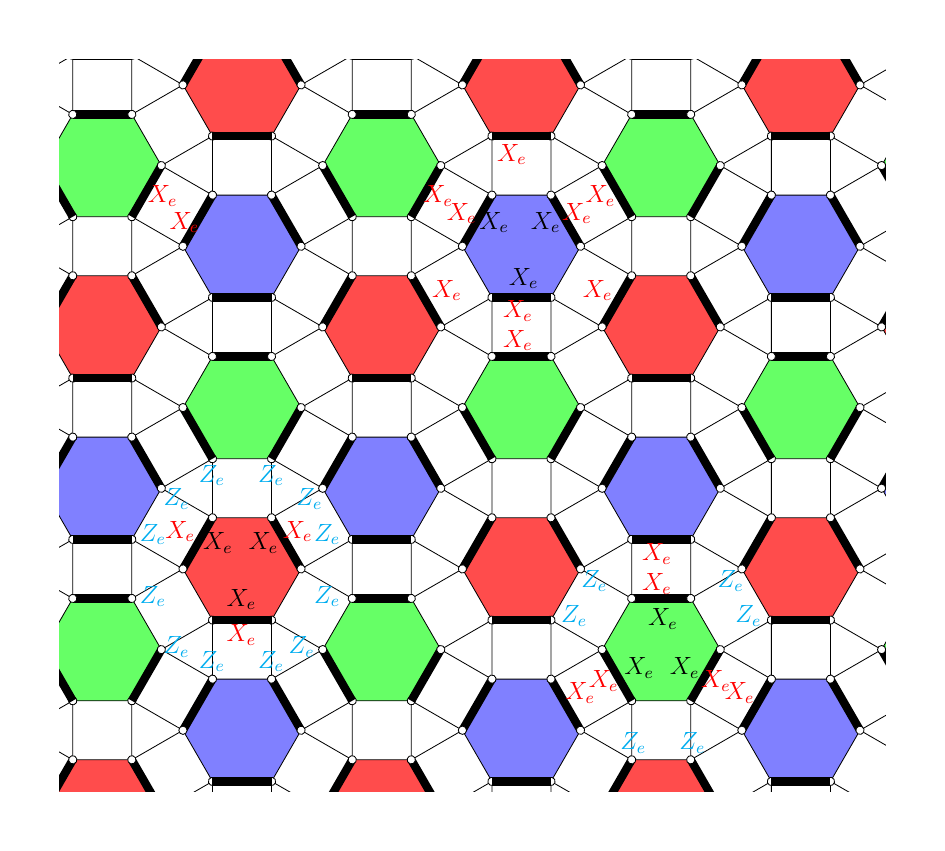}
    \includegraphics[scale=0.45]{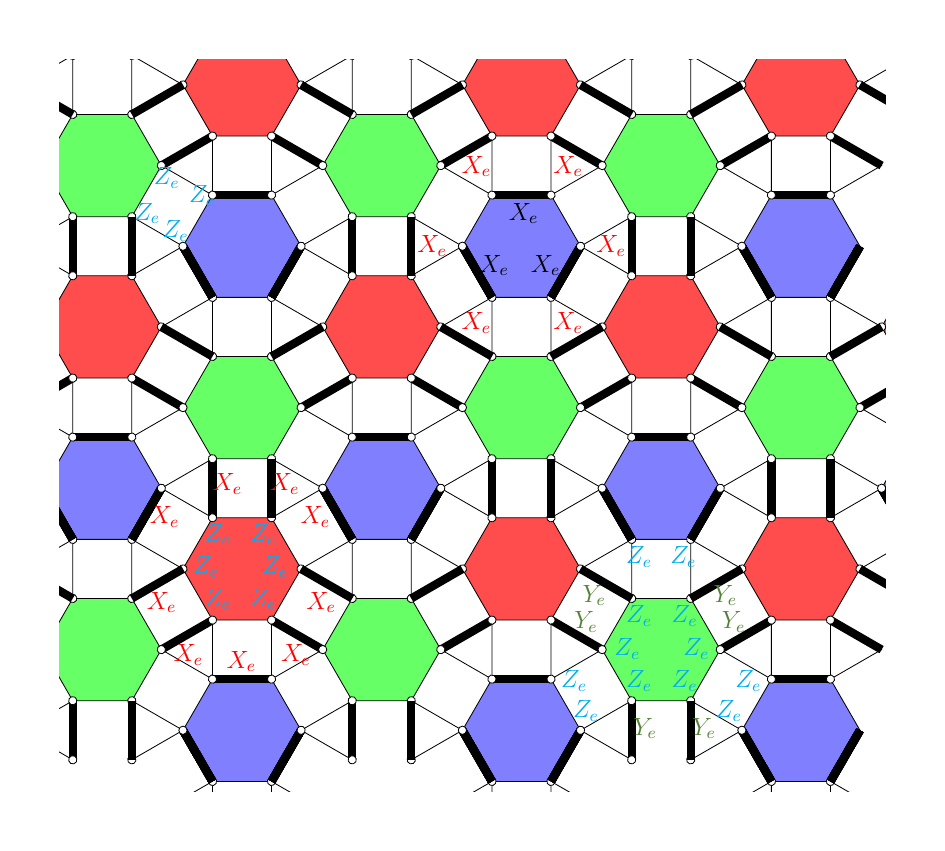}}\\
    \sidesubfloat[]{
\includegraphics[scale=0.12]{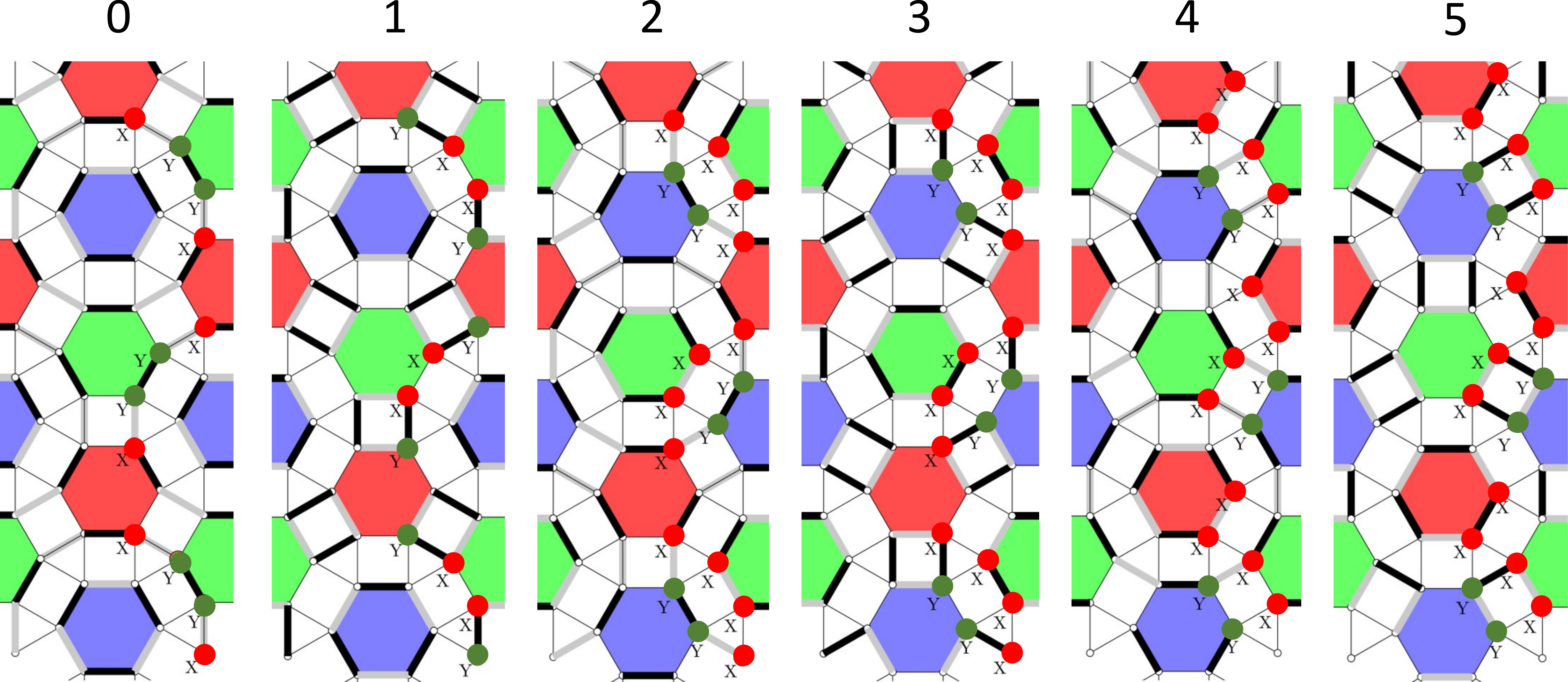}}
    \caption{(a) 6-round schedule that has the trivial automorphism of dynamically generated logical operators. (b) Effective color code ISG stabilizers after measurements of 0-round and 1-round, respectively, in the six-round Floquet color code with trivial automorphism. The thick black edges indicate the check operators measured in the current round. The six stabilizer generators, two for each inflated hexagon, are shown in terms of effective edge logical operators $X_e$ and $Z_e$. On an edge with $XX$ check, $X_e= ZZ, Z_e= XI$, on an edge with $YY$ check, $X_e= XX, Z_e= YI$ and on an edge with $ZZ$ check, $X_e= XX, Z_e= ZI$. The hexagon stabilizers are shown inside the hexagons, while all the Pauli operators outside the hexagon form the inflated hexagon stabilizer. Left: For 0-round, the stabilizer shown in brown is a $ZZZZ$ operator left invariant from the previous round of measurements and is expressed as $X_e X_e$ in terms of the effective edge Pauli operators. Right: For 1-round, the stabilizer shown in brown is a $ZZYY$ operator left invariant from the product of $ZZZZ$ and $IIXX$ measured in the previous two rounds of measurements; it is expressed as $Z_e Z_e Z_e Z_e$ in terms of the effective edge Pauli operators. (c) The trivial automorphism of a logical operator in the 6-round Floquet color code. The thick black edges indicate the check operators measured in the current round, and the gray edges indicate the checks measured in the next round.}
    \label{fig:6stepscheduleandeffcolorcode}
\end{figure*}

We now specify a variant of the Floquet color code that exhibits a trivial automorphism of the logical operator representations but is not a rewinding schedule. The schedule is depicted in Fig.~\ref{fig:6stepscheduleandeffcolorcode}(a). This schedule is designed so that the hexagon and inflated hexagon stabilizers of the blue plaquettes can be inferred from rounds 0 and 1. Similarly, the stabilizers associated with the green and red hexagons and inflated hexagons can be inferred from the 2,3 and 4,5 measurement rounds, respectively.

The ISGs after rounds 0 and 1 are shown in Fig.~\ref{fig:6stepscheduleandeffcolorcode}(b). The ISGs for rounds 2,3,4,5 can be determined by applying the lattice symmetries. Using entanglement renormalization, we find that each ISG is FDLQC-equivalent to the color code. The explicit FDLQCs, written using the polynomial representation~\cite{haah2013lattice,Dua2020ER}, can be found in the supplementary \texttt{Mathematica} files~\cite{FloquetSM}. 

The trivial automorphism of the logical operators is shown in Fig.~\ref{fig:6stepscheduleandeffcolorcode}(c).

\section{Counting of logical qubits in the 3D Floquet TC}
\label{sec:counting}

\begin{figure*}
    \centering
\includegraphics[scale=0.15]{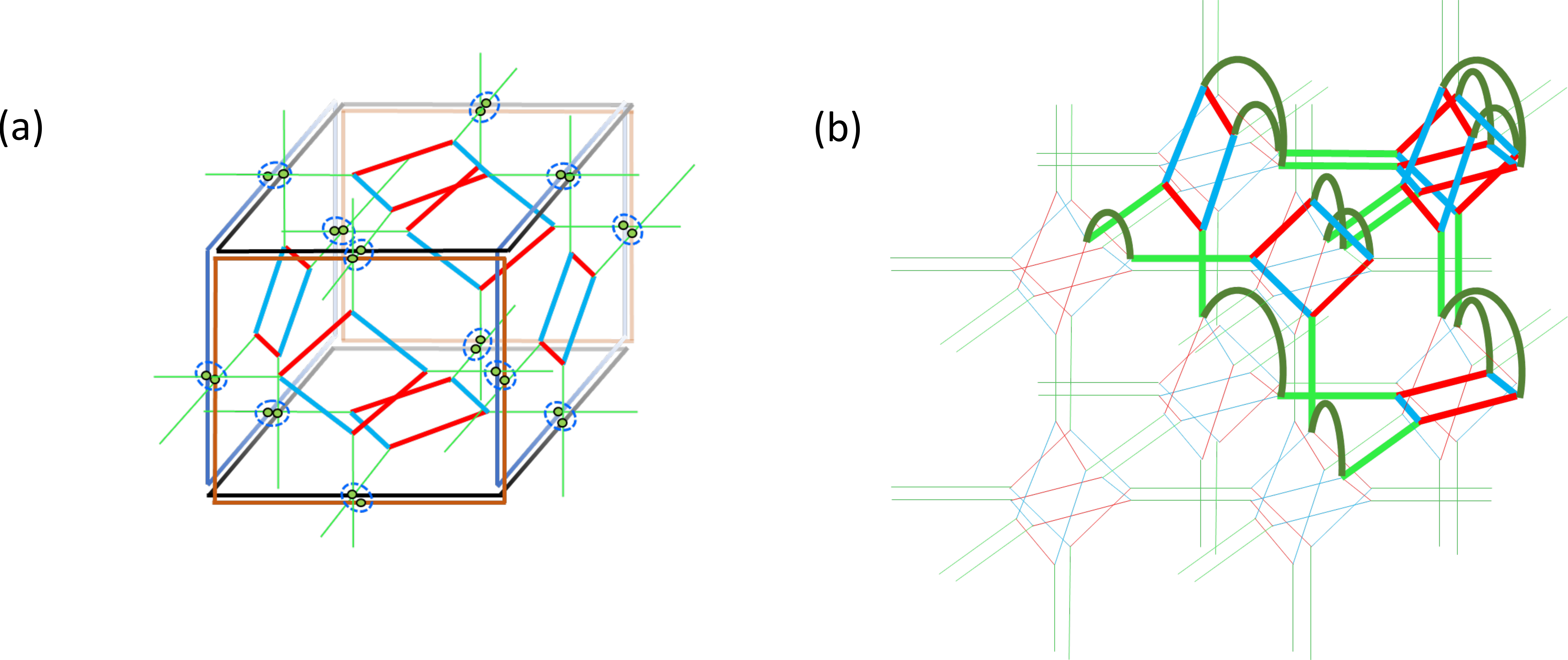}\hspace{5mm}
    \caption{(a) A local relation in the 3D TC stabilizer group obtained from condensation in stacks of square lattice TC; the product of the $ZZZZ$ plaquette operators of 3D TCs on the six plaquettes around a cube and the $ZZ$ condensation checks on the encircled pairs of qubits is equal to the identity. To understand the local relation in the G-ISG of the Floquet code, we also show the square octagon plaquettes corresponding to plaquettes involved in the local relation shown in (a).  (b) The corresponding local relation in the G-ISG of 3D Floquet TC on the associated lattice (stacks of the 2D square-octagon lattices). It is a product of condensation checks (shown using thick dark green arcs), green checks (shown in thick green), and the square plaquette operators (which are products of red and blue checks highlighted using thick lines).}
    \label{fig:localrelation}
\end{figure*}

The counting of logical qubits in the G-ISG is straightforward due to its mapping to the cubic lattice 3D TC up to concatenation with 2-qubit repetition codes. The local relations among plaquette stabilizers on the cubes in the cubic lattice 3D TC map to the local relation among checks as shown in Fig.~\ref{fig:localrelation}. Here, we state the counting of logical qubits in the cubic lattice 3D TC, which implies the same for the G-ISG. On a torus with linear size $L$, there are $3L^3$ physical qubits, $3L^3$ plaquette stabilizers with $L^3-2$ independent local relations among them, $L^3$ vertex stabilizers with $1$ global relation among them. Thus, we get three logical qubits.   

For the B-ISG, the evolved condensation checks are either local operators, as shown in Fig.~\ref{fig:evolution_condensation_op_3DTC} or nonlocal stabilizers, as shown in Fig.~\ref{fig:nonlocalstab}. We also have blue-green plaquettes as stabilizers in the ISG. Besides that, we have the stabilizers of the check group and the blue checks as stabilizers of the ISG. It is again useful to work in the effective picture of a 3-foliated stack of 2D rotated TCs and consider the condensation checks and nonlocal stabilizers on top of that. We start with the stack of 2D rotated TCs with periodic boundary conditions, and we now have the $X^{\otimes 4}$ condensation checks in the configuration as shown in Fig.~\ref{fig:sublattices_BISG}(a).
On every cube of this lattice, we have a $Z$ stabilizer as described above, i.e., it is the product of $Z$ stabilizers among three plaquettes sharing the canonical corner associated with the cube. These $Z$ stabilizers correspond to the vertex operators of the two copies of 3D TCs. We assume the linear system size $L$ to be even here and in the discussion on automorphism below. This is because in the B-ISG, having an odd number of cubes on a torus will lead to a spatial defect in the configuration of condensation checks shown in Fig.~\ref{fig:sublattices_BISG}(a), and we avoid such a scenario for simplicity. There are $L^3$ cubes and thus $L^3$ such $Z$ stabilizer operators. There are two relations among these $Z$ stabilizers, one on each sublattice, leaving us with $L^3-2$ independent $ Z$ stabilizers. There are $3L^2$ $X$ plaquette stabilizers and $3L \times L^2/4$ condensation stabilizers, where $3L$ is the number of planes. The number of relations among the $X$ stabilizers is $L^3/4+4$. The form of relations is complicated and is shown in the supplementary \texttt{Mathematica} file. 
Considering the three nonlocal stabilizers, we get $3$ logical qubits for the B-ISG. The B-ISG can be mapped via an explicit circuit to two copies of 3D TCs up to nonlocal stabilizers; see supplementary \texttt{Mathematica} file. The counting of logical qubits in the other ISGs is similar to the counting for G-ISG or B-ISG.

\section{Effective description of B-ISG of the X-cube Floquet code}
\label{sec:eff_description_and_counting_BISG_XC}
\begin{figure*}
    \centering
    \includegraphics[scale=0.165]{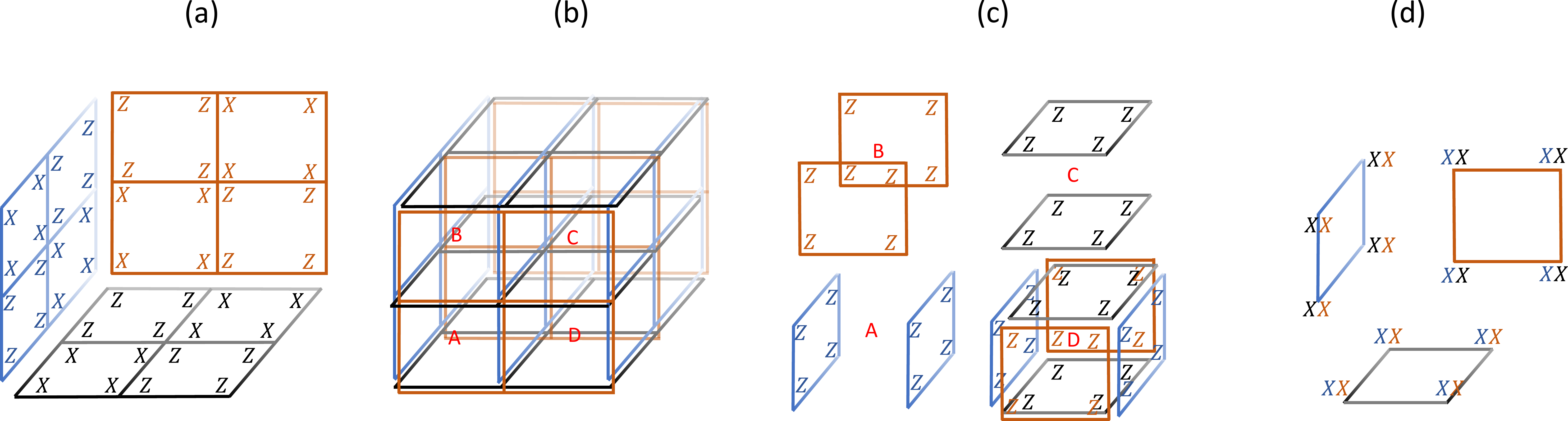}
    \caption{Effective description of the B-ISG of the X-cube Floquet code in terms of rotated 2D TC layers and evolved condensation checks that couple them. (a) The rotated 2D TC layers are shown along with the plaquette stabilizers. (b) The unit cell of the effective 3D lattice. There are four kinds of cubes labeled A, B, C and D. (c) For each cube, the product of $Z$-plaquette stabilizers of 2D TCs around it forms a stabilizer of the B-ISG. (d) The evolved condensation checks in the B-ISG are $X$-stabilizer operators. They are defined on the plaquettes where $Z$-stabilizers of 2D TC live but act on the qubits in the foliations orthogonal to the plaquette; see the difference of color in the Pauli operators and the edges of the plaquette.}
    \label{fig:B-ISG_stabilizers}
\end{figure*}

To calculate the number of logical qubits explicitly using the relations of stabilizers in the B-ISG, we use the effective description of a 3-foliated stack of rotated 2D TCs. The stabilizers of the B-ISG, including the evolved condensation checks, are shown in Fig.~\ref{fig:B-ISG_stabilizers}. 
The X-stabilizers of the rotated 2D TC layers are also $X$-stabilizers of the B-ISG. The product of Z-stabilizers of 2D TC layers around each cube of the 3D lattice forms a stabilizer of the B-ISG. Besides these, on every plaquette corresponding to the $Z$-stabilizer of the 2D TC layer, there lives a $X$-stabilizer acting on the qubits of the orthogonal foliations. These are the evolved condensation checks. 

For counting of logical qubits, we consider linear system size $L=2n$, where $L$ is the number of cubes along each lattice direction since the unit cell as shown in Fig.~\ref{fig:B-ISG_stabilizers}(b) consists of even number of cubes along each direction. We have $3L^3/2$ $X$-plaquette stabilizers, $L^3$ $Z$-cubic stabilizers and $3L^3/2$ evolved condensation stabilizers. This gives a total of $4L^3$ stabilizers. 
For the $X$-plaquette stabilizers of the 2D TCs, there is a relation in each plane, and hence, there are $3L$ planar relations. Besides these, we have $L^3$ local relations among the $X$-plaquette stabilizers and evolved condensation checks, which are also $X$-type stabilizers. There is a global relation formed from the product of cubic $Z$-stabilizers of each type (A, B, C, and D, respectively). However, only 3 of them are independent. There is a relation among the cubic stabilizers on every dual lattice plane. Considering the aforementioned global relations, there are $3(L-1)$ such independent planar relations. Thus, there are $3L+3+3(L-1)$ global relations and $L^3$ local relations, giving overall $6L+L^3$ relations among the stabilizers. 
Considering $3L^3$ physical qubits, we have $6L$ logical qubits. 

Due to the three non-local stabilizers as described in the main text, we have $6L-3$ logical qubits that are actually preserved. This is expected since the preceding ISG (G-ISG) is the X-cube model (up to concatenation) and has $6L-3$ logical qubits. 

\bibliographystyle{apsrev4-2}
\bibliography{references}
\end{document}